\documentclass[twocolumn]{aastex631}

\def\chandra{{\itshape Chandra\/}}
\def\erosita{{\itshape eROSITA\/}}

\def\astrosat{{\itshape AstroSAT\/}}
\def\hst{{\itshape HST\/}}
\def\jwst{{\itshape JWST\/}}
\def\spitzer{{\itshape Spitzer\/}}
\def\swift{{\itshape Swift\/}}
\def\herschel{{\itshape Herschel\/}}
\def\wise{{\itshape WISE\/}}
\def\galex{{\itshape GALEX\/}}
\def\xmm{{\itshape XMM-Newton\/}}

\def\xray{\hbox{X-ray}}
\def\pegase{{\ttfamily P\'EGASE}}
\def\bpass{{\ttfamily BPASS}}

\def\lgoh{$12 + \log({\rm O}/{\rm H})$}
\def\ltsima{$\; \buildrel < \over \sim \;$}
\def\simlt{\lower.5ex\hbox{\ltsima}}
\def\gtsima{$\; \buildrel > \over \sim \;$}
\def\simgt{\lower.5ex\hbox{\gtsima}}
\def\kms{\ifmmode{~{\rm km~s^{-1}}}\else{~km s$^{-1}$}\fi}
\def\lsim{\lower0.3em\hbox{$\,\buildrel <\over\sim\,$}}
\def\gsim{\lower0.3em\hbox{$\,\buildrel >\over\sim\,$}}

\def\flux{erg~cm$^{-2}$~s$^{-1}$}
\def\lum{erg~s$^{-1}$}

\def\sfr{$M_{\odot}$~yr$^{-1}$}

\def\aap{A\&A}

\def\apjl{ApJL}
\def\apjs{ApJS}

\def\lightning{{\ttfamily Lightning\/}}
\def\multilightning{{\ttfamily MultiLightning\/}}

\makeatletter
\DeclareRobustCommand{\HII}{%
  \mbox{H\check@mathfonts\fontsize\sf@size\z@\selectfont II}%
}
\makeatother

\def\ngal{88}
\def\nxps{6,432}
\def\nx{3,731}
\def\nrem{27}

\def\nms{63}
\def\pnull{0.047}

\usepackage{underscore}
\usepackage{longtable}
\usepackage{amsmath}
\usepackage{enumitem}
\usepackage{hyperref}

\begin{document}

\title{{\large An Empirical Framework Characterizing the Metallicity and Star-Formation History Dependence of X-ray Binary Population Formation and Emission in Galaxies}}

\correspondingauthor{Bret Lehmer}
\email{lehmer@uark.edu}

\author[0000-0003-2192-3296]{Bret~D.~Lehmer}
\affiliation{Department of Physics, University of Arkansas, 226 Physics Building, 825 West Dickson Street, Fayetteville, AR 72701, USA}
\affiliation{Arkansas Center for Space and Planetary Sciences, University of Arkansas, 332 N. Arkansas Avenue, Fayetteville, AR 72701, USA}

\author[0000-0001-8473-5140]{Erik~B. Monson}
\affiliation{Department of Astronomy and Astrophysics, 525 Davey Lab, The Pennsylvania State University, University Park, PA 16802, USA}

\author[0000-0002-2987-1796]{Rafael~T.~Eufrasio}
\affiliation{Department of Physics, University of Arkansas, 226 Physics Building, 825 West Dickson Street, Fayetteville, AR 72701, USA}

\author[0000-0002-8553-1964]{Amirnezam Amiri}
\affiliation{Department of Physics, University of Arkansas, 226 Physics Building, 825 West Dickson Street, Fayetteville, AR 72701, USA}

\author[0000-0001-5035-4016]{Keith Doore}
\affiliation{Department of Physics, University of Arkansas, 226 Physics Building, 825 West Dickson Street, Fayetteville, AR 72701, USA}
\affiliation{Arkansas Center for Space and Planetary Sciences, University of Arkansas, 332 N. Arkansas Avenue, Fayetteville, AR 72701, USA}

\author[0000-0001-8525-4920]{Antara Basu-Zych}
\affiliation{Center for Space Science and Technology, University of
Maryland Baltimore County, 1000 Hilltop Circle, Baltimore, MD 21250, USA}
\affiliation{NASA Goddard Space Flight Center, Code 662, Greenbelt, MD 20771, USA}

\author[0000-0002-9202-8689]{Kristen Garofali}
\affiliation{William H. Miller III Department of Physics and Astronomy, Johns Hopkins University, Baltimore, MD 21218, USA}
\affiliation{NASA Goddard Space Flight Center, Code 662, Greenbelt, MD 20771, USA}

\author[0000-0003-0708-4414]{Lidia Oskinova}
\affiliation{Institute for Physics and Astronomy
University of Potsdam, Karl-Liebknecht-Str. 24/25, D-14476 Potsdam, Germany}

\author[0000-0001-5261-3923]{Jeff J. Andrews}
\affiliation{Department of Physics, University of Florida, 2001 Museum Rd, Gainesville, FL 32611, USA}

\author[0000-0001-7539-1593]{Vallia Antoniou}
\affiliation{Department of Physics \& Astronomy, Texas Tech University, Box 41051, Lubbock, TX 79409-1051, USA}
\affiliation{Center for Astrophysics $|$ Harvard \& Smithsonian, 60 Garden Street, Cambridge MA 02138, USA}

\author[0000-0003-1509-9966]{Robel Geda}
\affiliation{Department of Astrophysical Sciences, Princeton University, 4 Ivy Lane, Princeton, NJ 08544, USA}

\author[0000-0002-5612-3427]{Jenny E. Greene}
\affiliation{Department of Astrophysical Sciences, Princeton University, 4 Ivy Lane, Princeton, NJ 08544, USA}

\author[0000-0003-3684-964X]{Konstantinos Kovlakas}
\affiliation{Institute of Space Sciences (ICE, CSIC), Campus UAB, Carrer de Can Magrans s/n, E-08193, Barcelona, Spain}
\affiliation{Institut d'Estudis Espacials de Catalunya (IEEC),  Edifici RDIT, Campus UPC, 08860 Castelldefels (Barcelona), Spain
}

\author[0000-0003-3252-352X]{Margaret Lazzarini}
\affiliation{Department of Physics \& Astronomy, California State University Los Angeles, Los Angeles, CA 90032, USA}

\author[0000-0002-3703-0719]{Chris T. Richardson}
\affiliation{Department of Physics \& Astronomy, Elon University, 100 Campus Drive, Elon, NC 27244, USA}



\begin{abstract}

We present a new empirical framework modeling the metallicity and
star-formation history (SFH) dependence of \xray\ luminous ($L \simgt
10^{36}$~\lum) point-source population luminosity functions (XLFs) in normal
galaxies. We expect the \xray\ point-source populations are dominated by
\xray\ binaries (XRBs), with contributions from supernova remnants near the low
luminosity end of our observations. Our framework is calibrated using the
collective statistical power of \nx\ \xray\ detected point-sources within \ngal\
\chandra-observed galaxies at $D \simlt$~40~Mpc that span broad ranges of
metallicity ($Z \approx$0.03--2~$Z_\odot$), SFH, and morphology (dwarf irregulars, late-types, and early-types). Our best-fitting models indicate that the XLF normalization per unit stellar mass declines by $\approx$2--3~dex from 10~Myr to 10~Gyr, with a slower age decline for low-metallicity populations.  The shape of the XLF for luminous X-ray sources ($L \simgt 10^{38}$~\lum) significantly steepens with increasing age and metallicity, while the lower-luminosity XLF appears to flatten with increasing age.  Integration of our models provide predictions for X-ray scaling relations that agree very well with past results presented in the literature, including, e.g., the $L_{\rm X}$-SFR-$Z$ relation for high-mass XRBs (HMXBs) in young stellar populations as well as the $L_{\rm X}/M_\star$ ratio observed in early-type galaxies that harbor old populations of low-mass XRBs (LMXBs). The model framework and data sets presented in this paper further provide unique benchmarks that can be used for calibrating binary population synthesis models. 

\end{abstract}


\vspace{-0.15in}


%
\section{Introduction}\label{sec:intro}
%

Observations of \xray\ emission from normal galaxies (i.e., those not dominated by luminous active galactic nuclei) outside the Local Group have been possible since the launch of the {\it Einstein Observatory}, allowing for studies that link high-energy phenomena and host-galaxy properties \citep[see, e.g.,][for a review]{Fab1989}. Over the last $\approx$25~yr, the {\it Chandra X-ray Observatory} (\chandra) 
has significantly opened up this field, providing resolved
views of the \xray\ emission in galaxies out to $\simgt$100~Mpc, as well as the X-ray detection of galaxy-integrated populations spanning the majority of cosmic history \citep[see, e.g.,][for reviews]{Fab2006,Fab2019,Gil2023}.

Within normal galaxies, \xray\ binaries (XRBs) dominate the $\simgt$1--2~keV emission 
across all morphological types. XRBs consist of either black holes or neutron
stars that are accreting from normal stellar companions. By their nature, XRB populations trace the demographics of populations of close binaries, massive
stars, compact-object remnants and accretion processes. XRB population emission has long been observed to scale with
physical properties of their host galaxies, including, e.g., star-formation rate
(SFR) and stellar mass ($M_\star$) \citep[][]{Gri2003,Gil2004a,Bor2011,Min2012a,Zha2012,Leh2019,Leh2020,Kou2020}. More recent studies have highlighted that XRB emission may provide an important
source of ionizing photons \citep[e.g.,
][]{Sch2019,Sen2019,Oli2021,Sim2021,Gar2024} to the interstellar mediums (ISMs) and the intergalactic medium of the high-redshift
universe \citep[e.g., ][]{Mes2013,Pac2014,Mad2017,Kov2022,Mun2022}.

XRBs are markers of the phases of close-binary evolution when mass-transfer is important and massive compact objects present, following the evolution of the most massive star. Evolutionary modeling of these systems involves many of the same physical prescriptions required for understanding the formation pathways of other astrophysically interesting objects,
including, e.g., SNe, millisecond pulsars, gravitational-wave emitting sources, and short GRBs, contextualizing the importance of XRB observations \citep[e.g.,][]{Gho2001,Zap2021,Bav2022a,Bav2022b,Bav2023, Xin2023, Fra2023, Kot2024}. Included among the various observational constraints of XRBs are population demographics connecting XRB populations and host-galaxy properties. 
Early \chandra\ studies of XRB populations in nearby galaxy samples
($\simlt$100~Mpc) showed initial support for ``universal'' linear scaling
relations in galaxy populations, linking the \xray\ power output of high-mass XRBs (HMXBs) to galaxy-integrated SFR, $L_{\rm X}^{\rm HMXB}$--SFR,  and low-mass XRBs (LMXBs)
to stellar mass, $L_{\rm X}^{\rm LXMB}$--$M_\star$
\citep[e.g.,][]{Ran2003,Col2004,Per2007,Leh2008,Leh2010}.  Additional studies quantified SFR and $M_\star$ scaled \xray\ luminosity functions (XLFs) for HMXB and LMXB
populations, respectively, using samples of late-type \citep[e.g.,][]{Gri2003,Min2012a,Leh2019} and early-type galaxies \citep[e.g.,][]{Gil2004b,Kim2004,Zha2012,Leh2019}.

More recently, several observations and theoretical considerations have
seriously challenged the universality of XRB luminosity and XLF scaling relations with SFR and $M_\star$ alone. For instance, mounting evidence has emerged supporting an
anti-correlation between the $L_{\rm X}$/SFR ratio and metallicity (or a $L_{\rm
X}$-SFR-$Z$ plane) from star-forming galaxies that are expected to be dominated by HMXB populations: overall $L_{\rm X}$/SFR is observed to decline with increasing metallicity \citep[e.g.,][]{Bas2013a,Bro2014,Bro2016,Dou2015,Vul2021,Kyr2024};
excess populations of HMXBs and ultraluminous \xray\ sources (ULXs; $L_{\rm X}
\simgt 10^{39}$~\lum) per SFR have been observed in low-metallicity galaxies
\citep[e.g.,][]{Cla1978,Dra2006,Map2010,Pre2013,Bas2016,Wol2018,Kov2020,Leh2021,Leh2022,Wal2022,Ged2024}; and a
rise in average $L_{\rm X}$/SFR with redshift has been observed
\citep[][]{Bas2013a,Fra2013a,Leh2016,Air2017,Wan2024}, consistent with being driven by the cosmic decline
in metallicity with redshift \citep{For2019,For2020}.  Such considerations have an important impact on cosmological models describing galaxy evolution \citep[see,. e.g.,][]{Vla2023}.

In addition to metallicity, XRB population emission has long been observed to vary with tracers of stellar population age (e.g., optical color and associations with galaxy spiral arms and bulges), pointing to enhancements in the XRB emission in young environments \citep[e.g., ][]{Fab1982,Kim1992,Wol1999,Ten2001,Zez2002,Sor2003,Col2004,Kil2005}.  More recent studies have made more explicit connections between XRB population demographics and galaxy star-formation history (SFH).  For instance, clear age trends in the HMXB formation efficacy (e.g.,
$N$(HMXB)/$M_\star$) as a function of age have been found in star-forming regions of the Magellanic Clouds,
M31, and M33 \citep[e.g.,][]{Sht2007,Ant2016,Gar2018,Ant2019,Laz2021,Laz2023}, showing evidence for peak HMXB activity at $\approx$30~Myr where HMXBs consisting of a NS with Be-type donor are expected
to be dominant. However, the majority of HMXBs within Local Group galaxies are comparatively low luminosity sources ($L \simlt 10^{38}$~\lum) that 
are fed by stellar winds and decretion disks of Be stars.  Hence, Local Group galaxies lack the powerful Roche-lobe overflow (RLO) HMXBs that are expected to power the most luminous sources \citep[$L \simgt 10^{38}$~\lum; see, e.g., ][]{Mis2024} observed in galaxies.  Such systems dominate the integrated \xray\ power output for many galaxies.

%
%
\begin{figure*}
\centerline{
\includegraphics[width=17.8cm]{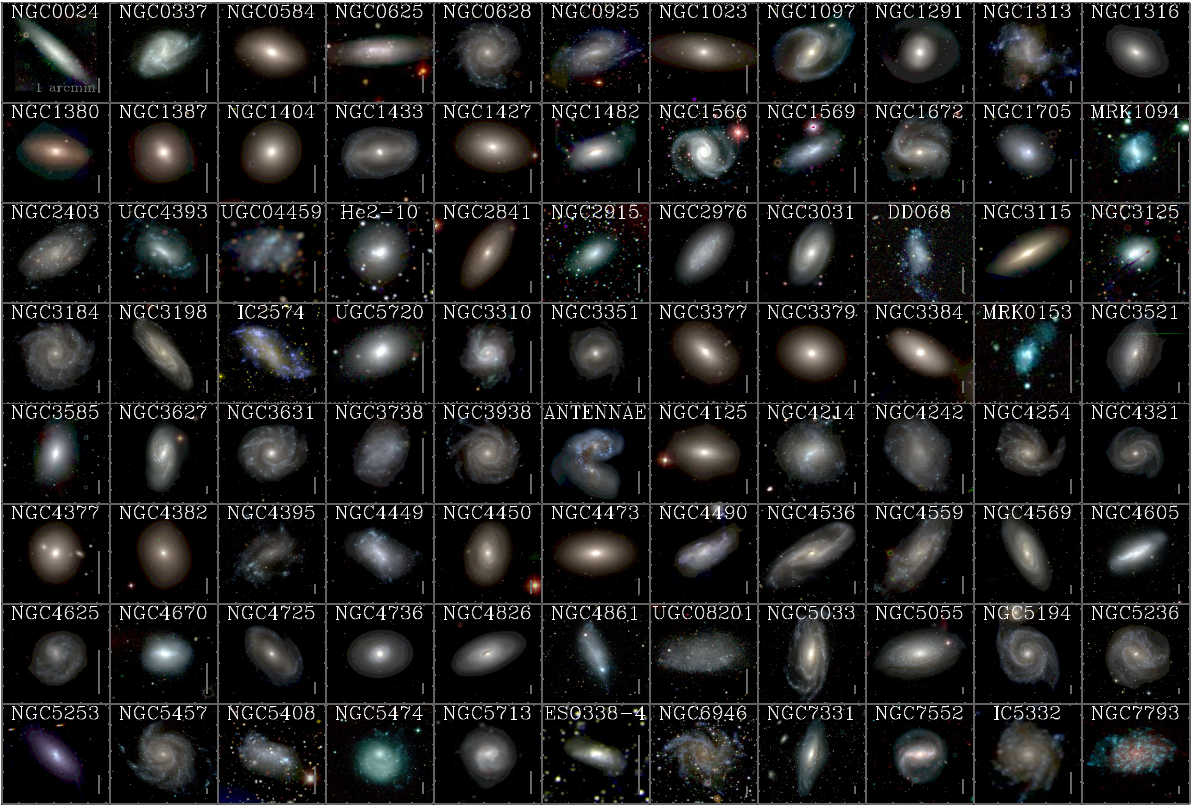}
}
\caption{
False-color optical/near-IR postage-stamp images for the \ngal\ galaxies in our sample.  The observatories used to create these postage stamps varies, but are mainly based on $g'$ (blue), $r'$ (green), and $i'$ (red) band data from SDSS or PanSTARRS.  Other postage stamps are based on best-matched filters from CTIO, KPNO, \swift\ UVOT, or \hst\ data. Image sizes are square in dimensions and have manually-determined scales ranging from 0.3--1.3~$\times$(2$a$), where $a$ is the semi-major axis provided in Table~\ref{tab:sam}. A 1~arcmin length vertical gray bar is provided in the lower-right corner of each image for scale.
}
\label{fig:img}
\end{figure*}

To assess the luminosity-dependent evolution of XRB populations,
\citet{Leh2017} utilized SFH maps and deep \chandra\ observations of the
relatively nearby ($D \approx 9$~Mpc) galaxy M51 to construct an age-dependent
XRB XLF model that was fit simultaneously to several independent populations
across subgalactic regions.  While highly uncertain, the best fitting model suggested that the stellar-mass normalized XRB XLF declined by a few orders of
magnitude in normalization and steepened in slope with increasing age. In support of these findings, subgalactic measurements of the XLF within NGC~300 indicate larger numbers of XRBs per unit SFR within younger populations in the galaxy \citep{Bin2024}.  The age-dependent evolution of XRB population emission has been more rigorously constrained by \citet{Gil2022}, who utilized SFH information for 344 normal galaxies in the Great Observatories Origins Deep Surveys (GOODS) fields,
with \chandra\ Deep Field \xray\ constraints, and showed in detail how $L_{\rm
X}$/$M_\star$ declines over 10~Myr to 10~Gyr timescales.

Binary population synthesis models that reproduce the above observed trends
have provided some insight into the physical processes that drive the
metallicity and SFH dependences of XRB population demographics and emission
\citep[e.g.,][]{Fra2013a,Zuo2014,Mis2023}.  On theoretical grounds, metallicity is expected to
impact stellar wind mass loss, which can affect the orbital evolution of
binaries and the resulting compact-object remnant mass distribution.
Specifically, low-metallicity binaries are expected to have relatively weak mass loss from stellar winds, resulting in less angular momentum loss from the systems, less binary widening on stellar evolutionary timescales, and more massive compact object remnants; effects that can yield more luminous XRB
populations \citep[e.g.,][]{Lin2010,Wik2017,Wik2019,Liu2024}. Similarly, stellar population age plays a primary role in determining the evolutionary stage of donor stars (e.g., RLO potential, strengths of their winds, and presence of an equatorial decretion disk) in XRBs that impacts the overall \xray\ power output of the populations.  In addition, LMXBs have lower formation efficiency since their accretion mode, Roche-lobe overflow, requires that they satisfy more strict evolutionary conditions \citep[e.g.,][]{Kal1998}. Young populations of XRBs (e.g., $\simlt$100~Myr) accrete
from relatively massive donor stars, and result in larger average mass-transfer rates compared to older populations that harbor low-mass donors, thus yielding more luminous XRBs per
stellar mass within young stellar populations \citep[see also, e.g., ][]{Fra2008,Mis2024}.

\begin{deluxetable*}{lccrrrrrrrc}
\tablewidth{1.0\columnwidth}
\tabletypesize{\footnotesize}
\tablecaption{Galaxy Sample and Basic Properties}
\tablehead{
\multicolumn{1}{c}{}  & \colhead{} & \colhead{} & \colhead{} & \colhead{} & \multicolumn{3}{c}{\sc Size Parameters} & \colhead{} & \colhead{} & \colhead{} \\
\vspace{-0.25in} \\
\multicolumn{1}{c}{\sc Galaxy} &  \multicolumn{2}{c}{\sc Central Position} & \colhead{$r_{\rm ex}$} & \colhead{$D$} & \colhead{$a$} & \colhead{$b$} & \colhead{PA} & \colhead{$\log M_\star$} &  \colhead{$\log$~SFR} & \colhead{$12 + \log({\rm O/H})$}  \\ 
\vspace{-0.25in} \\
\multicolumn{1}{c}{\sc Name} &  \colhead{$\alpha_{\rm J2000}$} & \colhead{$\delta_{\rm J2000}$} & \colhead{(arcsec)} & \colhead{(Mpc)} & \multicolumn{2}{c}{(arcmin)} & \colhead{(deg)} &  \colhead{($M_\odot$)}   & \colhead{(\sfr)} &  \colhead{(dex)} \\ 
\vspace{-0.25in} \\
\multicolumn{1}{c}{(1)} & \multicolumn{1}{c}{(2)} & \multicolumn{1}{c}{(3)} & \colhead{(4)} & \colhead{(5)} & \colhead{(6)} & \colhead{(7)} & \colhead{(8)} & \colhead{(9)} & \colhead{(10)} & \colhead{(11)}}
\startdata
             NGC0024\dotfill &     00 09 56.5 & $-$24 57 47.3 &                         \ldots &                       7.30 (1) &   1.38 &   0.39 &   43.5 (1) &    8.85$^{+0.09}_{-0.09}$ & $-$1.19$^{+0.26}_{-0.30}$ &                  8.59 (1) \\
             NGC0337\dotfill &     00 59 50.1 & $-$07 34 40.7 &                         \ldots &                       22.4 (1) &   0.87 &   0.49 &  157.5 (1) &    9.65$^{+0.11}_{-0.10}$ &    0.35$^{+0.21}_{-0.23}$ &                  8.44 (1) \\
             NGC0584\dotfill &     01 31 20.8 & $-$06 52 05.0 &                         \ldots &                       20.1 (1) &   1.47 &   0.91 &   62.5 (1) &   10.68$^{+0.03}_{-0.05}$ & $-$1.90$^{+0.31}_{-0.33}$ &                 8.76 (22) \\
             NGC0625\dotfill &     01 35 04.2 & $-$41 26 15.0 &                         \ldots &                       4.10 (9) &   1.43 &   0.47 &   92.0 (3) &    8.31$^{+0.17}_{-0.10}$ & $-$1.42$^{+0.45}_{-0.31}$ &                  8.10 (7) \\
       NGC0628 (M74)\dotfill &       01 36 41.8 & +15 47 00.5 &                              3 &                       7.30 (1) &   2.10 &   1.80 &   87.5 (1) &    9.69$^{+0.08}_{-0.07}$ & $-$0.26$^{+0.22}_{-0.27}$ &                  8.54 (1) \\
             NGC0925\dotfill &       02 27 16.9 & +33 34 44.0 &                         \ldots &                       9.12 (1) &   1.87 &   0.82 &  105.0 (1) &    9.32$^{+0.08}_{-0.09}$ & $-$0.50$^{+0.24}_{-0.28}$ &                  8.38 (1) \\
             NGC1023\dotfill &       02 40 24.0 & +39 03 47.7 &                              3 &                       11.4 (8) &   3.02 &   1.15 &   82.0 (1) &   10.61$^{+0.03}_{-0.06}$ & $-$1.35$^{+0.51}_{-0.40}$ &                 8.78 (22) \\
             NGC1097\dotfill &     02 46 19.1 & $-$30 16 29.7 &                              5 &                       17.1 (1) &   2.63 &   1.44 &  145.0 (1) &   10.65$^{+0.09}_{-0.08}$ &    0.92$^{+0.21}_{-0.26}$ &                  8.83 (1) \\
             NGC1291\dotfill &     03 17 18.6 & $-$41 06 29.1 &                              2 &                       10.8 (1) &   2.39 &   1.70 &  170.0 (1) &   10.78$^{+0.03}_{-0.04}$ & $-$0.95$^{+0.26}_{-0.33}$ &                 8.78 (22) \\
             NGC1313\dotfill &     03 18 15.8 & $-$66 29 53.0 &                         \ldots &                       4.20 (9) &   2.15 &   1.63 &   40.0 (3) &    8.93$^{+0.09}_{-0.09}$ & $-$0.57$^{+0.26}_{-0.28}$ &                  8.40 (8) \\
\enddata
\tablecomments{The full version of this table contains information for all \ngal\ galaxies from our sample. An abbreviated version of the table is displayed here to illustrate its form and content.  Col.(1): Adopted galaxy designation with Messier designation, if applicable. Col.(2) and (3): Right ascension and declination of the galactic center. Col.(4): Radius of the central region, in units of arcseconds, excluded from consideration due to the presence of an AGN or significant X-ray source crowding.  Col.(5): Adopted distance in units of Mpc and reference in parentheses.  Col.(6)--(8): Isophotal ellipse parameters, including, respectively, semi-major axis, $a$, semi-minor axis, $b$, and position angle east from north, PA.  In parentheses, we include a flag denoting the origin of the adopted ellipse parameters: 1 = $K_{20}$ isophotal region, 2 = HyperLEDA D25, 3 = FUV-based ellipse, and 4 = manually defined region (see detailed description in \S~\ref{sec:sfh}). Col.(9): Logarithm of the galactic stellar mass, $M_\star$, within the regions defined.  Col.(10): Star-formation rate within the defined regions.  The values in Col.(9) and (10) were derived from our SED fitting procedures, as described in $\S$\ref{sub:sed}.  Col.(11): Adopted estimate of the average oxygen abundances, 12+$\log ({\rm O/H})$, and references (in parentheses).  For consistency with other studies of XRB scaling relations that include metallicity, we have converted all strong-line abundance measurements to the \citet[][]{Pet2004} calibration following the prescriptions in \citet{Kew2008}.\\
\\
\noindent
{\bf Distance References.}-- 1 = \citet{Mou2010}; 2 = \citet{Eng2008}; 3 = \citet{Sac2016}; 4 = \citet{McQ2016}; 5 = \citet{Tul2013}; 6 = \citet{Fre2001}; 7 = \citet{Nat2015}; 8 = \citet{Har2013}; 9 = \citet{Lee2009}; 10 = \citet{Lee2023}; 11 = \citet{Kov2021}; 12 = HyperLEDA.
\\
\\
\noindent
{\bf Metallicity References.}-- 1 = \citet{Mou2010}; 2 = \citet{Eng2008}; 3 = \citet{Izo2007}; 4 = \citet{Bre2009}; 5 = \citet{Mon2012}; 6 = \citet{Hu2018}; 7 = \citet{Ski2003}; 8 = \citet{Wal1997}; 9 = \citet{Cro2009}; 10 = \citet{McQ2019}; 11 = \citet{Pil2014}; 12 = \citet{Berg2012}; 13 = \citet{Pil2007}; 14 = \citet{Mou2006}; 15 = \citet{Gom2021}; 16 = \citet{Gro2023}; 17 = \citet{Gan2022}; 18 = \citet{Mad2013}; 19 = \citet{Est2014}; 20 = \citet{Shi2005}; 21 = \citet{Tad2015}; 22 = Mass-metallicity relation from \citet{Kew2008}.
\label{tab:sam}}
\end{deluxetable*}

As discussed above, collective empirical constraints and theoretical models have identified metallicity and SFH as key physical factors that impact XRB and ULX populations.  However, a clear synthesis of the observational data that illustrates the dependence of both of these factors has yet to be realized. A highly desirable collective constraint for galaxy populations would be a
quantitative assessment for how XRB XLFs vary simultaneously with metallicity and age; the primary goal of this paper.  Compared to simple galaxy-integrated $L_{\rm X}$ scaling relations, XRB
{\it XLF scaling relations} describe the shapes and normalizations of XLFs vary with galaxy physical properties (e.g., SFR and $M_\star$), providing
several additional degrees of freedom to more precisely test theoretical models
and provide insight into the evolution of close-binary systems. Furthermore, constraints on XLF
variation with metallicity and SFH are of fundamental importance, as they can be integrated directly to infer $L_{\rm
X}$ scaling relations on a more generalized basis. 

In our previous works studying XLFs in nearby ($D<30$~Mpc) galaxies
\citep{Leh2019,Leh2020,Leh2021}, we found that XLF scaling relations involving only stellar mass, SFR, globular cluster content, and metallicity were insufficient
for universally modeling the XLFs of all local galaxies.  In particular,
\citet{Leh2021} found that the galaxies that were most poorly fit by scaled models
are low-mass starburst galaxies with SFHs that are bursty or rising to the present day,
implying that they host HMXB populations that differ from relatively large
galaxies that have smoother SFHs over $\approx$100~Myr timescales.  
Similarly, studies at subgalactic scales have explicitly found enhancements and variability in the XRB formation rate on relatively short timescales within the first 100~Myr following star-formation events \citep[see, e.g.,][]{Ant2016,Leh2017,Gar2018,Ant2019,Laz2023,Bin2023,Bin2024}.  These age variations have been predicted in population synthesis models and are also expected to vary with metallicity \citep[see, e.g.,][]{Lin2010,Wik2017,Wik2019}. Thus, an important goal is to characterize the key variations of the XLF as functions of both metallicity and SFH.

In this paper, we make use of the literature and large multiwavelength data archives to explicitly determine metallicities and SFHs for a sample of \ngal\ nearby galaxies with \chandra\ constraints on \xray\ point-source populations (see Figure~\ref{fig:img}).  Our goal is to build an empirically-calibrated model for how the XLF shapes and normalizations per stellar mass vary as a function of stellar-population metallicity and age.
We have organized our paper around the steps required to achieve this goal.  In $\S$\ref{sec:samp}, we construct our galaxy sample and compile metallicity information from the literature.  In $\S$\ref{sec:sfh}, we cull several FUV-to-FIR data sets and perform SED fitting to derive their SFHs.   In $\S$\ref{sec:xray}, we analyze the \chandra\ X-ray point-source data and derive observed constraints on their XLFs.  In $\S$\ref{sec:res}, we utilize the metallicity, SFHs, and XLF measurements as a basis for constructing  our metallicity and age dependent XLF modeling framework.
In $\S$\ref{sec:dis} we discuss our model framework in detail, providing broader context of our XLF model predictions for galaxy-integrated scaling relations and comparisons with past results and population synthesis models. We further discuss caveats to our model, methods for using the data presented here to constrain binary population synthesis models, and future observations and studies that could improve constraints on these results.  Finally, the key results are summarized in $\S$\ref{sec:sum}.

Throughout this paper, we make reference to \xray\ and multiwavelength fluxes and luminosities (or luminosity densities) that have been
corrected for Galactic absorption from gas and dust, but not host-galaxy absorption.  Unless stated otherwise, we quote \xray\ fluxes and luminosities in reference to the 0.5--8~keV bandpass. We adopt a \citet{Kro2001} initial mass
function (IMF) when performing multiwavelength UV--to--IR SED modeling, and we utilize a $\Lambda$CDM cosmology, with values of $H_0$ = 70~\hbox{km s$^{-1}$
Mpc$^{-1}$}, $\Omega_{\rm M}$ = 0.3, and $\Omega_{\Lambda}$ = 0.7 adopted
\citep[e.g.,][]{Spe2003}.

\section{Sample Selection and Properties}\label{sec:samp}
%

We began by culling well-studied samples of relatively nearby galaxies ($D
\simlt 40$~Mpc) with a wealth of \chandra\ and multiwavelength data.  We made use of previous samples that have been used for studying XLF relations in nearby galaxies, including
\citet{Leh2019,Leh2020,Leh2021} and \citet{Ged2024}, as well as a sample of four low-metallicity (\lgoh~$\approx$~7.8--8.0) star-forming galaxies with new \chandra\ Cycle~23 exposures that we present here (PI: B.~Lehmer). The samples within these studies were constructed from a variety of resources, including the {\it
Spitzer} Infrared Nearby Galaxies Survey \citep[SINGS;][]{Ken2003}, the Legacy ExtraGalactic UV Survey \citep[LEGUS;][]{Sab2018}, the STARBurst IRregular Dwarf Survey \citep[STARBIRDS;][]{McQ2018}, a subsample of galaxies from the Physics at High Angular resolution in Nearby GalaxieS (PHANGS) survey with \jwst\ coverage \citep{Lee2023}, the star-forming galaxy sample from \citet{Min2012a}, the \citet{Har2013} sample of early-type galaxies
with well-measured GC populations, and other miscellaneous studies \citep[e.g.,][]{Eng2008}.  

In our sample selection, we excluded galaxies that have high galactic inclinations ($i \simgt 70$~deg) that may harbor highly-absorbed \xray\
point-source populations that deviate from typical XLFs due to unmodeled
orientation effects.  We further excluded galaxies from the \citet{Leh2020}
ellipticals sample that had GC specific frequencies $S_N > 2$,\footnote{Here,
$S_N \equiv N_{\rm GC} 10^{0.4(M_{V}^T + 15)}$, where $N_{\rm GC}$ is the
number of GCs in a galaxy with galaxy-wide absolute $V$-band magnitude
$M_V^T$.} to avoid significant contributions from LMXB populations that form dynamically in GC environments and exhibit different XLF shapes \citep[see, e.g.,
][]{Irw2005,Hum2008,Bor2011,Leh2020}. To ensure good constraints on galaxy SFHs, we also excluded galaxies that lacked high-quality multiwavelength data spanning FUV-to-FIR; we discuss the requirements on these data in more detail in $\S$\ref{sub:phot} below.  Rejection from this category was based on a variety of reasons, including, e.g., the presence of very bright/optically-saturated Galactic foreground stars that make photometry difficult, or the multiwavelength data covering a small fraction of the galaxy footprint.

Consideration of the above selection criteria resulted in a sample of \ngal\ galaxies. In Figure~\ref{fig:img}, we show false-color optical/near-IR image postage stamps for the galaxy sample, and in Table~\ref{tab:sam}, we list the galaxies in our sample, along with their basic properties.  By construction, our sample spans a broad range of morphological types with good representation across the Hubble sequence: dwarf irregulars, peculiars, spirals, and ellipticals.  However, our sample includes only small numbers of major mergers, due to the volume limit, and excludes highly-inclined disk galaxies, by construction.

%
%
\begin{figure*}
\centerline{
\includegraphics[width=8.5cm]{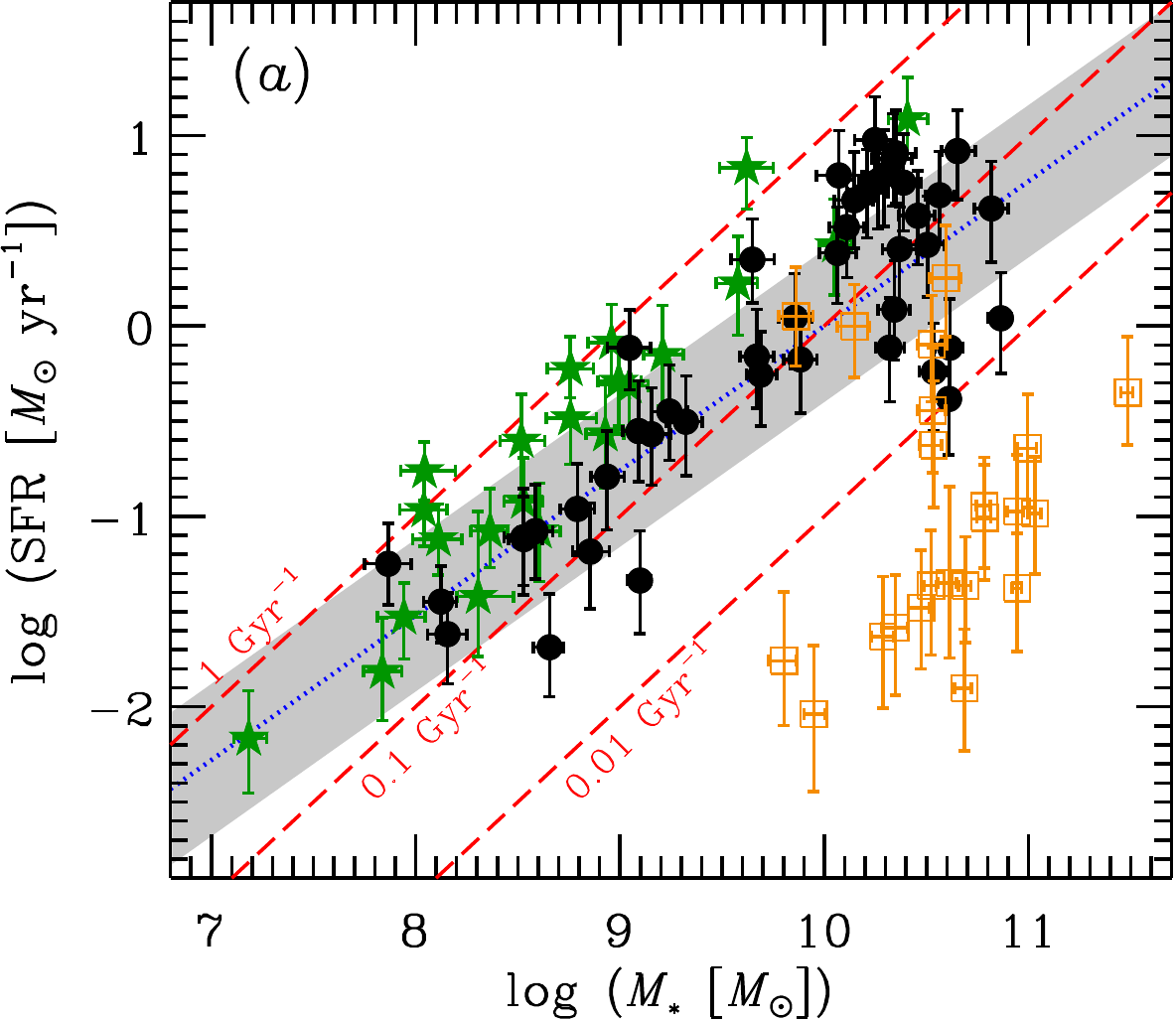}
\hfill
\includegraphics[width=8.5cm]{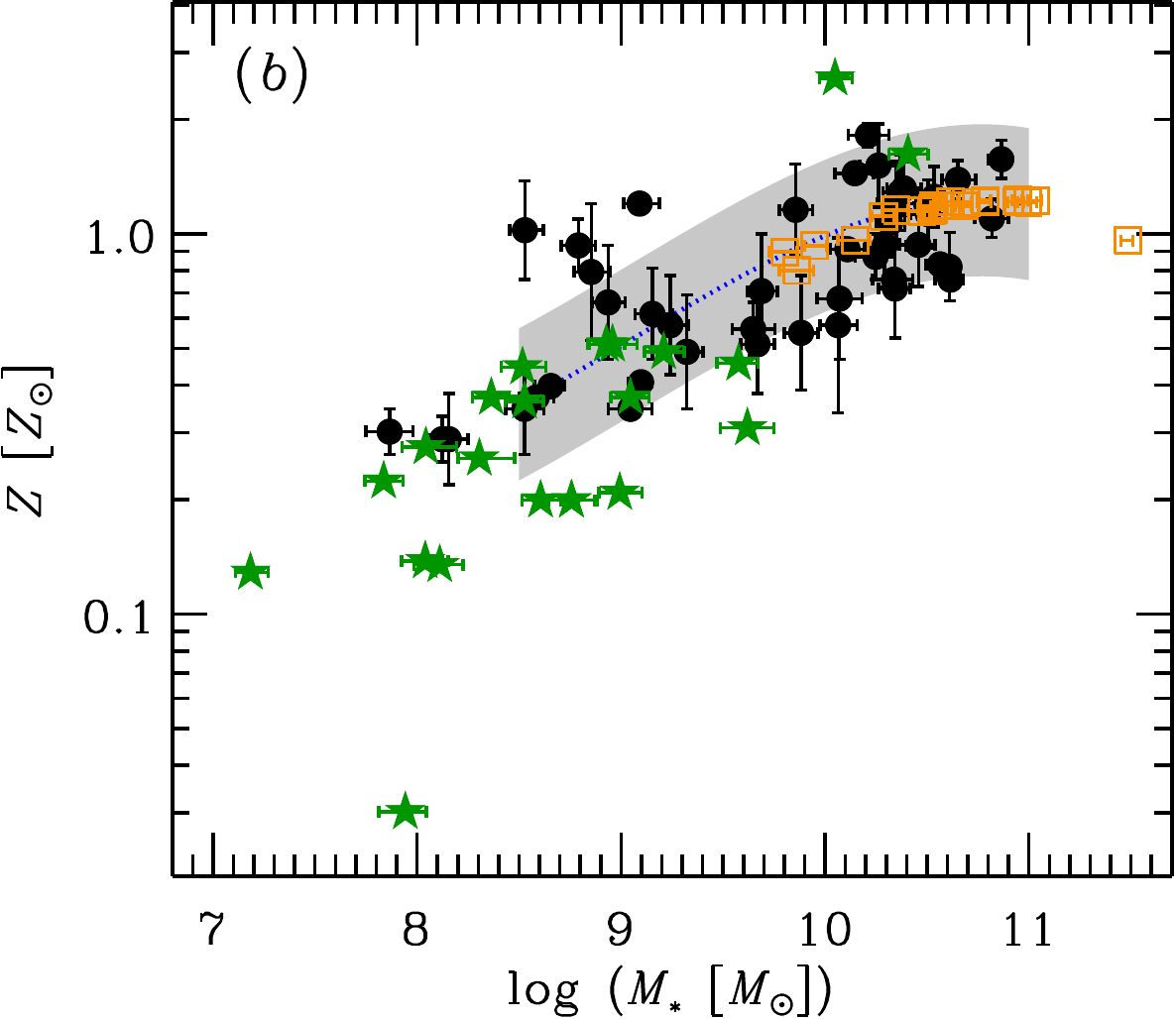}
}
\caption{
($a$) SFR versus $M_\star$ for the galaxy sample used in our study with 1$\sigma$ uncertainties shown.  Different symbol types correspond to sources with metallicity measurements
based on strong-line calibrations ({\it black filled circles}), direct-method calibration ({\it
filled green stars}), and the $M_\star$-$Z$ relation ({\it open orange squares}).  The gray shaded region represents the galaxy main
sequence, as defined by \citet{Air2019}, and lines of constant sSFR have been
overlaid for reference ({\it red dashed lines\/}). Our galaxy sample spans a
relatively broad range of SFHs, ranging from massive elliptical galaxies
(sSFR~$\simlt$~0.01~Gyr$^{-1}$) to starburst galaxies
(sSFR~$\simgt$~1~Gyr$^{-1}$).
($b$) Gas-phase metallicity, $Z$, versus $M_\star$ for our sample galaxies.  The
$M_\star$-$Z$ relation from Table~2 of \citet{Kew2008} and its 1$\sigma$
scatter are displayed as a dotted curve and gray shaded region,
respectively.
}
\label{fig:prop}
\end{figure*}

Figure~\ref{fig:prop}$a$ displays the SFR versus $M_\star$ values for the sources in our sample.  Values of SFR and $M_\star$ were derived from SED fitting results, which we describe in \S\ref{sec:sfh} (see Eqns.~\eqref{eqn:sfr} and \eqref{eqn:mstar}).  For comparison, we overlay the location of the galaxy main sequence (grey band), as defined in Eqn.~8 of \citet{Air2019}, and lines of constant specific-SFR (sSFR~$\equiv$~SFR/$M_\star$).  To first order, sSFR provides a proxy for galaxy SFH, with high sSFR representing galaxies with current active star-formation and galaxies with low sSFR representative of early-type galaxies dominated by old stellar populations. Our sample contains galaxies above the main-sequence, main-sequence galaxies, and many sub-main sequence and quiescent objects.

Following the procedure in \citet{Leh2021}, we culled gas-phase metallicity measurements\footnote{Throughout this paper, we quote metallicities in terms of either total mass-weighted abundances, $Z$, relative to the solar value, $Z_\odot = 0.02$, or gas-phase oxygen abundances, \lgoh, and take the solar value to be $12 + \log({\rm O}/{\rm H})_\odot = 8.69$ \citep{All2001,Asp2009}. However, all abundances are derived from the gas-phase oxygen abundances, and when relevant, we therefore assume $Z = 10^{\log({\rm O}/{\rm H}) + 3.31} Z_\odot$.} from nebular emission lines using either strong-line calibrations or ``direct method'' electron-temperature-based theoretical calibrations.  The strong-line measurements are based primarily on the \citet{Pet2004} relation using either the emission line ratios $R_{23} =$~([O~{\small II}]$\lambda$3727 + [O~{\small III}] $\lambda \lambda$4959, 5007)/H$\beta$ and [O~{\small III}]$\lambda$5007/[O~{\small II}]$\lambda$3727 or ([O~{\small III}]$\lambda$5007/H$\beta$)/([N~{\small II}]$\lambda$6584/H$\alpha$) (see references in Table~\ref{tab:sam}).  This relation is empirically calibrated against the direct method, which uses the weak-line ratio [O~{\small III}]$\lambda\lambda$4959, 5003/[O~{\small III}]$\lambda$4363, a more sensitive measure of the electron temperature and oxygen abundance \citep[e.g., ][]{Dav2017,Cur2017,Mai2019}.  For the subset of four galaxies that were drawn from the PHANGS sample, we adopted the $S$-cal strong-line-based metallicity measurements from \citet{Gro2023}, as presented in \citet{Lee2023}.  The $S$-cal method utilizes relations between [N~{\small II}]$\lambda\lambda$6548, 6584/H$\beta$, [S~{\small II}]$\lambda\lambda$6717, 6731/H$\beta$, and [O~{\small III}]$\lambda\lambda$4959, 5007/H$\beta$ with metallicity that are calibrated to using direct-method estimates \citep[see][for details]{Pil2016}.

For the case of most early-type galaxies that do not have oxygen abundance measurements due to a lack of star-formation activity, we chose to adopt the \citet{Pet2004} ``O3N2'' mass-metallicity relation (hereafter, $M_\star$-$Z$ relation) specified by Table~2 of \citet{Kew2008}. We note that our choice to adopt nebular-based abundances is one of consistency, and results in metallicity estimates that are expected to be applicable to the youngest populations ($\simlt$100s~Myr). We expect that relatively old stellar populations within all galaxies in our sample will have systematically lower abundances than those traced by the nebulae.  For example, in the extreme case of elliptical galaxies, the dominant $\sim$10~Gyr old stellar population metallicities have been measured to be systematically lower than the $M_\star$-$Z$ relation by $\approx$0.3--1~dex \citep[e.g.,][]{Gal2005,Pan2008,Loo2024}.  Since this issue will be present in all galaxies, regardless of morphological type, we caution that our quoted metallicities should be interpreted as relevant for the most recently formed stars at the time of observation. Future studies should address the impact of metallicity evolution within the galaxy populations; however, this detail is beyond the scope of the present paper.

In Figure~\ref{fig:prop}b, we display the $M_\star$ and $Z$ estimates for the galaxies in our sample.  Our galaxies mainly follow the $M_\star$-$Z$ relation from \citet{Kew2008} over the range of applicability (gray shaded region), and cover a broad metallicity range of $Z \approx$~0.03--2~$Z_\odot$, albeit with the vast majority of sources having $Z \simgt 0.1$.

In summary, our sample of \ngal\ galaxies have 44, 22, and 22 metallicities
estimated from strong-line, direct, and $M_\star$-$Z$ relations, respectively.  The symbol colors and styles in Figure~\ref{fig:prop} vary dependent on the methods used.

%
\section{Star-Formation History Derivations}\label{sec:sfh}
%

To achieve our goal of modeling metallicity and SFH dependent point-source XLFs
in galaxies, we first needed to obtain detailed characterizations of the SFHs of our galaxies.  We accomplished this by (1) gathering and analyzing multiwavelength FUV-to-FIR data from a variety of facility archives; (2) constructing galaxy-integrated SEDs across this vast range of wavelengths; and (3) performing SED fitting using the {\ttfamily Lightning} SED fitting code \citep[][]{Euf2017,Doo2023,Mon2024} to derive SFH solutions and posterior distributions.  {\ttfamily Lightning} v.~2024.0.1\footnote{Available at \url{https://github.com/ebmonson/lightningpy}. This work uses an early version, v.~2024.0.1, available at \url{https://github.com/ebmonson/lightningpy/releases/tag/v2024.0.1}} is a {\ttfamily python} code, building on the older IDL version of {\ttfamily Lightning}, that makes use of stellar population synthesis models over a range of metallicities, in combination with nebular effects and dust attenuation and emission, to fit spectrophotometric data spanning X-ray--to--far-IR wavelengths. When relevant, {\ttfamily Lightning} also has capabilities for including contributions from AGN and XRB populations \citep[see, e.g.,][for further details]{Doo2023, Mon2023}.  However, since our goal is to quantify and calibrate the relationship between XRB population XLFs with SFH and metallicity in normal galaxies, we do not use the AGN and XRB models here.  Future versions of \lightning\ will incorporate the new constraints afforded by the present study.  Our approach is to gather as much well-calibrated broad-to-narrow band FUV-to-FIR photometry as possible for each galaxy in our sample, and fit the SEDs of these galaxies with metallicities fixed at the values obtained in the literature, which are primarily based on spectroscopy.

%
%
\begin{figure*}
\centerline{
\includegraphics[width=17.8cm]{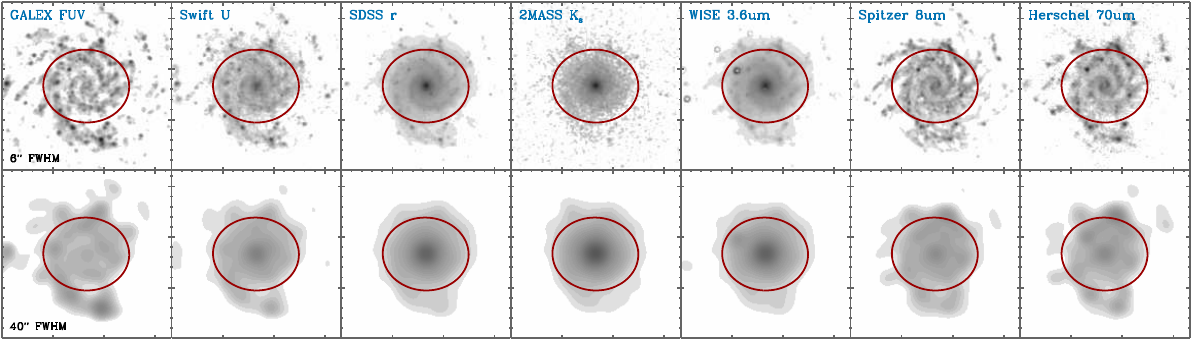}
}
\caption{
Example PSF-matched image sets for NGC~0628. These image sets span the FUV-to-IR (see annotations) and have been convolved to Gaussian PSFs with FWHM values of 6~arcsec ({\it top row\/}) and 40~arcsec ({\it bottom row\/}).  The red ellipse estimates the $K_s$-band 20~mag~arcsec$^{-2}$ isophotal contours. We used such apertures, along with the 40~arcsec FWHM PSF image sets to extract galaxy-integrated photometry for FUV-to-FIR data sets.  For \nrem\ galaxies, we made use of the 6~arcsec FWHM PSF image sets to extract much smaller photometry associated with the galaxy centers, where we exclude X-ray data and local SFH contributions from our analyses.  For NGC~0628, we excluded a very small 6~arcsec diameter circular region from our analysis (not shown here due to its relatively small size).
}
\label{fig:psf}
\end{figure*}

\subsection{Multiwavelength Data Cube Construction}\label{sub:phot}

Following the procedures in \citet{Euf2017}, we gathered FUV-to-FIR photometry
using a variety of resources. We made extensive use of public archives for downloading calibrated data sets.  These archives include the NASA/IPAC Infrared Science Archive (IRSA)\footnote{\url{https://irsa.ipac.caltech.edu/}} for 2MASS, \spitzer, \wise, and \herschel\ data sets; the PHANGS team site\footnote{\url{https://sites.google.com/view/phangs/home/data/astrosat}} for \astrosat\ data \citep{Has2024}; the Barbara A. Mikulski Archive for Space Telescopes (MAST)\footnote{\url{https://mast.stsci.edu}} for \galex, \swift, \hst, and \jwst\ data sets; the PanSTARRS-1 Image Access portal\footnote{\url{https://ps1images.stsci.edu/cgi-bin/ps1cutouts}} for PanSTARRS images; the Sloan Digital Sky Survey (SDSS) DR12 Science Archive Server (SAS)\footnote{\url{https://dr12.sdss.org/mosaics}} for SDSS images; and the Astro Data Lab server from the NSF NOIRLab\footnote{\url{https://datalab.noirlab.edu/sia.php}}; and the NASA/IPAC Extragalactic Database (NED)\footnote{\url{https://ned.ipac.caltech.edu/}} for additional data sets from the Cerro Tololo Inter-American Observatory (CTIO).  Since our galaxy samples were drawn from a variety of sources, with galaxies located across the entire sky, we do not have uniform sensitivity limits for our entire sample.  

For a given galaxy, we calibrated all available UV-to-IR photometry to common units of flux density per sky area (MJy~sr$^{-1}$).  Data sets with wavelengths $\simlt$5~$\mu$m were screened for bright foreground stars using the methods highlighted in \citet{Euf2017}, and stars with significant emission within or near the galactic footprints were masked and filled in with pixel values based on the local background. We applied this procedure in each bandpass independently and were conservative in selection, removing only bright stars that may have a significant impact on the global photometry of a given galaxy in a given bandpass. We followed a progressive scheme for determining the sizes of the foreground star regions that we masked, based on the signal-to-noise ratio (S/N) of the star in the given band.  Specifically, for most data sets, we used circular masking regions with radii of 2, 5, 7, and 10 times the half-width at half max (HWHM) of the PSF for S/N = 10--500, 500--1000, 1000--2500, and $>$2500, respectively.  For \hst\ data, however, we made use of a larger circular masking region of uniform radius $\approx$3~arcsec, as this was most effective in removing the signatures of bright stars without significant impact on the galaxy photometry.


\begin{deluxetable}{llrcc}
\tablewidth{1.0\columnwidth}
\tabletypesize{\footnotesize}
\tablecaption{Multiwavelength Coverage Used in SED Fitting}
\tablehead{
\multicolumn{1}{c}{\sc Galaxy} & \multicolumn{1}{c}{\sc } & \colhead{$\log \lambda$} & \colhead{$\log F_\nu$} & \colhead{$\log \nu L_\nu$} \\
\vspace{-0.25in} \\
\multicolumn{1}{c}{\sc Name} & \multicolumn{1}{c}{\sc Band} & \colhead{($\mu$m)} & \colhead{(Jy)} & \colhead{($L_\odot$)}   \\
\vspace{-0.25in} \\
\multicolumn{1}{c}{(1)} & \multicolumn{1}{c}{(2)} & \multicolumn{1}{c}{(3)} & \colhead{(4)} & \colhead{(5)}}
\startdata
NGC0024 & $A_V^{\rm Gal} = 0.0521$ & & & \\
  & {\ttfamily GALEX_FUV     } & $-$0.82 & $-$2.45$\pm$0.06 & 8.07$\pm$0.061  \\
  & {\ttfamily UVOT_UVW2     } & $-$0.70 & $-$2.31$\pm$0.02 & 8.09$\pm$0.021  \\
  & {\ttfamily UVOT_UVM2     } & $-$0.65 & $-$2.30$\pm$0.02 & 8.05$\pm$0.021  \\
  & {\ttfamily GALEX_NUV     } & $-$0.64 & $-$2.30$\pm$0.06 & 8.04$\pm$0.061  \\
  & {\ttfamily UVOT_UVW1     } & $-$0.60 & $-$2.20$\pm$0.02 & 8.11$\pm$0.021  \\
  & {\ttfamily 0.9m_B} & $-$0.36 & $-$1.54$\pm$0.02 & 8.52$\pm$0.021  \\
  & {\ttfamily Pan-STARRS_gp1} & $-$0.32 & $-$1.42$\pm$0.02 & 8.60$\pm$0.021  \\
  & {\ttfamily 0.9m_V} & $-$0.26 & $-$1.30$\pm$0.02 & 8.66$\pm$0.021  \\
  & {\ttfamily Pan-STARRS_rp1} & $-$0.21 & $-$1.23$\pm$0.02 & 8.68$\pm$0.021  \\
  & {\ttfamily 0.9m_R} & $-$0.19 & $-$1.27$\pm$0.02 & 8.62$\pm$0.021  \\
  & {\ttfamily Pan-STARRS_ip1} & $-$0.12 & $-$1.13$\pm$0.02 & 8.69$\pm$0.021  \\
  & {\ttfamily 0.9m_I} & $-$0.10 & $-$0.99$\pm$0.02 & 8.82$\pm$0.021  \\
  & {\ttfamily Pan-STARRS_zp1} & $-$0.06 & $-$1.06$\pm$0.02 & 8.70$\pm$0.021  \\
  & {\ttfamily Pan-STARRS_yp1} & $-$0.02 & $-$0.99$\pm$0.02 & 8.72$\pm$0.021  \\
  & {\ttfamily 2MASS_J       } & 0.09 & $-$0.94$\pm$0.04 & 8.67$\pm$0.041  \\
  & {\ttfamily 2MASS_H       } & 0.22 & $-$0.88$\pm$0.04 & 8.60$\pm$0.041  \\
  & {\ttfamily 2MASS_Ks      } & 0.33 & $-$1.00$\pm$0.04 & 8.37$\pm$0.041  \\
  & {\ttfamily WISE_W1       } & 0.53 & $-$1.27$\pm$0.01 & 7.91$\pm$0.010  \\
  & {\ttfamily IRAC_CH1      } & 0.55 & $-$1.25$\pm$0.02 & 7.90$\pm$0.021  \\
  & {\ttfamily IRAC_CH2      } & 0.65 & $-$1.44$\pm$0.02 & 7.61$\pm$0.021  \\
  & {\ttfamily WISE_W2       } & 0.66 & $-$1.51$\pm$0.01 & 7.52$\pm$0.012  \\
  & {\ttfamily IRAC_CH3      } & 0.76 & $-$1.28$\pm$0.02 & 7.66$\pm$0.021  \\
  & {\ttfamily IRAC_CH4      } & 0.90 & $-$1.03$\pm$0.02 & 7.78$\pm$0.021  \\
  & {\ttfamily WISE_W3       } & 1.06 & $-$1.16$\pm$0.05 & 7.48$\pm$0.051  \\
  & {\ttfamily WISE_W4       } & 1.34 & $-$1.20$\pm$0.02 & 7.15$\pm$0.024  \\
  & {\ttfamily MIPS_CH1      } & 1.37 & $-$1.15$\pm$0.02 & 7.17$\pm$0.021  \\
  & {\ttfamily MIPS_CH2      } & 1.85 & 0.11$\pm$0.07 & 7.95$\pm$0.072  \\
\enddata
\tablecomments{Col.(1) provides the galaxy name in the first row corresponding to the start of that galaxy's data.  The adopted Galactic extinction, $A_V^{\rm Gal}$, is also listed next to the galaxy's name.  All flux measurements have been corrected for Galactic extinction. Col.(2) lists the filter using the notation provided in \lightning\ (see $\S$~\ref{sub:sfh} for details). Col.(3)--(5) provide the base-10 logarithm of the central wavelength of the filter, the flux density and 1$\sigma$ error in units of Jy, and the monochromatic luminosity and 1$\sigma$ error, respectively.  The fluxes and luminosities are appropriate for the regions described in $\S$\ref{sub:sed}.  Only a portion of the table is shown here to illustrate photometric content.  The information in this table for all \ngal\ galaxies and photometric bandpasses is provided in the electronic edition of this paper.}
\label{tab:phot}
\end{deluxetable}

Next, for each galaxy, we convolved all images, using foreground-star-subtracted images when relevant, to common PSF and pixel scales to form data cubes.  We constructed two such data cubes with 6~arcsec and 40~arcsec FWHM Gaussian PSFs (both at 3~arcsec pixel scale).  A given data cube contains only data sets with native resolution that is sharper than the FWHM of the convolved images.  As such, the 6~arcsec PSF data cubes contain many bands spanning 0.1--10~$\mu$m (e.g., \galex, \swift, \hst, SDSS, PanSTARRS, CTIO, 2MASS, \spitzer\ IRAC, \jwst, \wise\ band 1) and \herschel\ PACS 70$\mu$m, when available, while the 40~arcsec PSF data cubes provides expanded coverage in the IR, adding all \wise\ bands, \spitzer\ MIPS (but not 160$\mu$m), as well as \herschel\ PACS and SPIRE.  In Figure~\ref{fig:psf}, we show example images from data cubes extracted for NGC~0628, a galaxy in our sample in which the central circular region of 6~arcsec diameter was flagged for exclusion due to the presence of a bright X-ray source that may be an AGN.  This shows the form and quality of the data cubes used in this paper (see also Fig.~\ref{fig:img} for images of all galaxies built from our data cubes).  

\subsection{Spectral Energy Distribution (SED) Extractions}\label{sub:sed}

For all galaxies, we made use of the 40~arcsec PSF data cubes for extracting galaxy-integrated SEDs.  To do this, we used elliptical
apertures that were chosen to be both large enough to encompass large fractions of
the stellar content of the galaxies, but also small enough to limit contributions
from \xray\ point-source populations that are unrelated to the galaxies
themselves (i.e., background AGN and galaxies and foreground Galactic stars).
For 65 galaxies, we found that the $K_s$-band
20~mag~arcsec$^{-2}$ isophotal ellipses, based on 2MASS $K_s$-band data \citep{Jar2003}, achieved such a compromise.  However, for actively star-forming galaxies with high-sSFRs, in particular dwarf starbursts, the $K_s$-band 20~mag~arcsec$^{-2}$ semi-major axes were small relative to the extents of the galaxies in bluer bands.  For such cases, we chose to adopt positions and sizes from the HyperLeda database\footnote{
\url{http://leda.univ-lyon1.fr/}} (based on $B$-band 25~mag~arcsec$^{-2}$ isophotes; 9 galaxies), the \galex\ FUV-based size parameters presented in \citet{Ged2024} (9 galaxies), or through manual construction, by eye (5 galaxies).  The latter sizes were constructed using \galex\ FUV or SDSS $u$-band images when other galactic footprints from the literature did not clearly encompass all obvious galactic structures or when the morphologies were complex (e.g., the Antennae).  As an example, the apertures used to extract photometry from NGC~0628 are displayed in Figure~\ref{fig:psf} as red ellipses.

Background levels were measured using image median values from elliptical annuli that were chosen to have inner and outer radii that were scaled factors of the elliptical photometry extraction regions. The scales of these factors were manually determined based on image inspection and ranged from $\approx$1.5--5 times the extraction radius.  In most cases, the resulting photometry was not highly sensitive to our choice of these apertures, provided they were well outside regions of the most intense emission and the S/N of the galaxy was high.  

For a given bandpass, the background-subtracted galaxy-integrated flux density was calculated by summing all $N$ pixels within the elliptical regions following:
\begin{equation}\label{eqn:flux}
F_\nu = \xi_\nu^{\rm cal} \xi_\nu^{\rm col} \sum_{i=1}^{N} (\phi_\nu^i - \langle \phi_\nu^{\rm bkg} \rangle ),
\end{equation}
where $\xi_\nu^{\rm cal}$ represents the calibration constant that converts instrument units, $\phi$, to flux-density units, $\xi_\nu^{\rm col}$ represents any color-corrections applied (see below), and $\langle \phi_\nu^{\rm bkg} \rangle$ is the median value of the background intensity, as determined from $M$ pixels in the background elliptical annuli. The uncertainty on the background-subtracted flux density was calculated as
\begin{equation}\label{eqn:unc}
\frac{\sigma_\nu^2}{F_\nu^2} = \left(N + \frac{N^2}{M}\right) \left(\frac{\xi_\nu^{\rm cal}\sigma_{\rm bkg}}{F_\nu}\right)^2 + \left(\frac{\sigma_{\rm cal}}{\xi_\nu^{\rm cal}}\right)^2 + \left(\frac{\sigma_{\rm col}}{\xi_\nu^{\rm col}}\right)^2,
\end{equation}
which contains the variance on $\phi_\nu^{\rm bkg}$ from the $M$ background pixels, $\sigma_{\rm bkg}^2$, the calibration uncertainty, $\sigma_{\rm cal}$, as a fraction of the source flux density, and a color-correction uncertainty term, $\sigma_{\rm col}$, when relevant.

For the majority of the bandpasses, we set $\xi_\nu^{\rm col} = 1$ and $\sigma_{\rm col} = 0$.  However, for \wise\ W3 (11~$\mu$m) and \spitzer\ M2 (70$\mu$m), the wavelength width (full widths of $\approx$10~$\mu$m and $\approx$16~$\mu$m, respectively) of the bandpasses and relatively wide variations of spectral shapes across these bandpasses (as well as differences with adopted calibrators) called for color-dependent corrections to be applied.\footnote{See, e.g., discussions at the \href{https://wise2.ipac.caltech.edu/docs/release/allsky/expsup/sec4_4h.html}{WISE Data Processing} cite related to \wise\ Band~3 \citep{Wri2010}.} These color corrections were determined by performing a first-pass SED fitting for all galaxies in the sample, as described below in $\S$\ref{sec:sfh}, and assessing how the residuals to the fits (as a ratio of $F_\nu^{\rm model}/F_\nu^{\rm obs}$) varied as functions of spectral shape around the bands, using color proxies.  Specifically, we investigated the relationship of the residuals for \wise\ 11$\mu$m versus observed \wise\ $f_\nu$[22$\mu$m]/$f_\nu$[11$\mu$m] and \spitzer\ 70$\mu$m residuals versus observed \spitzer\ $f_\nu$[70$\mu$m]/$f_\nu$[24$\mu$m] for our galaxy sample. We found clear relationships between residual and color, and quantified these relationships using least-squares fitting to arrive at the following color correction formulae:
$$f_{\rm W3}^{\rm corr} = f_{\rm W3}^{\rm orig} \left[ 1.067 + 0.249 \left( \frac{f_{\rm W4}^{\rm orig}}{f_{\rm W3}^{\rm orig}} \right)  -0.023 \left( \frac{f_{\rm W4}^{\rm orig}}{f_{\rm W3}^{\rm orig}} \right)^2 \right]$$
\begin{equation}\label{eqn:color}
    f_{\rm M2}^{\rm corr} = f_{\rm M2}^{\rm orig} \left[ 1.474 - 0.026 \left( \frac{f_{\rm M2}^{\rm orig}}{f_{\rm M1}^{\rm orig}} \right)   \right]
\end{equation}
where $f_{\nu}^{\rm corr}$ represents the color-corrected flux density, and $f_{\nu}^{\rm orig}$ represents the observed flux density prior to correction.  The fractional color corrections, as used in Eqns.~\ref{eqn:flux} and \ref{eqn:unc}, are specified as $\xi_\nu^{\rm col} = f_\nu^{\rm corr}/f_\nu^{\rm orig}$, and their uncertainties are calculated as the residual scatter to the relations in Eqn.~\ref{eqn:color}, which are $\sigma_{\rm col} = 0.16$ and 0.15 for \wise\ W3 and \spitzer\ M2, respectively.

For \nrem\ of the galaxies in our sample, the presence of a low-luminosity AGN or extreme X-ray source crowding make XLF analyses in the central regions intractable, and we chose to exclude X-ray data in these small regions (typically $\approx$6--20~arcsec diameter circles; Col.(4) in Table~\ref{tab:sam}) from our analyses.  Thus, for these cases, we restricted our analyses to ``annuli,'' which were constructed using ``total'' elliptical regions with circular aperture ``centers'' excluded.
To estimate the contributions of the centers to the total SEDs across the full FUV-to-FIR spectral range using the common 40~arcsec PSF photometry is not tenable, due to the PSF being larger than the centers themselves. We therefore chose to take a forward-modeling approach, in which we used the higher-resolution 6~arcsec PSF data cubes to estimate the SEDs from the centers for a subset of the bands and forward-modeled the center SED across all bands (including those at poorer resolution) to extract its contribution to the total SED across the full FUV-to-FIR spectral range. We describe this procedure in full detail in the next section. The detailed parameters of our elliptical extraction regions and the radii of excluded central regions are provided in Table~\ref{tab:sam}.

%
%
\begin{figure*}
\centerline{
\includegraphics[width=18cm]{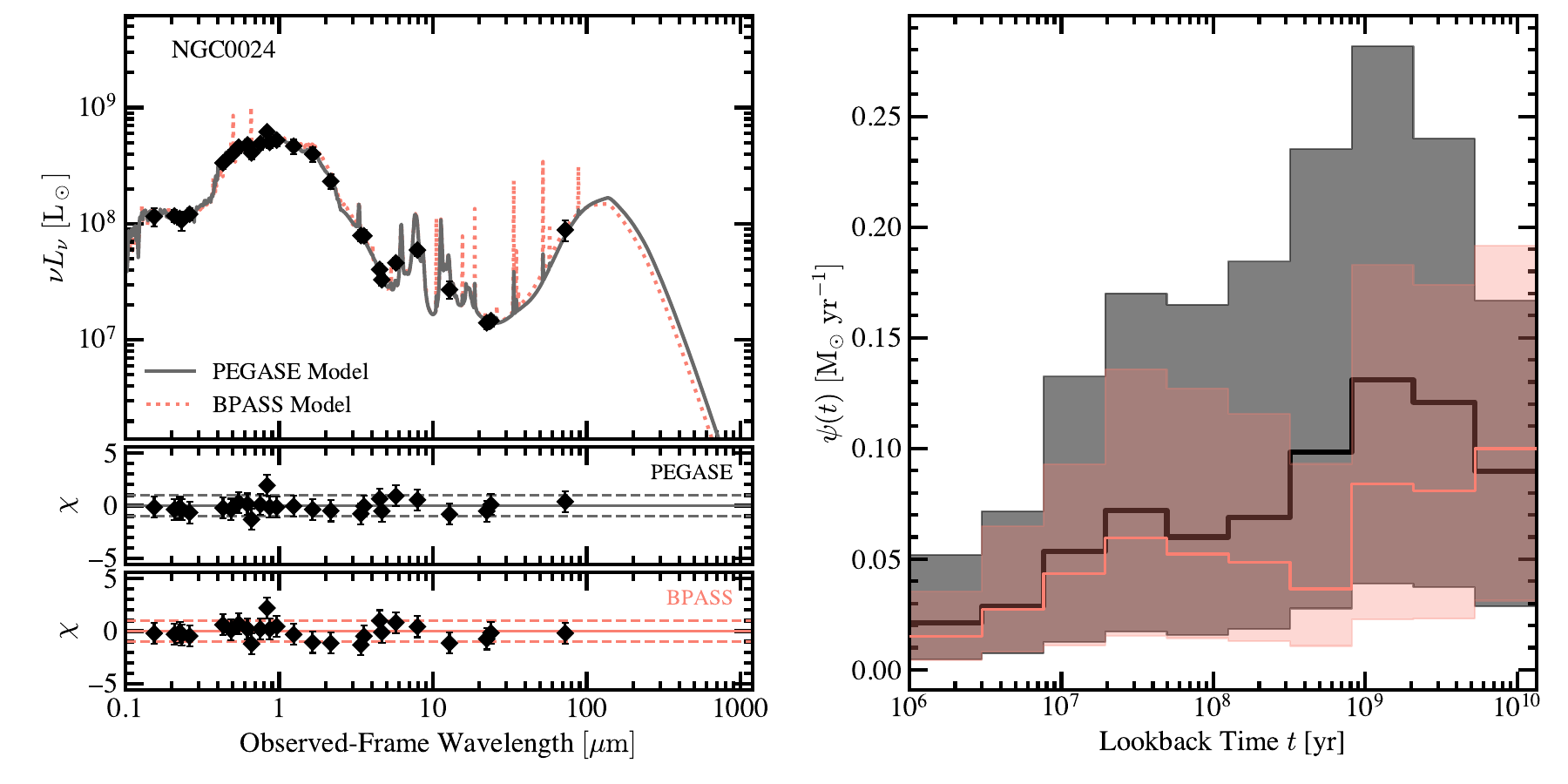}
}
\centerline{
\includegraphics[width=18cm]{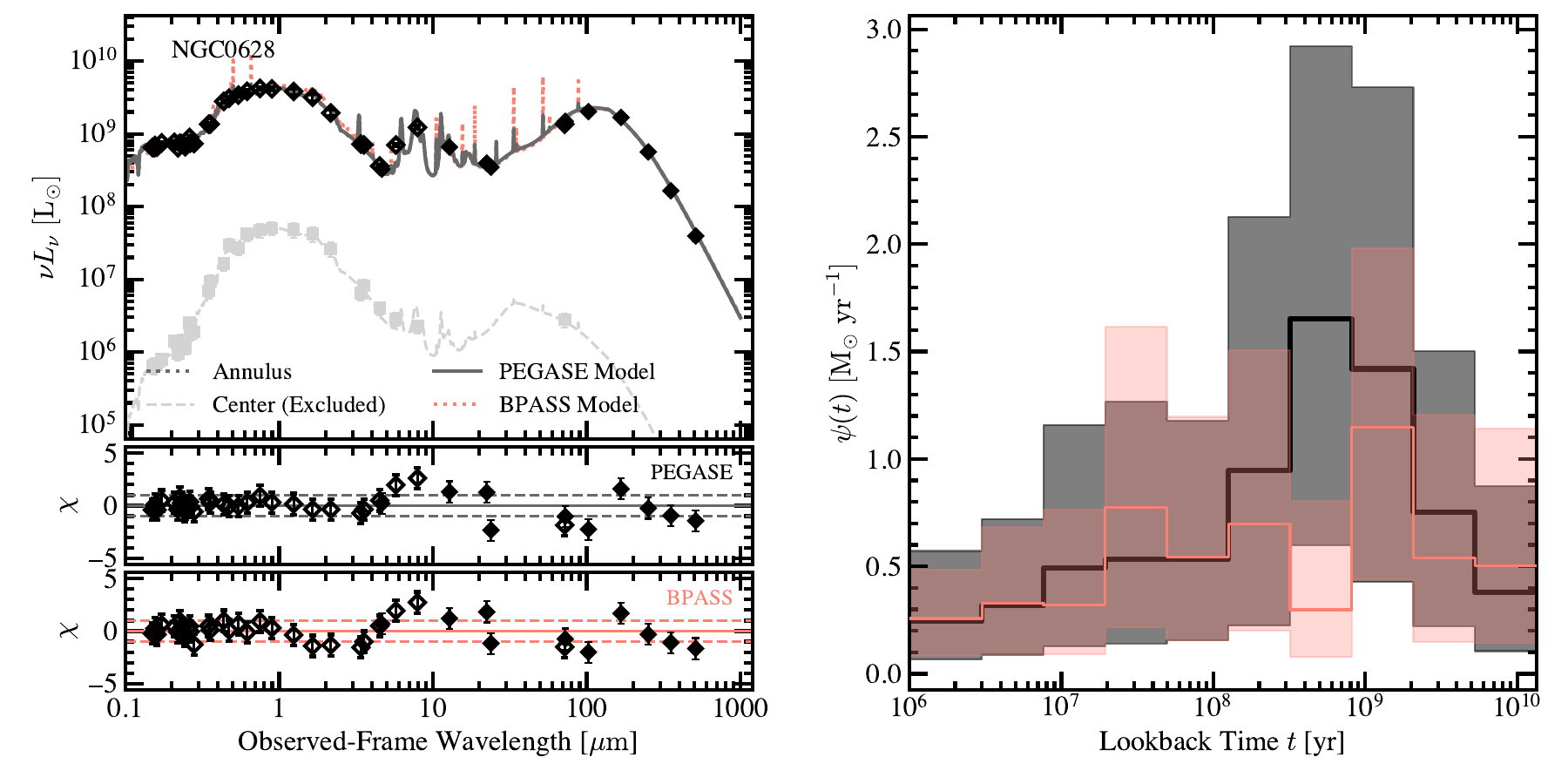}
}
\caption{
({\it left panels\/}) Example SEDs ($\nu L_\nu$ versus wavelength) and fit residuals for NGC~0024 (top) and NGC~0628 (bottom), the first galaxies in our sample when sorted by ascending right ascension that use the single \lightning\ and \multilightning\ procedures, respectively.
Photometric constraints on ``total'' regions are shown
as {\it black filled diamonds\/}, ``annuli'' as {\it open black diamonds\/}, and centers with {\it light-grey filled squares\/} with 1$\sigma$ uncertainties displayed for all.  Our
best-fitting models from {\ttfamily Lightning} are overlaid 
as solid-black and red-dashed curves for fits that use \pegase\ and \bpass, respectively, for the underlying stellar models. 
The bottom panels show residuals to the best fit models in units of $\chi = $~(data$-$model)$/\sigma$.
({\it right panels\/}) Resulting 10-step SFHs for both galaxies, in terms of average SFR ($\psi$) per age step ($t$), as derived by \lightning\ (NGC~0024) and \multilightning\ (annulus region for NGC~0628), assuming \pegase\ ({\it black with gray shading\/}) and \bpass\ models ({\it red with transparent red shading\/}). Each shaded region represents the 16--84\% confidence interval with the dark lines providing the median estimate on the SFHs.  Clear differences between \pegase\ and \bpass\ solutions are observed near $\approx$100~Myr to $\approx$2~Gyr, which is systematic across all galaxies (see Appendix~\ref{sec:appA} for further details).   Results presented throughout this paper are based on the SED fits from \pegase.
}
\label{fig:sed}
\end{figure*}

We corrected the photometry of each filter for Galactic (Milky Way)
extinction, using a \citet{Fit1999} reddening law with total-to-selective extinction ratio $R_V = 3.1$.  Values of $A_V$ for each galaxy were taken from the IRSA Galactic Dust Reddening and Extinction tool,\footnote{\url{https://irsa.ipac.caltech.edu/applications/DUST/}} which uses the \citet{Sch2011} recalibration of the \citet{Sch1998}
Cosmic Background Explorer (COBE) Diffuse Infrared Back-
ground Experiment (DIRBE) and Infrared Astronomical
Satellite (IRAS) Sky Survey Atlas (ISSA) dust maps.

In Table~\ref{tab:phot}, we provide the resulting photometry, as extracted following the procedure discussed above.  All values provided are corrected for extinction and are used in our \lightning\ SED fitting procedure to extract SFHs.

\subsection{SED Fitting Procedure and Resulting SFHs}\label{sub:sfh}

When fitting the SEDs, we adopted a piecewise-continuous SFH model in \lightning, which consists of the summation of spectral contributions from stellar populations formed in $n_{\rm SFH}$ independent constant-SFR time steps.  When linked together, the time steps span all cosmic lookback times, which we take here as 0--13.4~Gyr.  In this procedure we tested separately results based on stellar SEDs from the spectral population synthesis libraries from \pegase\ \citep{Fio1997} and \bpass\ \citep[v2.1;][]{Eld2017}.  For the latter, we made use of the \citet{Cha2003} IMF models ({\ttfamily imf_chab300}) that include binary stars.  In \lightning\ both $n_{\rm SFH}$ and the specific time intervals can be chosen by the user.  We experimented with choices of these intervals, and found that $n_{\rm SFH} = 10$ nearly logarithmically-spaced time intervals (see Table~\ref{tab:sed} for details) provided good characterizations of the SEDs with well-converged posterior distributions on fitting parameters.  Stellar population metallicities were fixed to the values provided in Table~\ref{tab:sam}.

Nebular emission associated with \HII\ regions was modeled in \lightning\ using {\ttfamily Cloudy} \citep{Fer1993,Fer2013}.
In this work, we utilize the photoionization modeling code {\ttfamily Cloudy} \citep{Fer2017}
to generate synthetic spectra of \HII\ regions. {\ttfamily Cloudy} calculates the full radiative transfer through the gas cloud, so each individual \HII\ region model has internal structure, with radial variations in ionization state and temperature, which in turn affects the location within the nebula where various emission lines are produced.

The key parameters of the photoionized gas in {\ttfamily Cloudy} and their range of variation are as follows. The ionization parameter ($\mathcal{U}$) is defined as the ratio of ionizing photon density to hydrogen density. We compute models for values of $\log \mathcal{U}$ in the range between $-4.0$ to $-1.0$, in steps of 0.5 dex. We compute models for hydrogen densities of the ionized gas ($n_{\rm H}$) for the values $\log n_{\rm H}/$cm$^3$ = 2 and 3.5, corresponding roughly to the observed values of electron densities in extragalactic \HII\ regions and star-forming galaxies. The shape of ionizing radiation field produced by a star depends on the age of the stellar population and on its metallicity. We use stellar population models from both the \bpass\ spectral synthesis code \citep{Eld2017} and \pegase\ \citep{Fio1997}, in a wide range of values of both stellar ages (1 to 40 Myr in steps of 0.1 dex) and stellar metallicities ($Z=6\times10^{-4}$ to 0.025 in steps of 0.1 dex). 
In each generated model, we set the metallicity of the stellar population model equal to the gas phase metallicity of the nebula. Throughout this work, we assumed the gas nebula without dust grains and consider models at constant pressure. 

We renormalized the transmitted continuum output by {\ttfamily Cloudy} to the stellar mass of the input stellar population, and we used these new templates in our SED fitting procedure. Where the stellar age is older than 40 Myr, we fall back on the source spectral templates from \bpass\ and \pegase, implicitly assuming that populations older than 40 Myr make no contribution to the nebular emission. Future updates to \lightning\ will introduce the functionality to fit observed line ratios, providing an additional handle on the SFH, but we do not directly use line fluxes from {\ttfamily Cloudy} in this work.  

In our fitting procedure, we assumed a fixed nebular density at $\log (n_{\rm H}/{\rm cm^{3}}) = 3.5$, ionization parameter at $\log \mathcal{U} = -2$, and stellar metallicities tied to the nebular metallicity listed in Col.(11) of Table~\ref{tab:sam}.  Given that the majority of our data is broadband photometry without strong constraints on nebular features, and that metallicity values are constrained elsewhere (often times by spectroscopy), our choices of nebular density and ionization parameter do not have a material impact on the resulting SFHs.

We adopt the \lightning\ implementations of the \citet{Nol2009} dust attenuation and \citet{Dra2007} dust emission models, as described in \citet{Doo2023}. We chose to vary the two parameters of the \citet{Nol2009} ($\tau_{V,{\rm diff}}$ and $\delta$) and three parameters of the \citet{Dra2007} model ($q_{\rm PAH}$, $U_{\rm min}$, and $\gamma$).  Energy balance is enforced, such that the attenuated power from the stellar model exactly balances the integrated dust emission model power.

\begin{deluxetable*}{lllr}[t]
\tabletypesize{\footnotesize}
\tablecolumns{4}
\tablecaption{\label{tab:sed}Summary of SED Fitting Parameters.}
\tablehead{\colhead{Model Component} & \colhead{Parameter} & \colhead{Parameter Description} & \colhead{Value/Range$^a$}}
\startdata
Stellar Population & $\{\psi\}_{i=1}^{n_{\rm SFH}}$                & Star formation history coefficients in \sfr, with $n_{\rm SFH} = 10$ age bins:       & $[0, 1000]$           \\
 &   & \multicolumn{2}{l}{$\{\log t_i [{\rm yr}] \} =$~\{ [$<6.47$], [6.47--6.88], [6.88--7.29], [7.29--8.10], [8.10--8.50], }\\
  &  & \multicolumn{2}{l}{\phantom{$\log t_j [{\rm yr}] =$~} [8.91--9.32], [8.50--8.91],  [9.32--9.72], [9.72--10.12] \} } \\
  & $Z$ & Metallicity of stellar population & Col.(11) Table~\ref{tab:sam} \\
Nebular Effects & $\log \mathcal{U}$ & Ionization parameter & $-2$ \\
 & $Z_{\rm neb}$ & Metallicity of ISM & tied($Z$) \\
 & $\log (n_{\rm H}/{\rm cm}^3)$ & Density of the ionized ISM & 3.5 \\
Dust Attenuation & $\tau_{V, {\rm Diff}}$  & Optical depth of diffuse dust in the $V$ band         & $[0, 3]$           \\
 & $\delta$                & Attenuation curve power-law slope deviation from \citet{Cal2000} law  & $[-2.3, 0.4]$           \\
 & $\tau_{V, {\rm BC}}$    &  Optical depth of birth-cloud dust in the $V$ band                        & $0$                \\
Dust Emission & $\alpha$                & Power-law slope of intensity distribution                        & $2$                \\
 & $U_{\rm min}$           & Intensity distribution minimum &  [0.1, 25]      \\
 & $U_{\rm max}$           & Intensity distribution maximum &  $3 \times 10^5$    \\
 & $\gamma$             & Mass fraction of dust exposed to intensity distribution   & $[0, 1]$         \\
 & $q_{\rm PAH}$           & Mass fraction of PAHs in dust mixture & $[0.0047, 0.0458]$  \\
\enddata
\vspace{0.1in}
$^{a}$Free parameters are indicated as such by the parameter ranges in square brackets. Priors on all free parameters follow uniform distributions.  Fixed parameters and their values are indicated as single numbers.
\label{tab:param}
\end{deluxetable*}

Table~\ref{tab:sed} provides a summary of the parameters used in our SED fitting procedure, including assumed priors on the parameters.
In total, our fits are typically based on 15 free parameters (ten SFH parameters plus
five dust attenuation and emission parameters); however, as we describe below, when the nuclear region is excluded from our analyses additional parameters are used to model the nuclear region SED.
We fit all parameters using flat priors with parameter ranges constrained to the domains of their applicability.  
The posterior distribution functions (PDFs) of the fit parameters were sampled
using the {\ttfamily emcee} procedure in {\ttfamily python} \citep{For2013}, which is a Markov Chain Monte Carlo (MCMC) sampler based on the \citet{Goo2010} Affine-invariant algorithm.  For the majority of the SED fits, we utilized 60 walkers, run over 50,000 MCMC steps. Final MCMC parameter chains were built by discarding all but the last 20,000 steps, and thinning the final walker-combined chains by a factor of 600.  For this setup, we found good convergence in MCMC parameter chains for the majority of our galaxies. However, a small number of galaxies required longer MCMC runs (up to $1.5 \times 10^5$ steps) to achieve full convergence.

For the subsample of \nrem\ galaxies that had central regions excluded from our analyses, we performed SED fitting using a joint spectral fitting procedure, which we hereafter refer to as \multilightning\footnote{\url{https://github.com/ebmonson/multilightning}}.  The goal of this procedure is to decompose the total SED solution into central region and elliptical annular region contributions, given data sets with widely different angular resolution. Thus, \multilightning\ will construct nuclear and annular region SED models with independent SFH parameters that for a given galaxy simultaneously fit (1) the nuclear region spectrum using the 6~arcsec data cube, (2) the annular region using the 6~arcsec data cube, and (3) the total region using the 40~arcsec data cube bands with PSFs too large to be contained within the 6~arcsec data cube. For the majority of these \nrem\ galaxies, we found that the nuclear region provides nearly negligible contributions to the overall SEDs (typically $\simlt$10\%).  As such, for the nuclear regions, we applied simpler SED models, which contained half the number of SFH steps and often times fewer dust emission parameters, as we have varying constraints on the $>$6$\mu$m SEDs within the 6~arcsec data cubes.  Specifically, when the longest wavelength constraint for the nuclear region was $\lambda_{\rm max}^{\rm nuc} < 6$~$\mu$m, all nuclear-region dust emission parameters ($U_{\rm min}$, $\gamma$, and $q_{\rm PAH}$) were linked to the total model values; when $\lambda_{\rm max}^{\rm nuc} =$~6--10~$\mu$m, we allowed $q_{\rm PAH}$ for the nuclear region model to vary independently and linked all other parameters to the total model values; and when $\lambda_{\rm max}^{\rm nuc} > 10$~$\mu$m, we allowed all nuclear-region dust emission parameters (i.e., $U_{\rm min}$, $\gamma$, and $q_{\rm PAH}$) to vary independently in the fits.

In the bottom panels of Figure~\ref{fig:sed}, we provide a \multilightning\ example for the galaxy NGC~0628, the first galaxy in our sample in R.A. order that we excluded a nuclear region.  This example is representative of the typical level of contributions that the nuclear regions make to the SEDs.  For most galaxies, the central contribution is very-low to negligible compared to the galaxy-integrated SED, with the exception of NGC~7552, which contains an excluded central circumnuclear starburst that dominates the galaxy-integrated infrared emission \cite[see extended materials, and][for further details]{Wes2023}. In this example, black open diamonds and filled diamonds represent the elliptical annular and total galaxy regions, respectively, while the gray squares show the contribution from the nuclear region. Each bandpass with photometric constraints on the nuclear region also has equivalent estimates of the photometry on the elliptical annular region.

\begin{deluxetable*}{lrrrrrrrrr}
\tablewidth{1.0\columnwidth}
\tabletypesize{\footnotesize}
\tablecaption{SED Fitting Parameter Results}
\tablehead{
\multicolumn{1}{c}{\sc Galaxy} & \multicolumn{9}{c}{SFH parameters, $\log \psi_i$ (\sfr)} \\
\vspace{-0.25in} \\
\multicolumn{1}{c}{\sc Name} & \colhead{$\log \psi_1$} & \colhead{$\log \psi_2$} & \colhead{$\log \psi_3$} & \colhead{$\log \psi_4$} & \colhead{$\log \psi_5$} & \colhead{$\log \psi_6$} & \colhead{$\log \psi_7$} & \colhead{$\log \psi_8$} & \colhead{$\log \psi_9$} \\
\vspace{-0.25in} \\
\multicolumn{1}{c}{(1)} & \multicolumn{1}{c}{(2)} & \multicolumn{1}{c}{(3)} & \colhead{(4)} & \colhead{(5)} & \colhead{(6)} & \colhead{(7)} & \colhead{(8)} & \colhead{(9)} & \colhead{(10)} }
\startdata
NGC0024 & $-$1.67$^{+0.39}_{-0.65}$ & $-$1.54$^{+0.40}_{-0.59}$ & $-$1.27$^{+0.40}_{-0.62}$ & $-$1.14$^{+0.37}_{-0.62}$ & $-$1.22$^{+0.44}_{-0.58}$ & $-$1.16$^{+0.43}_{-0.58}$ & $-$1.01$^{+0.38}_{-0.55}$ & $-$0.88$^{+0.33}_{-0.53}$ & $-$0.92$^{+0.30}_{-0.51}$ \\ 
NGC0337 & $-$0.06$^{+0.33}_{-0.56}$ & 0.21$^{+0.30}_{-0.53}$ & 0.29$^{+0.35}_{-0.55}$ & 0.35$^{+0.34}_{-0.55}$ & 0.14$^{+0.39}_{-0.59}$ & 0.02$^{+0.38}_{-0.64}$ & $-$0.07$^{+0.36}_{-0.57}$ & $-$0.13$^{+0.34}_{-0.53}$ & $-$0.22$^{+0.35}_{-0.56}$ \\ 
NGC0584 & $-$2.41$^{+0.41}_{-0.61}$ & $-$2.21$^{+0.34}_{-0.63}$ & $-$2.05$^{+0.43}_{-0.58}$ & $-$1.88$^{+0.41}_{-0.57}$ & $-$1.73$^{+0.39}_{-0.57}$ & $-$1.48$^{+0.39}_{-0.60}$ & $-$0.66$^{+0.44}_{-0.59}$ & 0.12$^{+0.34}_{-0.48}$ & 0.74$^{+0.26}_{-0.46}$ \\ 
NGC0625 & $-$1.52$^{+0.53}_{-0.63}$ & $-$1.63$^{+0.57}_{-0.61}$ & $-$1.48$^{+0.57}_{-0.60}$ & $-$1.38$^{+0.56}_{-0.65}$ & $-$1.37$^{+0.46}_{-0.58}$ & $-$1.16$^{+0.47}_{-0.56}$ & $-$1.20$^{+0.39}_{-0.59}$ & $-$1.41$^{+0.36}_{-0.57}$ & $-$1.52$^{+0.39}_{-0.54}$ \\ 
NGC0628 & $-$0.61$^{+0.36}_{-0.52}$ & $-$0.49$^{+0.36}_{-0.56}$ & $-$0.32$^{+0.41}_{-0.52}$ & $-$0.30$^{+0.38}_{-0.54}$ & $-$0.25$^{+0.34}_{-0.54}$ & $-$0.06$^{+0.34}_{-0.65}$ & 0.23$^{+0.24}_{-0.44}$ & 0.13$^{+0.30}_{-0.51}$ & $-$0.13$^{+0.33}_{-0.52}$ \\ 
NGC0925 & $-$0.87$^{+0.36}_{-0.63}$ & $-$0.78$^{+0.36}_{-0.53}$ & $-$0.65$^{+0.37}_{-0.55}$ & $-$0.48$^{+0.35}_{-0.58}$ & $-$0.38$^{+0.34}_{-0.61}$ & $-$0.28$^{+0.32}_{-0.53}$ & $-$0.44$^{+0.38}_{-0.53}$ & $-$0.50$^{+0.37}_{-0.64}$ & $-$0.55$^{+0.32}_{-0.57}$ \\ 
NGC1023 & $-$1.81$^{+0.57}_{-0.58}$ & $-$1.62$^{+0.43}_{-0.61}$ & $-$1.42$^{+0.60}_{-0.58}$ & $-$1.36$^{+0.68}_{-0.61}$ & $-$1.13$^{+0.48}_{-0.66}$ & $-$0.85$^{+0.57}_{-0.54}$ & $-$0.55$^{+0.46}_{-0.78}$ & 0.07$^{+0.30}_{-0.54}$ & 0.51$^{+0.27}_{-0.66}$ \\ 
NGC1097 & 0.56$^{+0.29}_{-0.56}$ & 0.57$^{+0.34}_{-0.58}$ & 0.81$^{+0.39}_{-0.61}$ & 0.93$^{+0.35}_{-0.58}$ & 0.76$^{+0.38}_{-0.58}$ & 0.90$^{+0.41}_{-0.54}$ & 1.07$^{+0.28}_{-0.50}$ & 1.04$^{+0.29}_{-0.50}$ & 0.83$^{+0.31}_{-0.59}$ \\ 
NGC1291 & $-$1.44$^{+0.37}_{-0.57}$ & $-$1.41$^{+0.41}_{-0.55}$ & $-$1.08$^{+0.40}_{-0.54}$ & $-$0.90$^{+0.35}_{-0.53}$ & $-$0.84$^{+0.33}_{-0.59}$ & $-$0.59$^{+0.43}_{-0.54}$ & $-$0.22$^{+0.36}_{-0.55}$ & 0.10$^{+0.37}_{-0.60}$ & 0.52$^{+0.34}_{-0.55}$ \\ 
NGC1313 & $-$0.86$^{+0.40}_{-0.58}$ & $-$0.78$^{+0.38}_{-0.56}$ & $-$0.69$^{+0.38}_{-0.55}$ & $-$0.59$^{+0.43}_{-0.59}$ & $-$0.49$^{+0.36}_{-0.57}$ & $-$0.45$^{+0.35}_{-0.53}$ & $-$0.53$^{+0.32}_{-0.54}$ & $-$0.85$^{+0.38}_{-0.57}$ & $-$0.99$^{+0.32}_{-0.52}$ \\ 
\enddata
\tablecomments{The full version of this table contains parameter estimations for all 15 SED fitting parameters across all \ngal\ galaxies.  Only a portion of the table is shown here for illustration of form and content.  Quoted parameter values include median and 16--84\% confidence intervals on the parameter values. Col.~(1): galaxy name. Col.(2)--(11): SFH values at each time step. Col.(12): optical depth, $\tau_{V, {\rm diff}}$. Col.(13): Attenuation curve deviation from \citet{Cal2000} law, $\delta$. Col.(14): Dust-irradiation intensity distribution minimum $U_{\rm min}$. Col.(15): Mass fraction of dust exposed to intensity distribution, $\gamma$. Col.(16): Mass fraction in dust mixture as PAH, $q_{\rm PAH}$.}
\label{tab:sfit}
\end{deluxetable*}

In Table~\ref{tab:sfit}, we provide the resulting parameter best-fit values, medians, and 16--84\% confidence intervals.  From the SFH derivations, we derived the more commonly quoted properties of SFR and $M_\star$ following:
\begin{equation}\label{eqn:sfr}
{\rm SFR} = \frac{1}{125~{\rm Myr}}\sum_{i=1}^4 \psi_i \Delta t_i,
\end{equation}
where $\Delta t_i$ represents the time interval for the $i^{\rm th}$ SFH step, and
\begin{equation}\label{eqn:mstar}
M_\star = \sum_{i=1}^{10} {\cal R}_i \psi_i \Delta t_i,
\end{equation}
where ${\cal R}_i$ converts the total stellar mass formed in stars within the $\Delta t_i$ interval to surviving, present-day stellar mass.  The values of ${\cal R}_i$ depend both on the age bin of the stellar population and the metallicity of the model.  All values of SFR and $M_\star$ used throughout this paper are based on these calculations.

To illustrate the quality of the data and SED fits with \lightning, we created Figure~\ref{fig:sed}, which shows the data, models, residuals, and inferred SFHs for example cases of NGC~0024 and NGC~0628. These cases are the first examples in our sample (in R.A. order) where (1) the SED photometry across the full galactic extent was modeled using a single \lightning\ model (NGC~0024) and (2) SED photometry for the nuclear region and elliptical annular region were modeled as separate components using \multilightning\ (NGC~0628).  For both examples fits are presented based on \pegase\ (black curves) and \bpass\ (red curves) models for comparison.  The full catalog of these diagrams is provided in the electronic version of this article.

As we discuss in detail in Appendix~\ref{sec:appA}, we find that comparisons between \pegase\ and \bpass\ stellar population models provide highly-consistent SFH values for all galaxies, with the exception of the three SFH bins $\psi_6$--$\psi_8$ (spanning 0.13--2.1~Gyr), for which \bpass\ fits produce systematically lower values compared to \pegase. We find that this result is driven primarily by differences in the spectral models near 1.5--2.5~$\mu$m, for which \bpass\ produces much higher predictions of the stellar population fluxes than \pegase\ due to the \bpass\ treatment of AGB stellar population emission \citep{Sta2017}. As such, \bpass\ fits require lower values of $\psi_1$, and $\psi_6$--$\psi_8$ to reproduce the data.  Notably, these differences are mainly unrelated to the binary-star aspect to \bpass\ compared to \pegase, and we find that for the data considered here, the inclusion of binary-star prescriptions through \bpass\ do not impact the SFHs (see Appendix~\ref{sec:appA} for further details). 

When considering comparisons of fit quality, in terms of posterior probability, we find that \pegase\ model fits provide better statistical characterizations of the data for the whole sample, compared to \bpass\ model fits (see Appendix~\ref{sec:appA} and Figure~\ref{fig:bpasspeg} for details).  This is primarily due to excess residuals in the near-IR (see, e.g., examples in Fig.~\ref{fig:sed}).  Given that the \bpass\ and \pegase\ model fit results are in excellent agreement for the majority of the SFH bins, and that disagreements between fits are unrelated to binary-related phenomena, we hereafter choose to proceed using results from our \pegase\ fits when assessing the age and metallicity dependence of the XRB XLF (e.g., results in Tables~\ref{tab:sam} and \ref{tab:sfit} are based on \pegase\ models).  

%
\section{X-ray Luminosity Function Construction}\label{sec:xray}
%

By selection, all galaxies in our sample were observed by \chandra\ ACIS
(either ACIS-I or ACIS-S).  Our data analysis and point-source cataloging
procedure was performed following the procedures outlined in
\citet{Leh2019,Leh2020,Leh2021}, which we summarize in detail in Appendix~\ref{sec:appB}.

%
%
\begin{figure*}
\centerline{
\includegraphics[width=17cm]{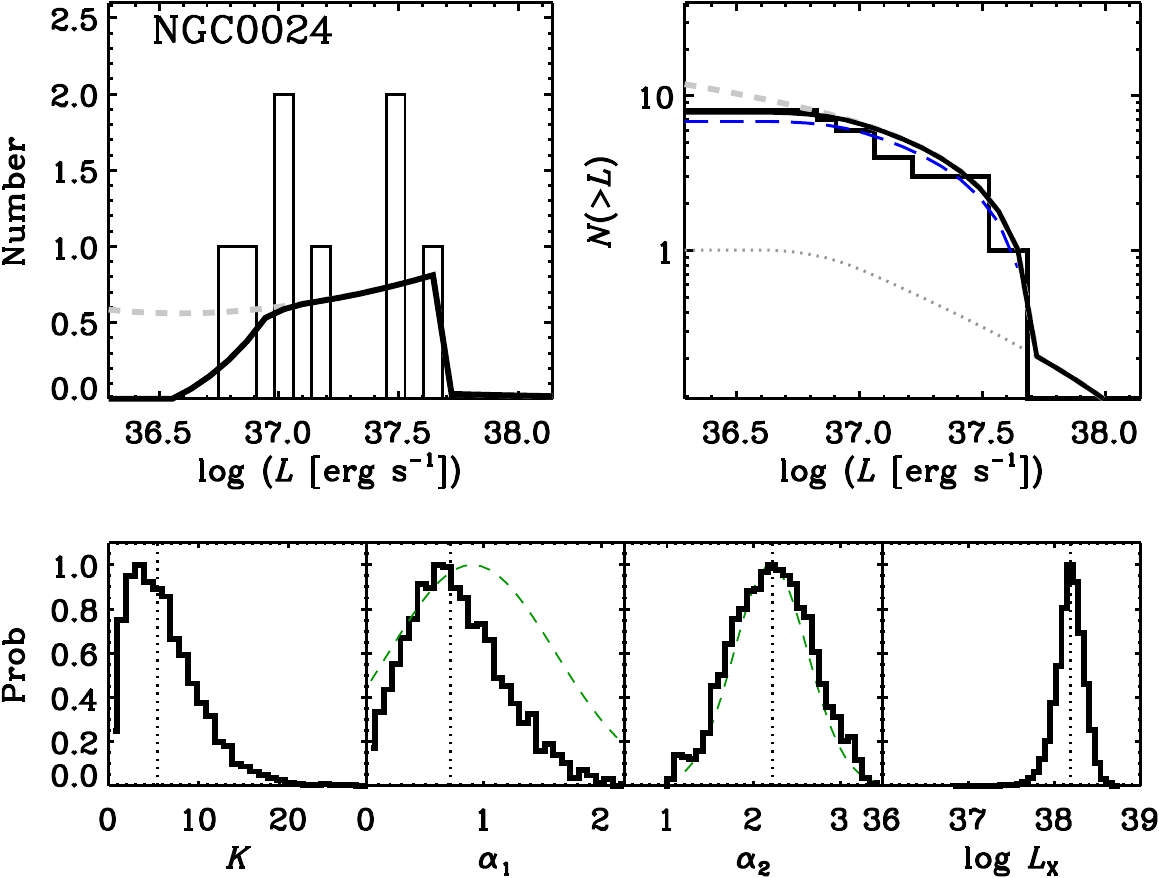}
}
\caption{
Example broken power-law XLF fitting procedure for NGC~0024.  Top panels: ({\it upper left\/}) Number of X-ray detected point sources as a function of luminosity, $L$, as observed in the galaxy ({\it histogram\/}) and best-fit model ({\it black curve\/}).  The best-fit model has been corrected for completeness, and the intrinsic uncorrected model is shown as a dashed gray curve. ({\it upper right\/}) Cumulative XLF ({\it step-like histogram\/}) and best-fitting observed ({\it black curve\/}) and intrinsic models ({\it dashed gray curve\/}).  Contributions from CXB sources ({\it dotted curve}) and sources intrinsic to the galaxy ({\it blue long-dashed curve\/}) are also shown. Bottom panels: ({\it left to right\/}) PDFs for broken power-law model normalization ($K$), low-luminosity slope ($\alpha_1$), high-luminosity slope ($\alpha_2$), and intrinsic population integrated X-ray luminosity ($L_{\rm X}$). Adopted priors on $\alpha_1$ and $\alpha_2$ are shown as green dashed curves.
}
\label{fig:bknpo}
\end{figure*}

One of the first goals of this paper is to calculate {\it intrinsic} X-ray point-source characteristics for the galaxies in our sample, including XLFs and galaxy-integrated point-source luminosities, $L_{\rm X}$.
To achieve this, we began by constructing {\it observed XLFs} for each galaxy, using
the point sources coincident with the areal extents of the galaxies, as
defined in Table~\ref{tab:sam} (i.e., within the galactic ellipses, excluding
any removed central regions).  The observed XLF of a given galaxy is comprised of a histogram of the number of point sources binned in $\log L$ space, where $L$ represents the point-source luminosity, assuming the distance to the galaxy. We adopted bins of width
$\Delta \log L= $~0.078~dex, which corresponds to the typical uncertainty on
$\log L$ for our point-sources, based on uncertainties related to galactic distance and
point-source counts.

In Figure~\ref{fig:bknpo}, we show an example observed XLF for NGC~0024 ({\it upper-left plot}). This representation of the data does not include corrections for incompleteness or unrelated background \xray\ point sources from the cosmic \xray\ background (CXB) and occasional foreground stars that are inevitably present across the extents of the galaxies \citep[e.g.,][]{Kim2007,Leh2012}. Given these factors, it is not always straightforward to measure accurately the {\it intrinsic XLF} of a given galaxy and its corresponding point-source-integrated flux and luminosity.  

To mitigate the above limitations, we fit the observed XLF of each galaxy following a forward-fitting approach, in
which we include contributions from the intrinsic \xray\ sources (the vast
majority of which we expect to be XRBs) and CXB sources, with incompleteness
folded into our models. 
For a given galaxy,
the \xray\ point-source luminosity distribution (i.e., the histogram from in Fig.~\ref{fig:bknpo}, upper-left) was modeled as:
\begin{equation}\label{eqn:mod}
M(L) = \xi(L) \Delta \log L \left[\frac{dN_{\rm int}}{d \log L} + {\rm CXB}(L) \right],
\end{equation}
where the $\xi(L)$ is the luminosity-dependent completeness function for the
galaxy, $dN_{\rm int}/d \log L$ is a model of the {\it intrinsic XLF}, and
CXB$(L$) is the differential number counts from CXB sources.

For the {\it intrinsic} point-source XLF, we 
fit the data using a broken power law form:
\begin{equation}\label{eqn:bkn}
\frac{dN_{\rm int}}{d \log L} = K \log e   \left \{ \begin{array}{lr} L^{-\alpha_1 + 1}  &
\;\;\;\;\;\;\;\;(L < L_b) \\ 
L_b^{\alpha_2-\alpha_1} L^{-\alpha_2 + 1},  & (L_b \le L < L_c) \\ 
0,  & (L \ge L_c) \\ 
\end{array}
  \right.
\end{equation}
where $K$, $\alpha_1$, $L_b$, and $\alpha_2$ are
the broken power-law normalization, low-luminosity slope, break luminosity, and
high-luminosity slope, respectively; both XLF models are truncated above,
$L_c$, the cut-off luminosity.  Throughout the remainder of this paper, we take
$L$, $L_b$, and $L_c$ to be in units of $10^{38}$~\lum, when quoting and
describing normalization values.  For a given galaxy, we fit the data to
determine all constants, except for the break and cut-off luminosities, which
we fix at $L_b = 10^{38}$~\lum\ and $L_c =$~max\{$L$\}$^{\rm gal}_{i}$ (the maximum luminosity of the point-sources in the galaxy).  The choice of fixing $L_c$ to the most luminous detected source ensures that integration of the XLF produces values of the galaxy-integrated luminosity, $L_{\rm X}$, that do not exceed the data constraints.

Also, for many of the galaxies, only a small number of sources are detected on either side of the $L_b$, making it difficult to constrain $\alpha_1$ and/or $\alpha_2$.  To mitigate poor constraints on these parameters in such cases, we adopted Gaussian priors with means and standard deviations of \{$\mu_{\alpha_1}$, $\sigma_{\alpha_1}$\} = \{0.9, 0.5\} and \{$\mu_{\alpha_2}$, $\sigma_{\alpha_2}$\} = \{2.2, 0.3\}, which are based on the XLF fitting results of \citet{Leh2019}.  As we show below, these priors impact the resulting fit when either the observational limits are shallow (e.g., limiting fluxes are larger than $L_b$) or the number of sources detected in the galaxies are small.

For the CXB contribution, we implemented a fixed form for the number counts,
provided by \citet{Kim2007}.  The \citet{Kim2007} extragalactic
number counts provide estimates of the number of sources per unit area versus
0.5--8~keV flux.  The best-fit function follows a broken power-law distribution
with parameters derived from the combined \chandra\ Multiwavelength Project
(ChaMP) and \chandra\ Deep Field-South (CDF-S) extragalactic survey data sets
\citep[see Table~4 of][]{Kim2007}.  For each galaxy, the number counts were
converted to an observed \hbox{0.5--8~keV} XLF contribution by multiplying
the number counts by the areal extent of the galaxy, as defined in Table~\ref{tab:sam}, and
converting CXB model fluxes to \xray\ luminosities, given the distance to the
galaxy.

To complete our model of the observed XLF, we fold both the intrinsic XLF and CXB model contributions through the completeness curve, $\xi(L)$, for the given galaxy. These completeness curves were modeled following the approach detailed in $\S$3.3 of L19, which uses Monte Carlo simulations to calculate the fraction of point sources recovered as a function of source counts ($L$) and location in the X-ray image.  These recovery fractions are weighted across the extent of the galaxy and averaged to obtain the global completeness curve, $\xi(L)$. For points
of reference, we utilize these completeness curves to obtain the 50\% completeness limit, $L_{50}$, which correspond to the point-source luminosity in which 50\% of input sources are recovered in our simulations.  These values
are tabulated in Table~\ref{tab:xlf}.

For each galaxy, we constructed the observed XLF
using 100 luminosity bins of constant $\Delta \log L$ that spanned the range of $L_{\rm min} = 10^{35}$~\lum\ to
$L_{\rm max} = 5 \times 10^{42}$~\lum, and we used only $L \ge L_{50}$ bins for our statistical analyses.  For most galaxies, the majority of the bins contained zero
sources, with other bins containing small numbers of sources.  As such, we
evaluated the goodness of fit using a modified version of the C-statistic
\citep[][]{Cas1979,Kaa2017}:
\begin{equation}
C = 2 \sum_{i=1}^{n_L} M_i - N_i + N_i \ln(N_i/M_i),
\end{equation}
where the summation takes place over the $n_L=100$ bins of \xray\ luminosity, and
$N_i$ and $M_i$ are the observed and model numbers of sources in the $i$th luminosity bin.  We note that when $N_i =
0$, $N_i \ln (N_i/M_i) = 0$, and when $M_i=0$ (e.g., beyond the cut-off
luminosity), the entire $i$th term in the summation is zero.

To identify best-fit values and sample posterior distributions of the model parameters, we made use of the MCMC procedure described in $\S$4.1 and 4.3 of L19.  This MCMC sampler uses a Metropolis-Hastings algorithm \citep{Has1970}, run with a single MCMC chain of 200,000 steps and a burn-in phase of 40,000 iterations.  Due to the simplicity of this problem, MCMC chains quickly converge.  

%
%
\begin{figure}
\centerline{
\includegraphics[width=8cm]{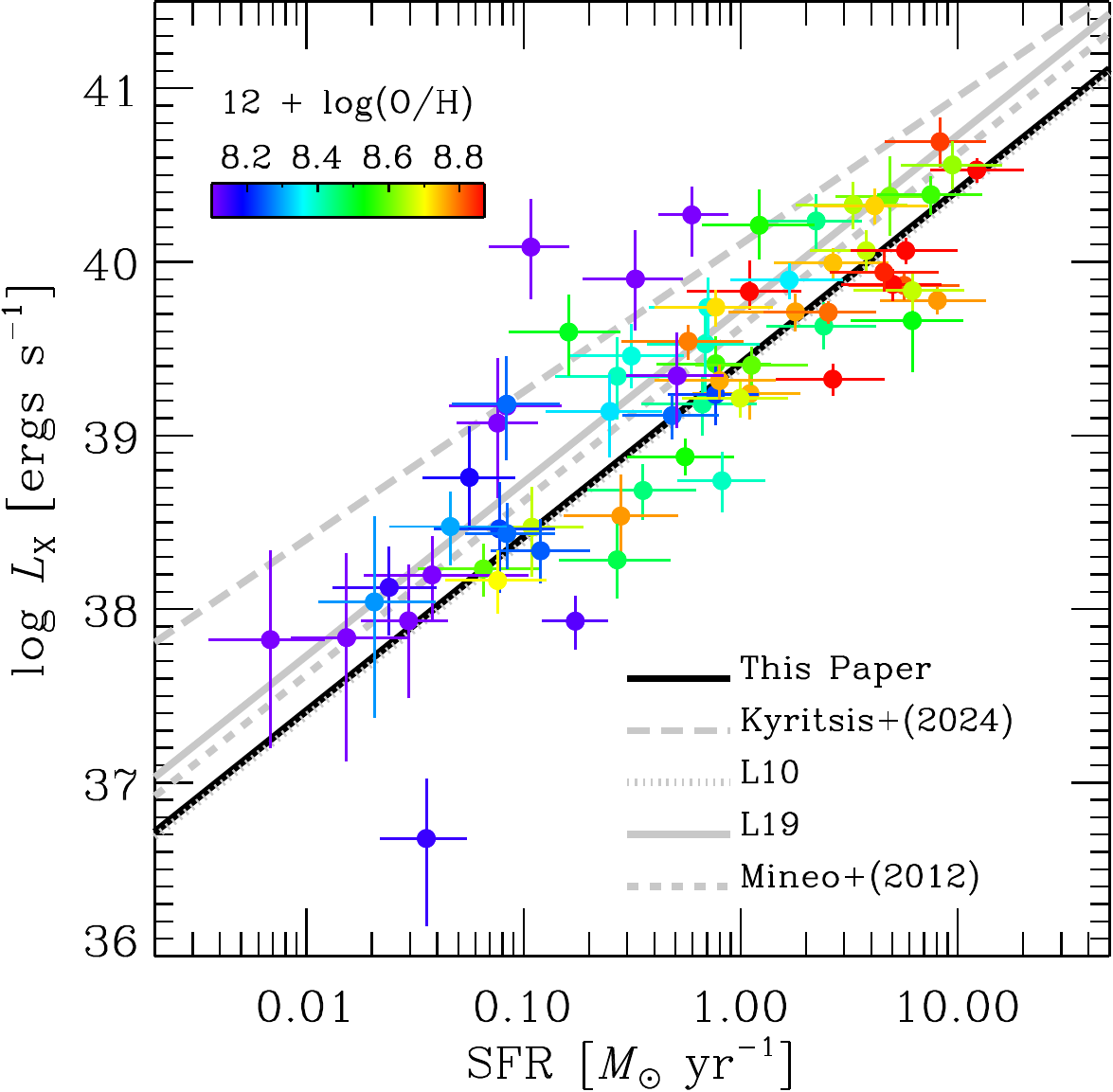}
}
\caption{
Galaxy-integrated intrinsic 0.5--8~keV luminosity, $L_{\rm X}$, versus SFR for main-sequence and starburst galaxies in our sample ({\it filled circles with 1$\sigma$ error bars\/}).  SFR values were calculated from our SED fitting posterior distributions (see $\S$\ref{sec:sfh}) and $L_{\rm X}$ was calculated from our broken power-law XLF modeling procedure (see $\S$\ref{sec:xray}) and Equation~\ref{eqn:lx}.  Our best-fit linear regression model to these data is shown as a solid black line (see Eqn.~\ref{eqn:lxsfr}) and comparisons from the literature are shown (see annotations).
}
\label{fig:lxsfr}
\end{figure}

%
%
\begin{figure*}
\centerline{
\includegraphics[width=17cm]{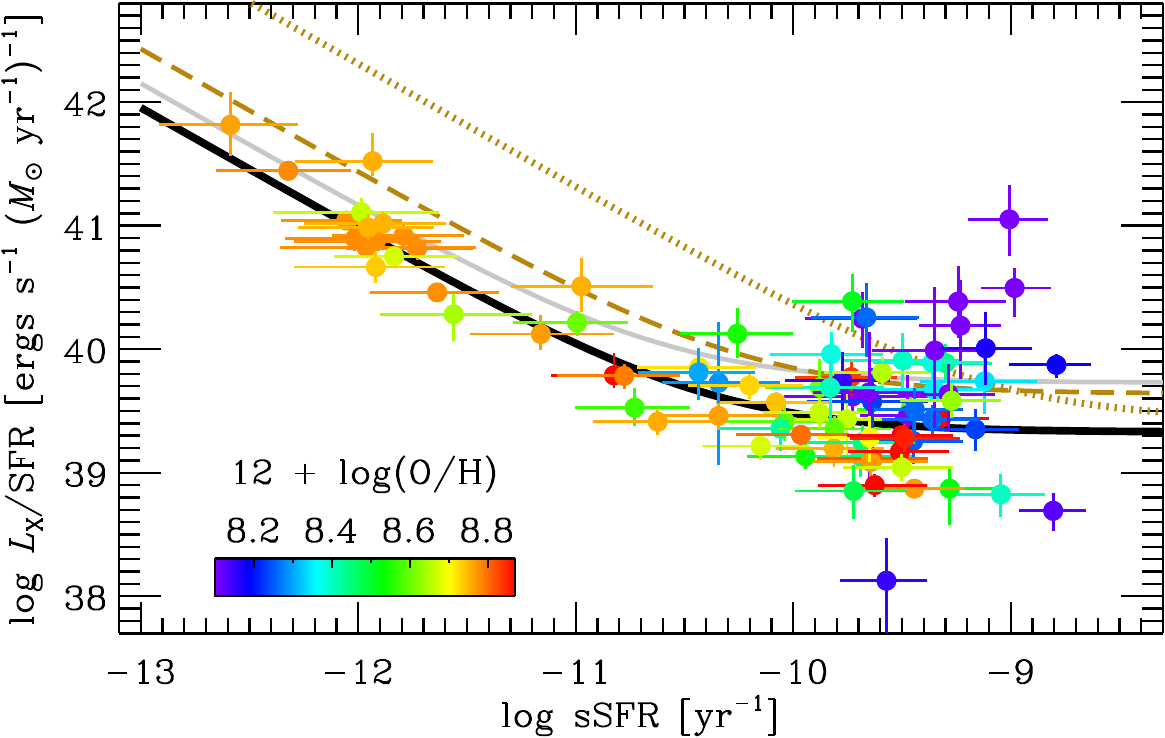}
}
\caption{
Galaxy-integrated intrinsic 0.5--8~keV luminosity per SFR, $L_{\rm X}$/SFR, versus specific SFR, sSFR, for the full sample of \ngal\ galaxies.  Each symbol is color-coded by gas-phase metallicity (see color bar) and our best-fit model for the dependence of $L_{\rm X}$ on SFR and $M_\star$ (see Eqn.~\eqref{eqn:ab}) is shown as a solid black curve.   For comparison, we have overlaid the best-fit from L19 {\it gray curve\/} and two relations from \citet{Kou2020} appropriate for regions of size 4$\times$4~kpc$^2$ ({\it gold dashed curve\/}) and 1$\times$1~kpc$^2$ ({\it gold dotted curve\/}).
}
\label{fig:lxalbe}
\end{figure*}

For illustrative purposes, we show in Figure~\ref{fig:bknpo} the fitting results for NGC~0024.  The black curve in the upper-left panel shows the best-fit broken power-law plus CXB model (Eqn.~\ref{eqn:mod}), and its intrinsic representation (i.e., without inclusion of completeness corrections) as a gray dashed curve.  The upper-right panel shows the cumulative XLF and model components, including the completeness-corrected CXB ({\it dotted curve\/}) and XLF ({\it blue long-dashed curve\/}) components, with the full model without completeness corrections shown as a gray dashed curve. The bottom panels include posterior distributions on modeled parameters ($K$, $\alpha_1$, and $\alpha_2$) and the integrated intrinsic point-source luminosity $L_{\rm X}$, which we calculated at each MCMC step following:
\begin{equation}\label{eqn:lx}
L_{\rm X} =  \int_{L_{\rm lo}}^{L_{\rm c}} L \frac{dN_{\rm int}}{dL} dL = \int_{L_{\rm lo}}^{L_{\rm c}} \frac{1}{\log e}\frac{dN_{\rm int}}{d \log L} dL,
\end{equation}
where we set $L_{\rm lo} = 10^{35}$~\lum.  For NGC~0024, and a subset of other galaxies in our sample, all X-ray detected sources reside at $L < L_b$ and thus $L_c < L_b$.  Equation~\eqref{eqn:bkn} therefore implies that $\alpha_2$ is irrelevant to the model in this particular case, and the model follows a single power-law distribution with slope $\alpha_1$.  As such, the posterior distribution of $\alpha_2$ follows directly the prior.  A full set of XLF fits for all galaxies can be found in the supplemental materials of this article.

%
\section{Analysis and Results}\label{sec:res}
%

\subsection{Basic X-ray Scaling Relations Revisited}
\label{sub:scal}

As discussed in $\S$1, there are numerous publications in the literature showing that galaxy-integrated X-ray luminosity, $L_{\rm X}$, scales linearly (or nearly linearly) with SFR \citep[e.g.,][]{Gri2003,Per2007,Leh2010,Min2012a,Leh2016,Leh2019,Vul2021,Kyr2024}, following
$\log L_{\rm X} = \log {\rm SFR} + \omega$.
In Figure~\ref{fig:lxsfr}, we show our version of the $L_{\rm X}$-SFR relation.  To avoid the impact of LMXB dominated systems (e.g., elliptical galaxies), we include only the \nms\ galaxies in our sample that have SFR values greater than the lower-bound of the galaxy main sequence shown in Figure~\ref{fig:prop}$a$ (above the lower-bound of the shaded region).   Symbols in Figure~\ref{fig:lxsfr} have been color-code by gas-phase metallicity to show its impact on the $L_{\rm X}$-SFR  relation.

For comparison, we overlay the $L_{\rm X}$-SFR relations from \citet{Min2012a,Leh2010,Leh2019} (see annotations).  For these comparisons, we adopted the L19 ``cleaned sample'' value of $\omega_{\rm L10} = 39.73_{-0.10}^{+0.15}$, which is based on global XLF fitting to subgalactic regions across a sample of 38 nearby galaxies.  We corrected the quoted \citet{Min2012a} relation from their assumed Salpeter IMF to our Kroupa IMF (multiplying their scaling relation by 1.6) to obtain $\omega_{\rm Mineo} = 39.62$, and corrected the L10 2--10~keV band to 0.5--8~keV band (multiplying their scaling relation by 1.5) to obtain $\omega_{\rm L10} = 39.38 \pm 0.06$.

We performed least-squares fitting to derive a relation based on the data shown in Figure~\ref{fig:lxsfr}, obtaining:
\begin{equation}\label{eqn:lxsfr}
\log L_{\rm X} = \log {\rm SFR} + (39.39 \pm 0.017).
\end{equation}
This value is nearly identical to that of L10, but with an uncertainty that is $\approx$3.5 times smaller due to the larger number of galaxies across the full range of SFR.  The residual scatter in the relation is 0.49~dex, which is a factor of $\approx$3.2 times larger than the median measurement error on $\log L_{\rm X}$ ($\approx$0.15~dex).  This indicates, as past studies have found, that there are additional physical dependencies on $L_{\rm X}$, as well as additional sources of scatter (e.g., XLF sampling uncertainties).  Indeed the most significant outliers to the relation appear to be low-metallicity galaxies with SFR~$\approx$~0.1--1~\sfr\ that have elevated $L_{\rm X}$ values, consistent with past studies (see discussion in $\S$\ref{sec:intro}).

While the $L_{\rm X}$-SFR relation is commonly thought to be driven by HMXB populations that dominate in star-forming galaxies, there is strong evidence for non-negligible contributions from LMXBs, which will more explicitly dominate low-sSFR galaxies.  We can directly model the impact of older populations of LMXBs using the combined relation:
\begin{equation}\label{eqn:ab}
L_{\rm X} = \alpha_{\rm LMXB} M_\star + \beta_{\rm HMXB} {\rm SFR},
\end{equation}
where $\alpha_{\rm LMXB}$ and $\beta_{\rm HMXB}$ are fitting constants accounting for scaling relations of LMXB and HMXB luminosities with $M_\star$ and SFR, respectively.  In Figure~\ref{fig:lxalbe}, we show $L_{\rm X}$/SFR versus sSFR for our full sample of \ngal\ galaxies.  As has been noted in several past studies, a clear trend of $L_{\rm X}$/SFR~$\propto$~sSFR$^{-1}$ is apparent across much of the sSFR range, as expected from Equation~\ref{eqn:ab}.

%
%
\begin{figure*}
\centerline{
\includegraphics[width=8.5cm]{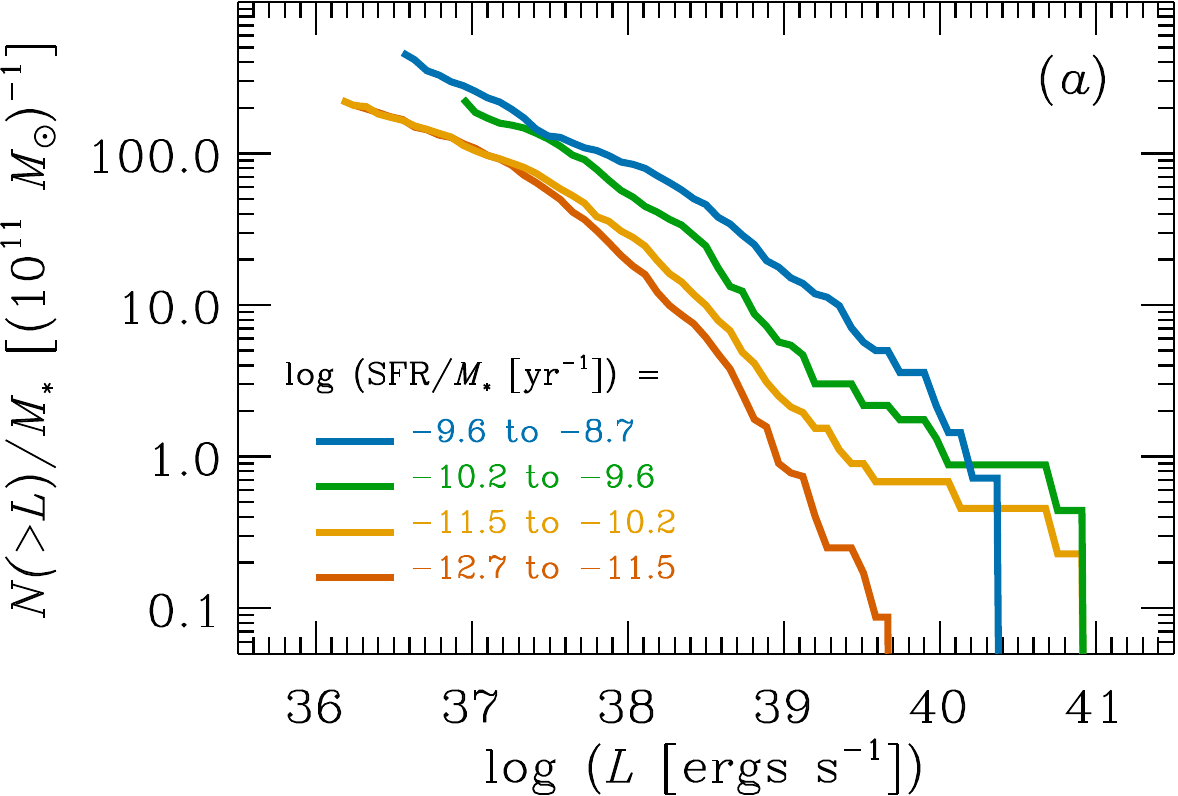}
\hfill
\includegraphics[width=8.5cm]{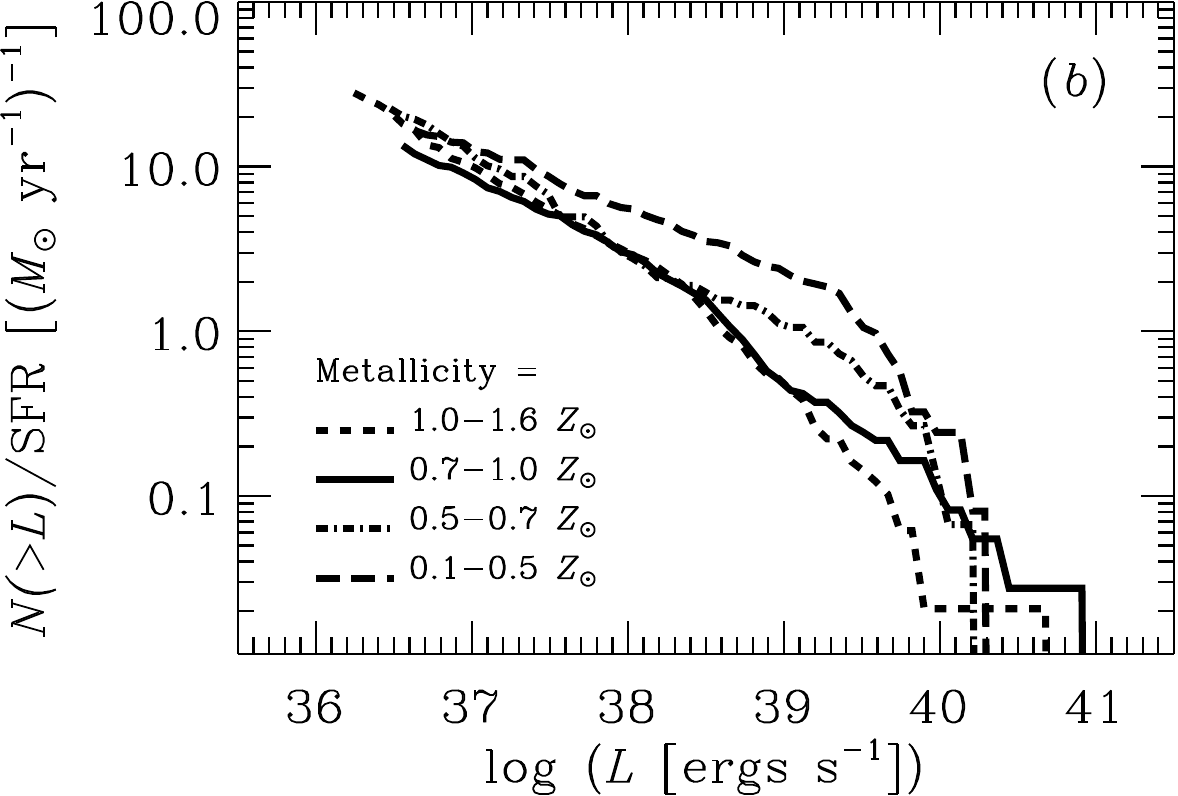}
}
\caption{
Empirical XLFs for galaxy subsamples with ($a$) metallicity in the range of $\approx$0.6--1.6~$Z_\odot$ and groupings based on sSFR, and ($b$) $\log$~sSFR~$\ge -10$ in groupings based on metallicity (see annotations).  These empirical XLFs have been corrected for completeness and estimates of CXB contributions have been subtracted. They provide proxies for how the XRB XLFs vary with stellar population age ($a$) and metallicity ($b$).  We find that with increasing stellar age, the XLF normalization per $M_\star$ declines, and the shape of the XLF transitions to becoming flatter at $L \simlt 10^{37.5}$~\lum\ and steeper at $L > 10^{37.5}$~\lum\ ($a$).  With increasing metallicity, we find that the high-sSFR XRB population XLFs become steeper and contain fewer $L > 10^{38}$~\lum\ sources ($b$). These observed trends motivate the construction of our age and metallicity dependent XLF modeling (see $\S$\ref{sub:mod} for details). 
}
\label{fig:xlfdep}
\end{figure*}

When fitting the data using Equation~\ref{eqn:ab}, we obtain the following best-fit values
\begin{equation} \label{eqn:albeval}
\begin{split}
\log (\alpha_{\rm LMXB}~[{\rm ergs~s^{-1}~M_\odot^{-1}}])  & = 29.957 \pm 0.004, \\
\log (\beta_{\rm HMXB}~[{\rm ergs~s^{-1}~(M_\odot~yr)^{-1}}]) &  = 39.303 \pm 0.004,
\end{split}
\end{equation}
in which the uncertainties correspond to 1$\sigma$ uncertainties on the fitting parameters. 
We overlay the best-fit model in Figure~\ref{fig:lxalbe} as a black curve, and provide the equivalent best-fit from L19 as a gray curve.  We find that the relation presented here provides systematically lower values than those of L19.  We can attribute the majority of the differences between these two studies to methodology.  For L19, $L_{\rm X}$ was computed assuming a universal value of $L_c$ that was higher than the most luminous detected source in a given galaxy.  As such, the larger $L_c$ upper limit of integration in Equation~\ref{eqn:lx} yielded systematically higher measured values of the intrinsic point-source $L_{\rm X}$ than those here.  The methods applied here should yield more realistic accounting of the point-source populations actually present within the galaxies, and we regard our updated measurements as superseding those of L19.

As is evident from Figure~\ref{fig:lxalbe}, the incorporation of scaling relations involving both SFR and $M_\star$ allows for reasonable predictions of $L_{\rm X}$ for the full sample that is not achievable by an $L_{\rm X}$-SFR scaling alone.  sSFR provides an important proxy for SFH, spanning galaxies dominated by old stellar populations of $\simgt$10~Gyr at the lowest sSFR to galaxies with very active ongoing star formation at the highest sSFR.  However, significant residual scatter of 0.43~dex remains, relative to the median $L_{\rm X}$ uncertainty of 0.15~dex, and this scatter is most evident for high-sSFR galaxies (sSFR~$\simgt 10^{-10}$~yr$^{-1}$) that are expected to be HMXB dominant.  While some of the scatter is expected to be due to stochastic scatter related to XLF sampling variations \citep[see, e.g., ][for further details]{Gil2004a,Leh2019,Leh2021}, visual inspection of the metallicity-based color-coding in Figure~\ref{fig:lxalbe} reveals a suggestive stratification of $L_{\rm X}$/SFR by metallicity in this sSFR regime, with the lowest metallicity galaxies having the highest $L_{\rm X}$/SFR values. Taken together, these observations support an age and metallicity dependence to the XRB population emission within galaxies, which we explore in detail in the next section.

\begin{deluxetable*}{lrcrrrrcccccccc}
\label{tab:xlf}
\tablewidth{0pt}
\tabletypesize{\scriptsize}
\tablecaption{X-ray Luminosity Function Fits By Galaxy}
\tablehead{
 \multicolumn{1}{c}{ } & \colhead{} &  \colhead{} & \multicolumn{8}{c}{\sc Broken Power Law$^\dagger$} &  \multicolumn{4}{c}{\sc Global Model$^\ddagger$} \\
\vspace{-0.3in} \\
\multicolumn{1}{c}{} &  \colhead{} & \colhead{}  &  \multicolumn{8}{c}{\rule{3.5in}{0.01in}} & \multicolumn{4}{c}{\rule{1.5in}{0.01in}} \\
\vspace{-0.25in} \\
\multicolumn{1}{c}{\sc Galaxy} &  \colhead{} & \colhead{$\log L_{50}$}  &  & & & \colhead{$\log L_{\rm X}$} & & & & & \multicolumn{4}{c}{} \\
\vspace{-0.25in} \\
\multicolumn{1}{c}{\sc Name} & \colhead{$N_{\rm src}$} & \colhead{(\lum)} & \colhead{$K$} & \colhead{$\alpha_1$} & \colhead{$\alpha_2$} & \colhead{(\lum)} & \colhead{$C$} & \colhead{$C_{\rm exp}$} & \colhead{$C_{\rm var}$} & \colhead{$p_{\rm null}$} & \colhead{$C_{i}^{\rm glob}$} & \colhead{$C_{{\rm exp},i}^{\rm glob}$} & \colhead{$C_{{\rm var},i}^{\rm glob}$} & \colhead{$p_{{\rm null},i}^{\rm glob}$} \\
\vspace{-0.25in} \\
\multicolumn{1}{c}{(1)} & \multicolumn{1}{c}{(2)} & \multicolumn{1}{c}{(3)} & \colhead{(4)} & \colhead{(5)} & \colhead{(6)} & \colhead{(7)} & \colhead{(8)} & \colhead{(9)} & \colhead{(10)} & \colhead{(11)} & \colhead{(12)} & \colhead{(13)} & \colhead{(14)} & \colhead{(15)}
}
\startdata
        NGC0024 &    8 & 36.8 &    3.76$^{+3.32}_{-2.05}$ &    0.89$^{+0.42}_{-0.39}$ &    2.22$^{+0.47}_{-0.48}$ &      38.2$^{+0.1}_{-0.2}$ &    13 &    15 &    20 &      0.650 &    27 &    12 &    42 &      0.021 \\ 
        NGC0337 &    6 & 38.4 &    5.19$^{+5.84}_{-2.57}$ &    0.95$^{+0.61}_{-0.58}$ &    1.42$^{+0.25}_{-0.20}$ &              40.2$\pm$0.2 &    20 &    19 &    20 &      0.881 &    22 &    16 &    26 &      0.234 \\ 
        NGC0584 &    8 & 38.4 &    30.4$^{+23.9}_{-17.2}$ &    0.91$^{+0.65}_{-0.56}$ &    2.22$^{+0.37}_{-0.45}$ &              39.9$\pm$0.3 &    10 &    11 &    14 &      0.854 &    13 &    12 &    18 &      0.877 \\ 
        NGC0625 &    4 & 36.1 &    1.45$^{+1.02}_{-0.70}$ &    0.32$^{+0.33}_{-0.22}$ &    2.04$^{+0.47}_{-0.44}$ &      38.2$^{+0.2}_{-0.3}$ &    13 &    16 &    26 &      0.501 &    18 &    18 &   112 &      0.997 \\ 
        NGC0628 &   51 & 36.3 &    4.90$^{+2.05}_{-1.51}$ &    1.20$^{+0.16}_{-0.17}$ &    2.33$^{+0.43}_{-0.46}$ &              38.9$\pm$0.1 &    25 &    34 &    61 &      0.241 &    39 &    43 &    70 &      0.658 \\ 
\\
        NGC0925 &    7 & 37.5 &    1.37$^{+0.87}_{-0.55}$ &    1.44$^{+0.89}_{-0.68}$ &    1.24$^{+0.25}_{-0.15}$ &              39.5$\pm$0.2 &    22 &    21 &    22 &      0.821 &    25 &    19 &    43 &      0.297 \\ 
        NGC1023 &   71 & 36.8 &      11.6$^{+3.2}_{-2.6}$ &             1.20$\pm$0.16 &    2.12$^{+0.33}_{-0.30}$ &              39.5$\pm$0.1 &    23 &    34 &    63 &      0.164 &    29 &    38 &    77 &      0.290 \\ 
        NGC1097 &   29 & 38.0 &      14.0$^{+4.6}_{-3.6}$ &    0.94$^{+0.65}_{-0.54}$ &    1.53$^{+0.12}_{-0.10}$ &              40.7$\pm$0.1 &    30 &    34 &    43 &      0.523 &    39 &    27 &    47 &      0.063 \\ 
        NGC1291 &   65 & 37.1 &      24.4$^{+5.2}_{-4.6}$ &             0.83$\pm$0.17 &    2.17$^{+0.24}_{-0.21}$ &              39.9$\pm$0.1 &    26 &    35 &    61 &      0.252 &    30 &    35 &    75 &      0.542 \\ 
        NGC1313 &   12 & 36.5 &    1.91$^{+1.09}_{-0.77}$ &    0.46$^{+0.39}_{-0.32}$ &    1.52$^{+0.27}_{-0.23}$ &      39.3$^{+0.2}_{-0.3}$ &    32 &    31 &    35 &      0.876 &    35 &    30 &    57 &      0.496 \\ 
\\
        NGC1316 &   84 & 37.9 &    64.0$^{+13.5}_{-10.9}$ &    0.82$^{+0.52}_{-0.47}$ &    1.85$^{+0.17}_{-0.14}$ &      40.4$^{+0.1}_{-0.0}$ &    23 &    27 &    55 &      0.643 &    39 &    31 &    77 &      0.374 \\ 
        NGC1380 &   37 & 37.6 &      21.2$^{+5.9}_{-4.7}$ &    1.28$^{+0.47}_{-0.45}$ &    2.19$^{+0.24}_{-0.21}$ &              39.9$\pm$0.1 &    20 &    26 &    43 &      0.375 &    24 &    26 &    52 &      0.784 \\ 
        NGC1387 &   16 & 37.8 &      11.9$^{+4.7}_{-3.5}$ &    1.01$^{+0.60}_{-0.54}$ &    2.06$^{+0.38}_{-0.34}$ &      39.5$^{+0.2}_{-0.1}$ &    13 &    20 &    30 &      0.229 &    18 &    23 &    41 &      0.461 \\ 
        NGC1404 &   74 & 37.4 &      23.1$^{+5.4}_{-4.4}$ &             1.19$\pm$0.27 &    1.74$^{+0.17}_{-0.14}$ &              40.1$\pm$0.1 &    26 &    34 &    61 &      0.313 &    38 &    33 &    70 &      0.531 \\ 
        NGC1433 &   16 & 37.8 &    7.57$^{+3.63}_{-2.45}$ &    0.96$^{+0.65}_{-0.56}$ &    1.63$^{+0.40}_{-0.31}$ &      39.4$^{+0.2}_{-0.1}$ &    19 &    20 &    30 &      0.794 &    25 &    27 &    47 &      0.742 \\ 
\\
        NGC1427 &   50 & 37.6 &      20.1$^{+5.6}_{-4.5}$ &    1.69$^{+0.47}_{-0.44}$ &    2.24$^{+0.26}_{-0.23}$ &      39.9$^{+0.2}_{-0.1}$ &    15 &    26 &    46 &      0.112 &    56 &    26 &    39 &   $<$0.001 \\ 
        NGC1482 &    9 & 37.9 &    1.30$^{+1.42}_{-0.72}$ &    1.02$^{+0.85}_{-0.64}$ &    1.30$^{+0.32}_{-0.26}$ &      39.7$^{+0.2}_{-0.3}$ &    17 &    18 &    24 &      0.983 &    34 &    30 &    36 &      0.541 \\ 
        NGC1566 &   31 & 37.9 &      10.3$^{+5.1}_{-3.6}$ &    0.96$^{+0.65}_{-0.58}$ &    1.67$^{+0.21}_{-0.16}$ &              40.4$\pm$0.2 &    27 &    31 &    47 &      0.567 &    36 &    32 &    58 &      0.612 \\ 
        NGC1569 &   18 & 35.4 &    2.68$^{+2.32}_{-1.31}$ &    0.94$^{+0.19}_{-0.20}$ &    2.23$^{+0.51}_{-0.50}$ &      37.9$^{+0.1}_{-0.2}$ &    30 &    30 &    25 &      0.982 &    49 &    27 &    41 &   $<$0.001 \\ 
\enddata
\tablecomments{All fits include the effects of incompleteness and model contributions from the CXB, following description in $\S$\ref{sub:xlf}.  Col.(1): Galaxy name, as reported in Table~\ref{tab:sam}.  Col.(2): Total number of \xray\ sources detected within the galactic boundaries defined in Table~\ref{tab:sam}.  Col.(3): Logarithm of the luminosities corresponding to the respective 50\% completeness limits. Col.(4)--(7): Median and 1$\sigma$ uncertainty values of the broken power-law normalization, slopes, and integrated X-ray luminosity, respectively.  Col.(8): C-statistic, $C$, associated with the best broken power-law model. Col.(9): Expected value of $C$ from model.  Col.(10): Expected variance on $C$ from model. Col.(11): Null-hypothesis probability of the best broken power-law model.  The null-hypothesis probability is calculated following the prescription in Eqn.~\ref{eqn:pnull} and is appropriate for the use of the C statistic. Col.(12)--(15): Respectively, C-statistic, $C_{\rm exp}$, $C_{\rm var}$, and null-hypothesis probability for the age and metallicity dependent XLF model described in $\S$\ref{sub:fit}. \\
$^\dagger$Broken power-law models are derived following Eqn.~\eqref{eqn:bkn} with priors on $\alpha_1$, $\alpha_2$, $L_b$, and $L_c$, as described in $\S$\ref{sub:xlf}.\\
$^\ddagger$The age and metallicity dependent ``global model'' provides a prediction for the galaxy XLF, given a SFH and metallicity estimate.  Details for how the global model is constructed are provided in $\S$\ref{sub:fit}.\\
}
\end{deluxetable*}

\subsection{Construction of the Metallicity and Age Dependent XLF Model}\label{sub:mod}

To begin to construct an age and metallicity dependent model of the XRB XLF within galaxies, we utilized a combination of observational constraints and population synthesis model expectations. To infer age-dependent variations in the XLF, we first constructed empirical XLFs in bins of sSFR (i.e., a proxy for average stellar age) for an isolated range of metallicity where the majority of our galaxies are observed (0.6--1.6~$Z_\odot$).  These XLFs were constructed by combining the completeness corrected, CXB-subtracted XLFs of all galaxies within a given sSFR (and metallicity) range.  We show the resulting stellar-mass normalized empirical XLFs in Figure~\ref{fig:xlfdep}$a$ for four sSFR bins. In this representation, the lowest-sSFR bin ({\it red curve\/}) can be thought to be dominated by old populations of LMXBs, with negligible contributions from young populations, and the highest-sSFR bin is expected to contain both old LMXBs (perhaps at the same baseline as the lowest-sSFR XLF) plus young populations of HMXBs that overwhelm the old population contributions. The progression across the sSFR can be taken as a proxy for an age progression that reveals how the XRB XLF shape and normalization evolve with age. We note, however, that each empirical XLF will contain contributions from populations across all cosmic look-back times, and so any one XLF cannot be taken as a pure representation of the XLF at any particular age.  Nonetheless, generalized trends can be inferred to allow us to build a model for the XLF evolution with age.

%
%
\begin{figure}
\centerline{
\includegraphics[width=8.5cm]{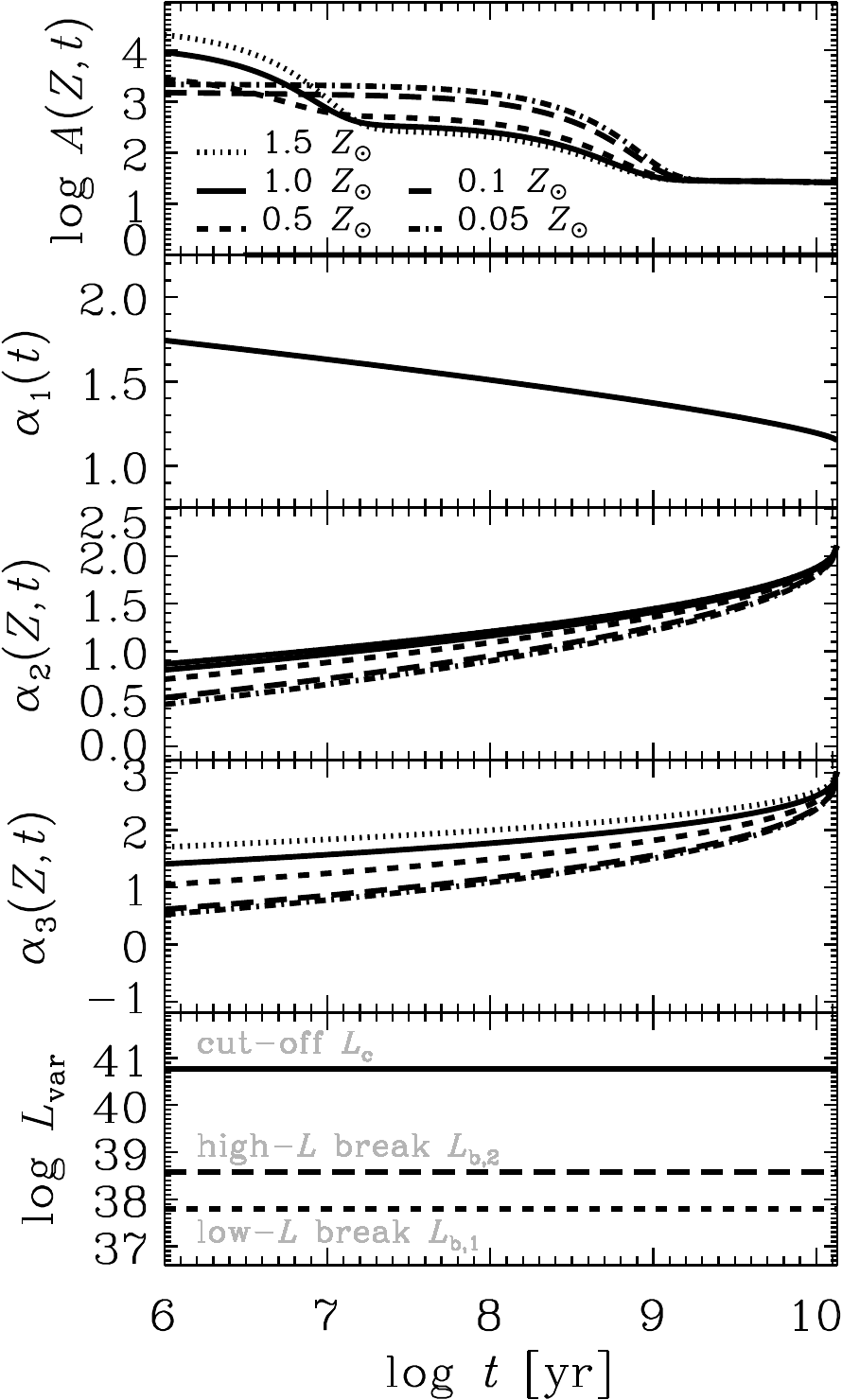}
}
\caption{
Best-fit age and metallicity dependent power-law fit parameters, based on the model defined in Equations~\eqref{eqn:xlfmod} and \eqref{eqn:par}.  These results constrain breaks and cut-off luminosities for the XRB XLFs (bottom panel) and indicate that (1) the XLF normalization, $A$, declines with age at a pace that is slower for lower-metallicity galaxies; (2) that the low-luminosity slope, $\alpha_1$, becomes shallower with increasing age; and (3) the medium and high luminosity XLF slopes, $\alpha_2$ and $\alpha_3$, increase with increasing age with lower metallicity galaxies having shallower high-$L$ XLF slopes.
}
\label{fig:par}
\end{figure}

Inspection of Figure~\ref{fig:xlfdep}$a$ indicates that the XRB XLF evolves in both shape and normalization with age.  Motivated by these observations and past studies of XLFs in galaxy samples, we infer the following changes in the XLF as stellar age increases from young-to-old populations: (1) the normalization (i.e., number of XRBs per stellar mass) decreases; (2) the high-luminosity slope ($L \simgt 10^{37.5}$~\lum) becomes steeper (power-law index increases); (3) the low-luminosity slope ($L \simlt 10^{37.5}$~\lum) becomes shallower; (4) the most luminous sources (i.e., the cut-off luminosity) are observed at $\approx$10$^{41}$~\lum\ for many of the sSFR bins; and (5) the shape transitions from an approximately two-slope broken power-law to a three-slope power-law.  These inferences are consistent with the results from \citet{Leh2019}, which show the normalization declines and the high-$L$ slope clearly steepens with decreasing sSFR, \citet{Gil2004a} and \citet{Zha2012}, which show that the early-type galaxy XLF can be well described by a three-slope power-law with a low-luminosity slope that is flatter than that of late-type galaxies \citep[see also][]{Leh2020}. Going forward, we will therefore contextualize XLFs in terms of a three-slope power-law shape that varies with age and metallicity.

Figure~\ref{fig:xlfdep}$b$ provides a complementary set of empirical XLFs, normalized by SFR, for high-sSFR galaxies ($\log$~sSFR~$\simgt -10$), selected in bins of metallicity.  The construction of these empirical XLFs is motivated by the observation that the majority of the variation in $L_{\rm X}$/SFR occurs at high-sSFR where we expect the XLFs are dominated by young HMXBs and appears to be metallicity dependent.  It is clear from Figure~\ref{fig:xlfdep}$b$ that the XLF is metallicity dependent, and in the context of a three-slope power-law, we make the following observations in the XLF shape as it progresses from low-to-high metallicity: (1) the normalization appears to decline; (2) the low-luminosity slope appears to be constant; (3) the mid-to-high-luminosity slope appears to increase, leading to a decline in high-$L$ sources at high metallicity; and (4) the maximum source luminosity (i.e., the cut-off luminosity) appears to be consistent across all metallicity bins.

Motivated by the above trends, we chose to build a ``global'' age and metallicity dependent XLF model as a three-slope power-law with variable parameters following:
\begin{equation}\label{eqn:xlfmod}
\begin{split}
\frac{dN(t, Z)}{d \log L \; dM_\star} = & A \; \exp{[-L/L_{\rm
c}]} \times \\
& \left \{ \begin{array}{lr} L^{-\alpha_1},  & \;\;\;\;\;\;\; (L < L_{b,1}) \\ 
L_{b,1}^{\alpha_2-\alpha_1} L^{-\alpha_2},  & (L = L_{b,1}-L_{b,2})  \\ 
L_{b,2}^{\alpha_3-\alpha_2} L_{b,1}^{\alpha_2-\alpha_1} L^{-\alpha_3},  & (L > L_{b,2}) \\
\end{array} 
  \right.
\end{split}
\end{equation}
where the power-law parameter sets \{$A$, $L_c$, $\alpha_1$, $L_{b,1}$, $\alpha_2$, $L_{b,2}$, $\alpha_3$\} are themselves continuous functions of age, $t$, and/or metallicity, $Z$, that are specified using a total of 21 parameters.  We hereafter refer to the $i$th parameter of this set as $p_i$. In the context of Equation~\ref{eqn:xlfmod}, these parameterizations are defined as:
\begin{multline}
\notag
A(t,Z) =  \\
\left \{ \begin{array}{lr} 0,  &  \;\; (t_9 < 0.03)\\
p_1 Z^{p_2} e^{-(t_9-t_{\rm ref})/p_3} + p_4 Z^{p_5} e^{-(t-t_{\rm ref})/p_6} + p_7 & \;\; (t_9 \ge 0.03) \\
\end{array}
\right.
\end{multline}
$$\alpha_1(t) = p_8 + p_{9} [{c_1} + c_2 \log t_9]^{p_{10}}$$
$$\log L_{b,1} = p_{11}$$
$$\alpha_2(t,Z) = p_{12} + [p_{13} + Z^{p_{14}}] [{c_1} + c_2 \log t_9]^{p_{15}}$$
$$\log L_{b,2} = p_{16}$$
$$\alpha_3(t,Z) = p_{17} + [p_{18} + Z^{p_{19}}] [{c_1} + c_2 \log t_9]^{p_{20}}$$
$$\log L_c = p_{21}$$
$$t_{\rm ref} = 0.003~{\rm Gyr}$$
$$c_1 = 0.308$$
\begin{equation}\label{eqn:par}
    c_2 = -0.274.
\end{equation}
For the above system of equations, $t_9$ is defined as the look-back time in units of Gyr, $Z$ is the metallicity in solar units, and all luminosities are taken to be in units of $10^{38}$~\lum.  We note that in our equations there are a few terms that are functionally identical.  For example, $A(t,Z)$ contains two age-dependent decays of the XLF normalization with age on timescales of $p_3$ and $p_6$.  Also, the age and metallicity dependent functional forms of $\alpha_2$ and $\alpha_3$ are the same (i.e., $p_{12}$--$p_{15}$ and $p_{17}$--$p_{20}$).  To differentiate their impact on the evolution of the XLF, we adopt uniform priors over unique parameter ranges to distinguish their functional dependencies. These parameter ranges were motivated by the observed trends identified in Figure~\ref{fig:xlfdep} (see discussion earlier in this section), however, they are broad enough to permit very wide ranges of possible best-fit outcomes. In Table~\ref{tab:xfit}, we summarize the parameters of the model and provide their range of uniform priors.

\begin{deluxetable}{lcrrr}
\tablewidth{1.0\columnwidth}
\tabletypesize{\footnotesize}
\tablecaption{Age and Metallicity Dependent Model Parameter Estimates}
\tablehead{
\multicolumn{1}{c}{\sc Param} & \multicolumn{1}{c}{\sc Units} & \multicolumn{1}{c}{\sc Prior} & \multicolumn{1}{c}{\sc Best} & \multicolumn{1}{c}{\sc Median$_{-16\%}^{+84\%}$} \\
\vspace{-0.25in} \\
\multicolumn{1}{c}{(1)} & \multicolumn{1}{c}{(2)} & \multicolumn{1}{c}{(3)} &  \multicolumn{1}{c}{(4)} & \colhead{(5)} }
\startdata
$p_{1}$ & ($10^{11} M_\odot$)$^{-1}$ & [0,$\infty$] & 4369 & 5696$^{+9652}_{-3633}$ \\
$p_{2}$ &  & [$-2$,2] & 2.00 & 1.90$^{+0.10}_{-0.82}$ \\
$p_{3}$ & Gyr & [0,0.015] & 0.0029 & 0.0117$^{+0.0033}_{-0.0062}$ \\
$p_{4}$ & ($10^{11} M_\odot$)$^{-1}$ & [0,$\infty$] & 331 & 357$^{+411}_{-201}$ \\
$p_{5}$ &  & [$-2$,2] & $-$0.62 & $-$2.00$\pm$0.00 \\
$p_{6}$ & Gyr & [0.02,3] & 0.24 & 0.51$^{+0.90}_{-0.34}$ \\
$p_{7}$ & ($10^{11} M_\odot$)$^{-1}$ & [0,$\infty$] & 28 & 33$^{+7}_{-4}$ \\
$p_{8}$ &  & [$-3$,3] & 1.15 & 1.16$^{+0.10}_{-0.17}$ \\
$p_{9}$ &  & [$-2$,2] & 0.54 & 0.85$^{+0.31}_{-0.16}$ \\
$p_{10}$ &  & [$-2$,2] & 0.76 & 1.17$^{+0.83}_{-0.63}$ \\
$p_{11}$ & $\log$~\lum & [37,38] & 37.80 & 37.70$^{+0.08}_{-0.11}$ \\
$p_{12}$ &  & [$-3$,3] & 2.11 & 2.11$^{+0.45}_{-0.24}$ \\
$p_{13}$ &  & [$-5$,5] & $-$2.23 & $-$4.39$\pm$0.05 \\
$p_{14}$ &  & [$-2$,2] & 0.14 & 0.31$^{+0.15}_{-0.11}$ \\
$p_{15}$ &  & [$-2$,2] & 0.48 & 1.12$^{+0.77}_{-0.52}$ \\
$p_{16}$ & $\log$~\lum & [38.1,39] & 38.58 & 38.50$\pm$0.08 \\
$p_{17}$ &  & [$-5$,5] & 3.03 & 3.40$^{+1.60}_{-0.55}$ \\
$p_{18}$ &  & [$-5$,5] & $-$2.543 & $-$4.901$\pm$0.003 \\
$p_{19}$ &  & [$-2$,2] & 0.61 & 0.64$^{+0.23}_{-0.17}$ \\
$p_{20}$ &  & [$-2$,2] & 0.38 & 0.62$^{+0.92}_{-0.35}$ \\
$p_{21}$ & $\log$~\lum & [39,42] & 40.76 & 40.67$\pm$0.24 \\
\hline
\multicolumn{5}{c}{Statistical Fit Results}\\
\hline
$C_{\rm global}$ &  &  & & 2788.9 \\
$C_{\rm exp}$ &  &  & & 2646.6 \\
$\sqrt{C_{\rm var}}$ &  &  & & 71.5 \\
$p_{\rm null}$ &  &  &  &0.047 \\
\enddata
\tablecomments{Table of best-fit values for the model parameters defined in Equations~\eqref{eqn:xlfmod} and \eqref{eqn:par} (i.e., $p_1$--$p_{21}$), and statistical evaluation of the model goodness of fit (bottom quantities).  See $\S$\ref{sub:fit} for details.}
\label{tab:xfit}
\end{deluxetable}

%
%
\begin{figure*}
\centerline{
\includegraphics[width=18cm]{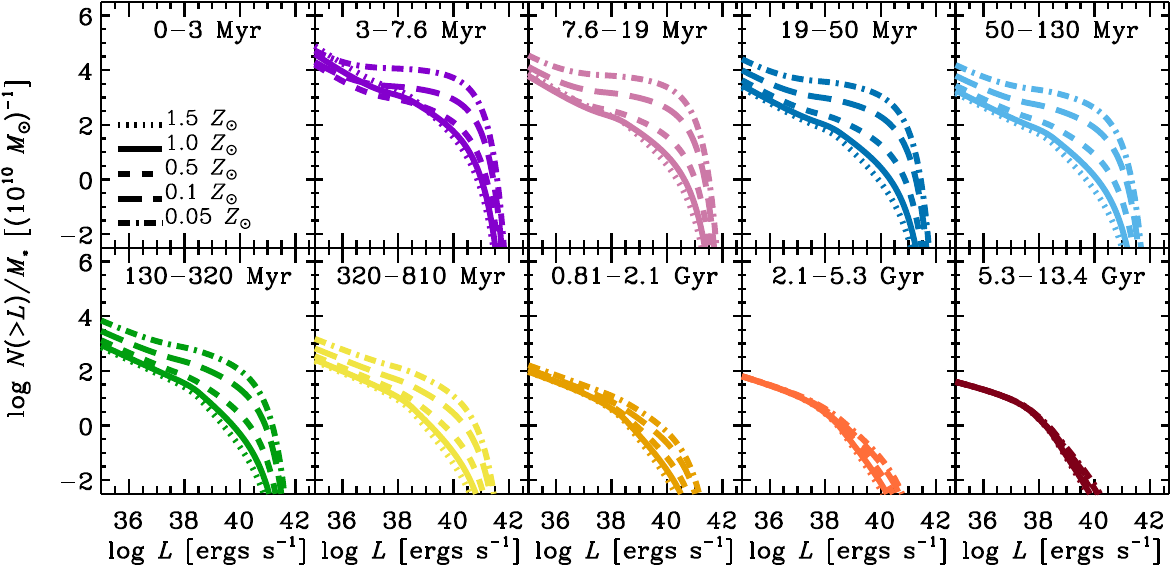}
}
\caption{
Best-fit stellar-mass normalized integrated XLF models from Equations~\eqref{eqn:xlfmod} and \eqref{eqn:par} separated into the 10 SFH age bins, as defined by our SED fitting (see Table~\ref{tab:sed}), and evaluated at 5 metallicity bins per age bin (see annotations).  Note that the first age bin shows no XLF curves due to our definition of $A(t_9 < 0.03,Z) = 0$ in Equation~\eqref{eqn:par}.
}
\label{fig:base}
\end{figure*}

\subsection{Model Optimization and Calculation of Uncertainties}\label{sub:fit}

Using the modeling framework above, we fit all \ngal\ galaxies XLF data simultaneously using the Poisson statistical framework described in $\S$\ref{sub:mod}.  For a given $k$th galaxy with metallicity $Z_k$, we can specify the SFH in terms of surviving stellar mass contributions as a function of $j$th age bin:
\begin{equation}
M_\star(t_j)_k = (\mathcal{R}_j \psi_j \Delta t_j)_k,
\end{equation}
where each term has the same meaning as it did in Equation~\ref{eqn:mstar}.
We can use this form of the SFH, along with Equation~\ref{eqn:xlfmod}, to provide a prediction for the intrinsic XLF of that galaxy following:
\begin{equation}\label{eqn:base}
    \left.\frac{dN_{\rm int}}{d \log L}\right|_k =  \sum_{j=1}^{n_{\rm SFH}} M_\star(t_j)_k \frac{dN(t_j, Z_k)}{d \log L \; dM_\star}.
\end{equation}
Given values of the SFHs (i.e., $M_\star(t_j)_k$), the model can be implanted into Equation~\ref{eqn:mod} and evaluated using the $C$ statistic applied across all galaxies and luminosity bins via
\begin{equation}\label{eqn:cglob}
C_{\rm global} = 2 \sum_{i=1}^{n_L} \sum_{k=1}^{n_{\rm gal}} M_{i,k} - N_{i,k} + N_{i,k} \ln(N_{i,k}/M_{i,k}).
\end{equation}
Using the best-fit SFHs for all galaxies, we minimized Equation~\ref{eqn:cglob} using the Levenberg-Marquardt optimizer {\ttfamily MPFIT} in {\ttfamily IDL} \citep{Mar2009} and identified corresponding best-fit values for the 21 parameters of our model.  As we will discuss below in quantitative detail, the optimized model provides XLF predictions that are statistically consistent with the observational data for every one of the \ngal\ galaxies.  In Table~\ref{tab:xfit}, we tabulate the optimized parameter values, and in Figure~\ref{fig:par}, we plot the age and metallicity dependence of the optimized power-law parameters, as defined in Equations~\eqref{eqn:xlfmod} and \eqref{eqn:par}.  In Figure~\ref{fig:base}, we display the corresponding base-function XLFs for each of the 10 age bins, in terms of the stellar-mass normalized cumulative XLF, $N(>L)/M_\star$, evaluated at 5 metallicities spanning 0.05--1.5~$Z_\odot$. 

Our solutions will be impacted by both Poisson uncertainties on the measured XLFs as well as the SFH uncertainties, which are often large and correlated between SFH bins.  To propagate these uncertainties to the derived parameters, and also assess the goodness of fit for our solution, we performed a posterior predictive check using a Monte Carlo resampling procedure.  In this procedure, we first drew 1,000 realizations of the SFHs from the MCMC posterior chains of each galaxy.  Next, we used Equation~\ref{eqn:base}, and the optimized values in Table~\ref{tab:xfit}, to specify XLF models for each realization.  We then drew simulated XLF data sets from these new model XLF realizations, incorporating the model XLF realization, the CXB contribution, and the completeness function following Equation~\ref{eqn:mod}.  Finally, each of the resulting simulated XLF data sets (i.e., the drawn SFHs and the simulated XLFs of all galaxies) was refit following the procedures described above to determine the best-fit statistic $C_{\rm global}$ (see Eqn.~\ref{eqn:cglob}) and values of the parameters for that draw.

%
%
\begin{figure}
\centerline{
\includegraphics[width=8.5cm]{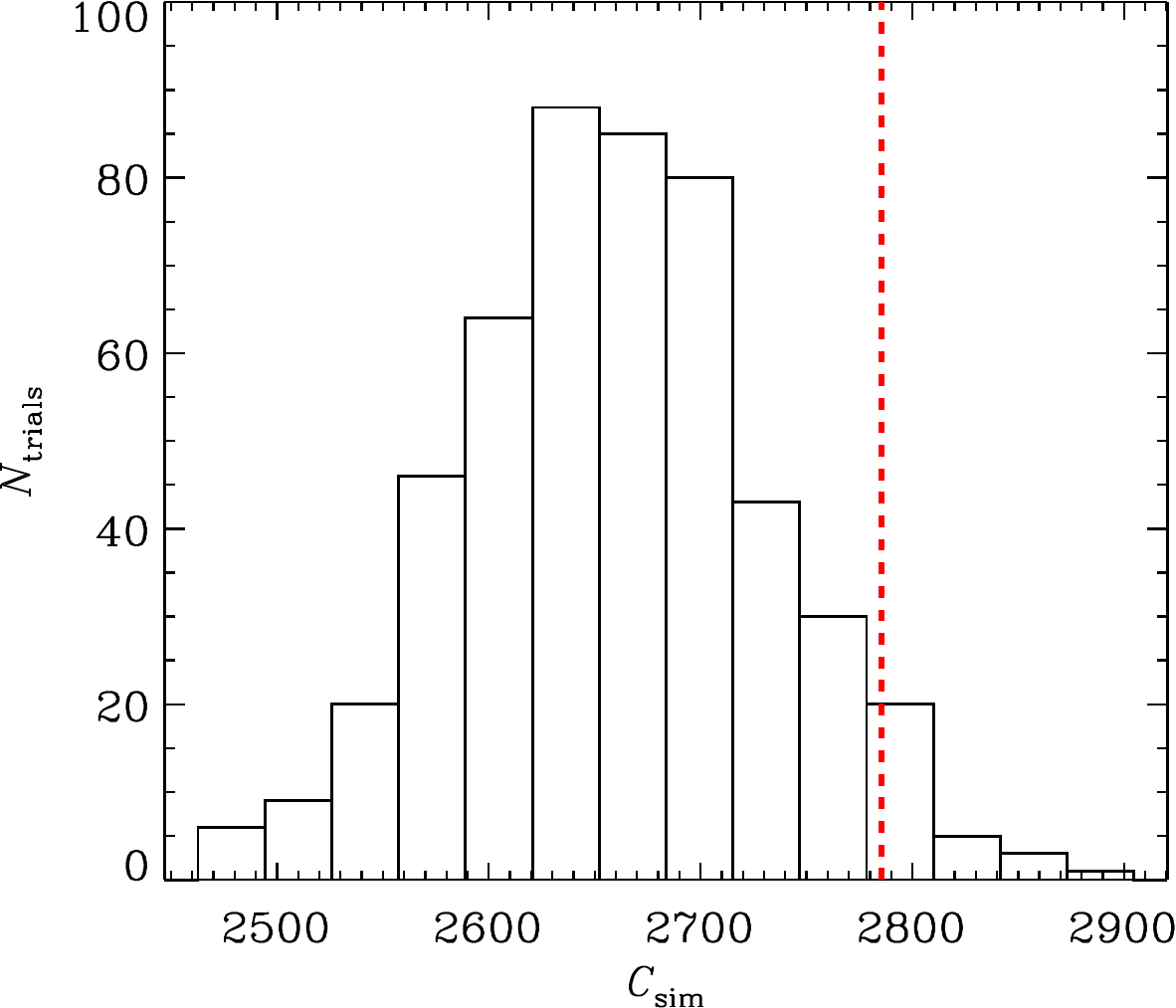}
}
\caption{
Distribution of simulated values of $C_{\rm sim}$, as obtained from the Monte Carlo procedure described in \S\ref{sub:fit}.  Our best-fit value of $C_{\rm global}$ is represented as a vertical red dashed line, which indicates that such a value is consistent with the expected distribution from our simulations.
}
\label{fig:cglob}
\end{figure}

%
%
\begin{figure*}[t]
\centerline{
\includegraphics[width=18cm]{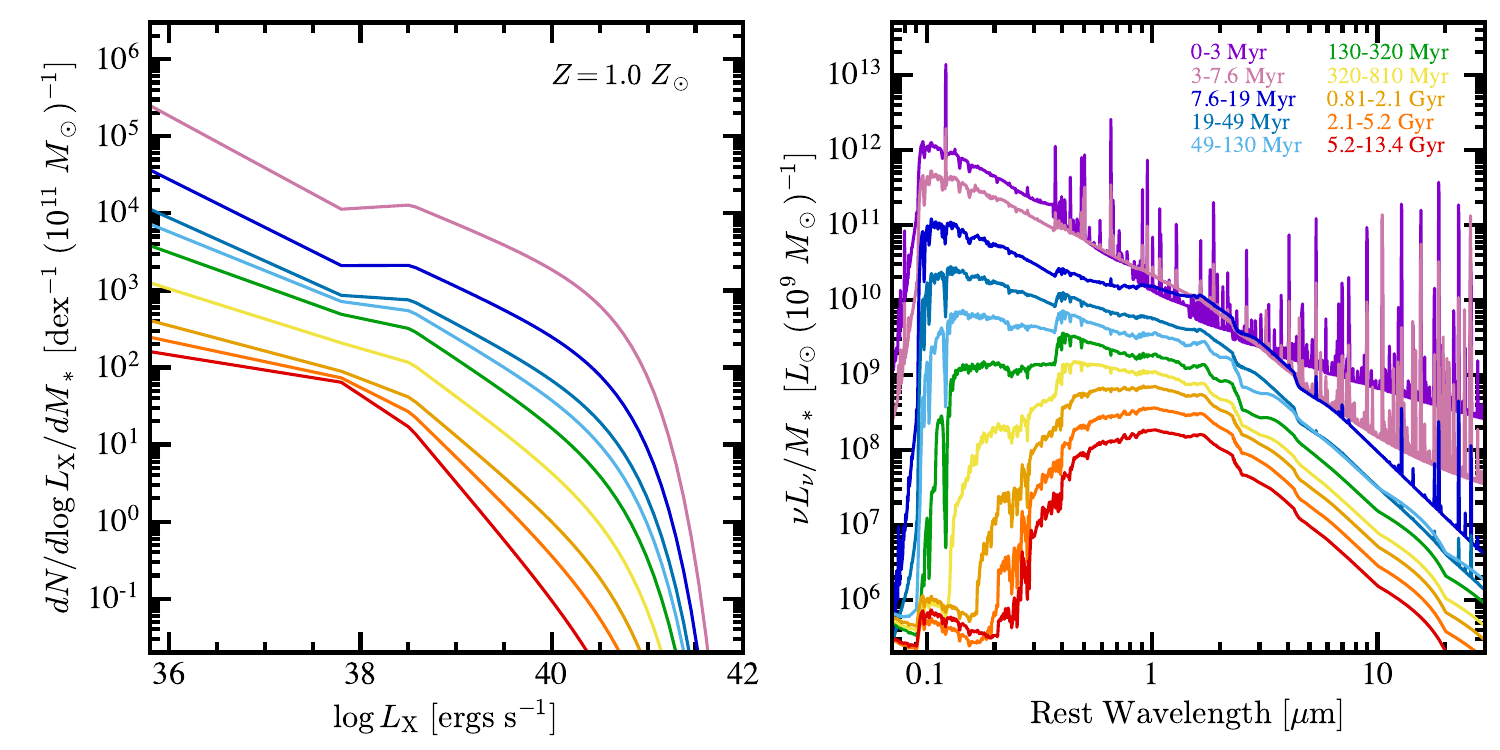}
}
\centerline{
\includegraphics[width=18cm]{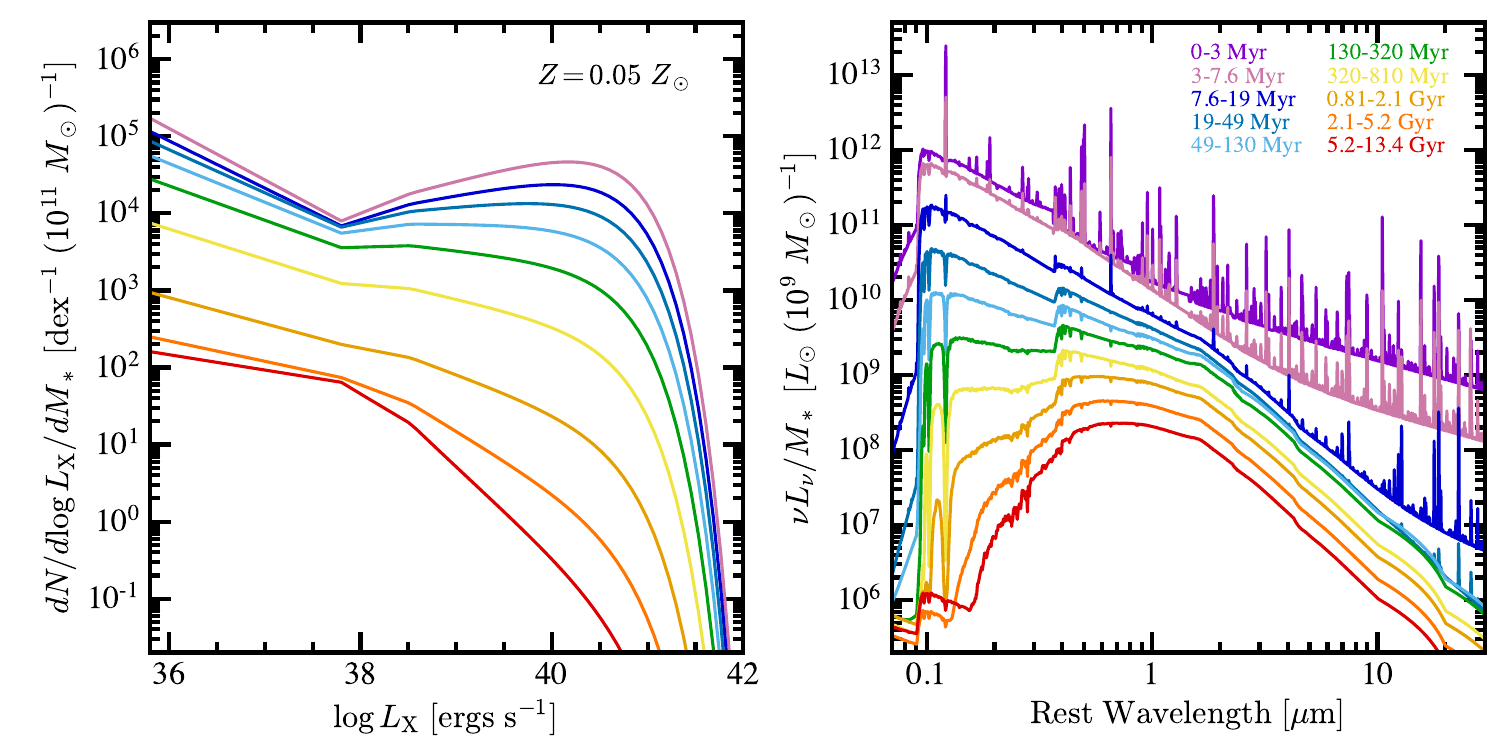}
}
\caption{
Example base function models for age-dependent XLFs ({\it left column\/}) and \pegase-based stellar population and nebula SEDs ({\it right column\/}) at solar metallicity ({\it top row\/}) and 0.05~$Z_\odot$ ({\it bottom row\/}).  For each plot, the color progression from purple-to-red designates a young-to-old age progression (see annotation for specific age ranges).  Note that the XLF models for stellar populations of ages 0--3~Myr is set to zero by construction to account for the timescale of the first SNe and compact-object formation.
}
\label{fig:xsbase}
\end{figure*}

In Figure~\ref{fig:cglob}, we show the distribution of $C_{\rm sim}$ for the 1,000 simulations, with the value of $C_{\rm global}$ for our best-fit indicated.  To quantify the goodness of fit for our best-fit model, we calculated the null-hypothesis probability as follows:
\begin{equation}\label{eqn:pnull}
p_{\rm null} = 1 - {\rm erf}\left( \sqrt{\frac{(C_{\rm global} - C_{\rm exp})^2}{2 \; C_{\rm
var}}} \right),
\end{equation}
where $C_{\rm exp}$ and $C_{\rm var}$ are the mean and variance of the $C_{\rm sim}$ distribution.  We find $p_{\rm null} =$~\pnull, suggesting that the model is statistically consistent with the data, albeit with some tension. We suspect that the use of a different model that has more flexibility to reproduce evolutionary features of the XLF may potentially improve our fits; however, we do not have clear ideas at present for the form of such a model.  As we discuss in $\S$\ref{sub:cav}, perhaps future physically-motivated binary population synthesis models could provide improved characterization of our data. 

Given that our model provides a statistically acceptable value of $C_{\rm global}$, we can use the parameter distribution values obtained in our Monte Carlo runs as estimates on their posterior distributions, with propagation of Poisson uncertainties and SFHs inherently carried along.  In Column~(5) of Table~\ref{tab:xfit}, we list the median and 16--84\% confidence intervals of the parameters. We caution that co-variances among parameters are certainly present and therefore advise against using the combined set of median values to extract a model from Equations~\ref{eqn:mod} and \ref{eqn:par}. Throughout the rest of this paper, we show uncertainties in quantities based on these Monte Carlo runs that properly account for parameter co-variances through calculations of such parameters at every Monte Carlo step.

%
%
\begin{figure*}
\centerline{
\includegraphics[width=5.8cm]{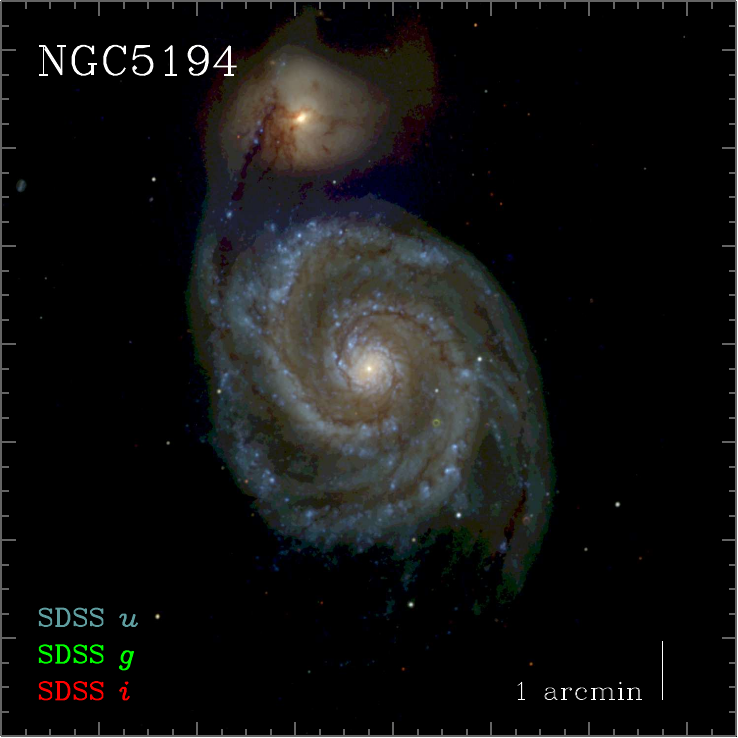}
\includegraphics[width=5.8cm]{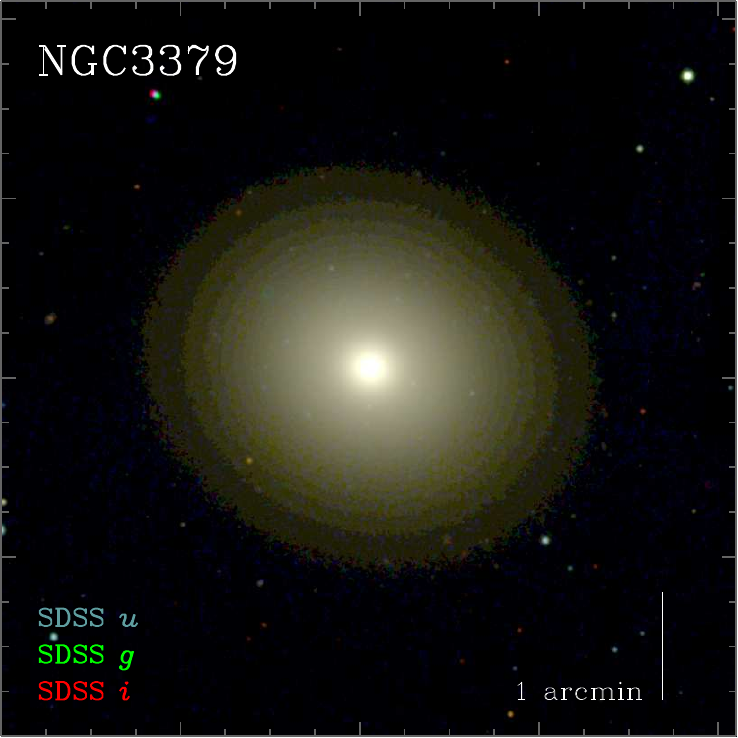}
\includegraphics[width=5.8cm]{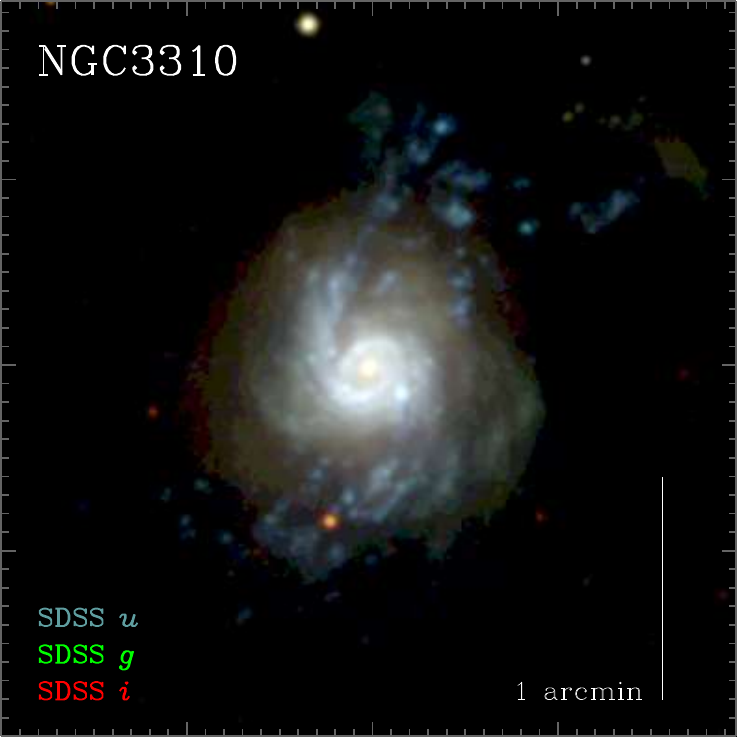}
}
\centerline{
\includegraphics[width=5.9cm]{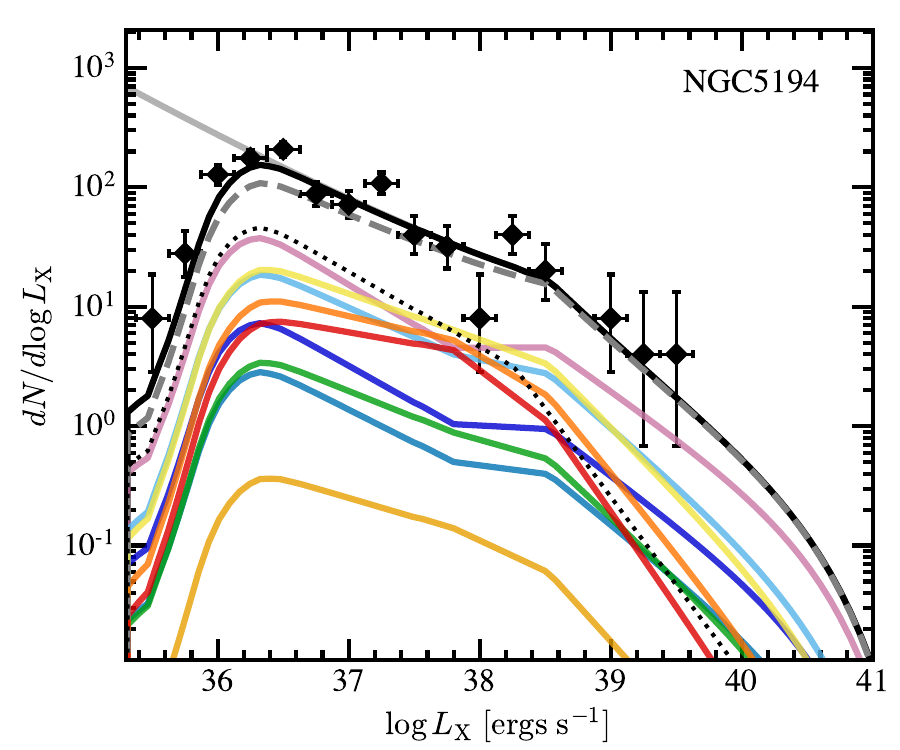}
\includegraphics[width=5.9cm]{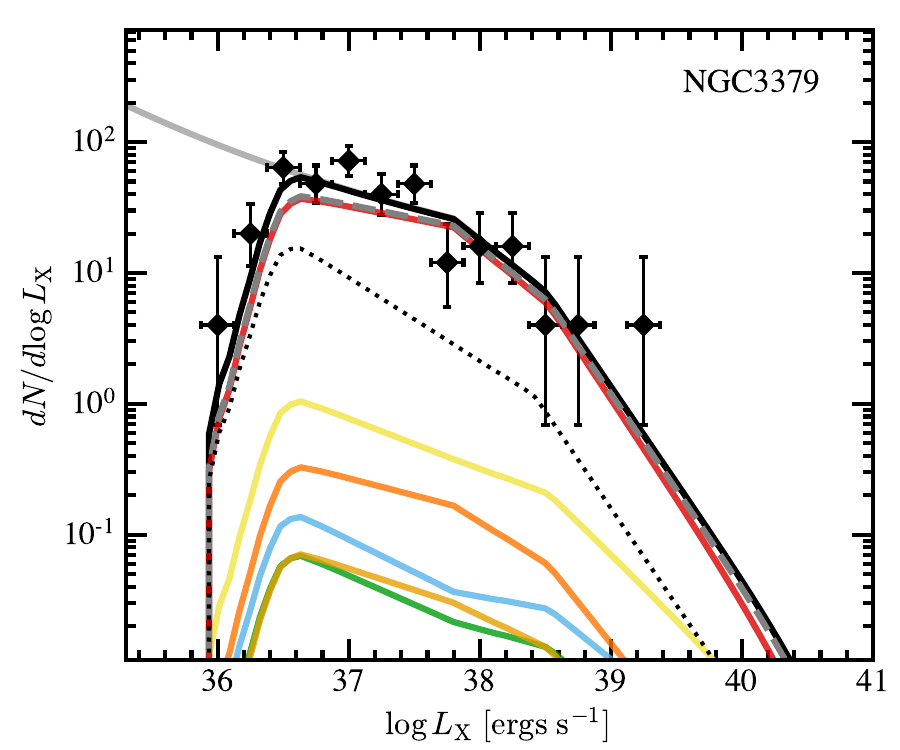}
\includegraphics[width=5.9cm]{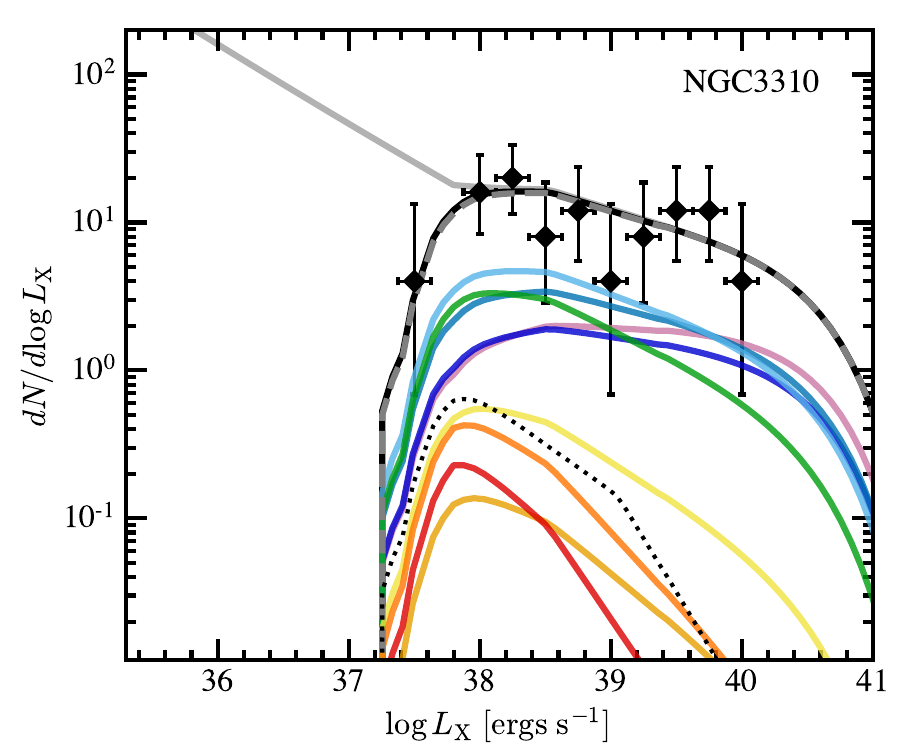}
}
\caption{
({\it top row}) SDSS $u' g' i'$ (blue, green, red) images of NGC~5194 (M51), NGC~3379, and NGC~3310, galaxies with relatively deep \chandra\ observations that represent examples of a normal galaxy with a mix of young and old stellar populations, an early-type galaxy with primarily old stellar populations, and a low-metallicity starburst galaxy, respectively.
({\it bottom row}) Binned XLFs and age-and-metallicity dependent model predictions for the three example galaxies.  Data are shown as filled circles with 16--84\% Poisson confidence intervals.  The full models are shown as black curves, which include contributions from CXB sources ({\it dotted curves\/}) and \xray\ point sources ({\it gray dashed curves\/}) that have been corrected for incompleteness -- the completeness-corrected total models are displayed as solid gray curves.  Note that the XLF models for stellar populations of ages 0--3~Myr are set to zero by construction to account for the timescale of the first SNe and compact-object formation.  The breakdown of contributions to the \xray\ point-source model from each of the 10-bin SFH intervals are shown as colored curves, with colors that follow the same scheme as that shown in the legends of Figure~\ref{fig:xsbase}, right panels.  We note that the model predictions are not explicit fits to a given galaxy's data, but are instead predicted directly from  metallicity and SFH information combined with our models described in $\S$\ref{sub:mod}.
}
\label{fig:xlfgal}
\end{figure*}

%
\section{Discussion}\label{sec:dis}
%

\subsection{Galaxy-by-Galaxy Model Predictions and Data Comparison}\label{sub:gal}

The success of the age-and-metallicity dependent framework presented here provides a notable improvement over past studies in its near ``universality'' across a broad range of galaxy types.  For example, the L19 SFR-and-$M_\star$ dependent XLF model failed to provide such universality, in part due to a number of low-metallicity galaxies with excess XRBs.  The expanded SFR-$M_\star$-metallicity XLF modeling for high-sSFR galaxies presented in L21 also failed to provide good XLF models for a number of galaxies that had bursty SFHs, many of which are included in the present study.  

The implication here is that our empirically-motivated XLF model provides a complete framework for characterizing the XLF of a galaxy, given a SFH and metallicity, akin to stellar population synthesis and dust emission modeling frameworks that are used to characterize UV-to-IR emission from galaxies.  While our XLF framework is empirically based, its data calibration methods can be applied to more physically-motivated binary population synthesis models in the future.  

In Figure~\ref{fig:xsbase}, we illustrate this viewpoint by showing the implied XLF base functions alongside the equivalent \pegase-based stellar population synthesis models that were assumed in the SED fitting and calibration procedures used in this paper.  In principle, the combination of our XLF models, and stellar population, nebular, and dust models could be used in tandem to holistically describe X-ray to IR normal-galaxy data.  

In Figure~\ref{fig:xlfgal}, we demonstrate this by showing the observed XLFs, and corresponding SFH and $Z$ XLF model predictions, for three galaxies with deep \chandra\ observations that span a broad range of environments.  These galaxies include the nearly solar metallicity star-forming galaxy NGC~5194, the elliptical galaxy NGC~3379, and the low-metallicity starburst galaxy NGC~3310.  In each of the XLF plots in the bottom row of Figure~\ref{fig:xlfgal}, the model XLFs shown are based solely on the best values of $Z$ and the SFH (from SED fitting UV-to-IR data) and are not adjusted to fit the individual galaxy's observed XLF.  The full set of equivalent XLF diagrams are shown for all \ngal\ galaxies in our sample in the extended materials. When propagating SFH uncertainties via our posterior predictive check procedure, as described in $\S$\ref{sub:fit}, we calculated $p_{\rm null}$ for each galaxy on an individual basis and found that all galaxy data were described well by our global model (see Col.~12--15 of Table~\ref{tab:xlf}) with $p_{\rm null}^{\rm gal} \simgt$~0.01 for all but three galaxies in our sample: NGC~1427, NGC~1569, and NGC~5408.  NGC~1569 and NGC~5408 are both very low-mass ($\simlt$10$^8$~$M_\odot$) dwarf galaxies with an excess of \xray\ point-sources, particularly at low luminosities ($\log L \approx$~36--37.5).  As such, they are likely to have very bursty SFHs that are not captured by our models. NGC~1427, by contrast is an early-type galaxy, which contains an excess of sources very close to the completeness limit.  It is possible that false detections (e.g., statistical fluctuations) or large luminosity uncertainties could impact these results near the sensitivity limits.

In future work, we plan to include a first version of our age-and-metallicity dependent XLF model into {\ttfamily Lightning} to provide a mechanism for simultaneously fitting XLF and SED data for individual galaxies. Such an XLF model framework can then be replaced with theoretical binary population synthesis models that reproduce X-ray and multiwavelength data in a manner consistent with the methods used here.

\subsection{The Age and Metallicity Dependent XRB XLF Constraint}\label{sub:tzxlf}

We find that the XRB XLF undergoes significant evolution of both age and metallicity. The normalization factor declines by 2--3~dex from $\approx$3~Myr to 10~Gyr, with the decline occurring more slowly with decreasing metallicity.  Although the true rates of decline are currently uncertain and impacted by our parameterization choices, these empirically determined trends have been predicted by population synthesis models \citep[see, e.g.,][]{Lin2010}.  The reason for slower declines at lower metallicity is the predicted excess of Roche-Lobe overflow HMXBs at low-metallicity, due to these relatively tight binaries with small stellar radii having enhanced survivability through the common-envelope phase that occurs in the Hertzsprung gap \citep[e.g.,][]{Bel2010}.  

As stellar population age advances, we find that the slopes of the XLF evolve.  Specifically, the low-luminosity slope, $\alpha_1$ decreases with age, while the medium and high luminosity XLF slopes, $\alpha_2$ and $\alpha_3$, both increase with age (i.e., steepen).  $\alpha_2$ and $\alpha_3$ also exhibit declines as metallicity decreases, resulting in larger numbers of luminous HMXBs and ULXs in low-metallicity galaxies.  As noted in $\S$\ref{sec:intro}, such a result has been commented on previously in the literature \citep[e.g.,][]{Map2010,Bas2016,Kov2020,Leh2021} and is predicted in population synthesis models \citep[e.g.,][]{Lin2010,Fra2013a,Wik2019,Liu2024}.

Our procedure also constrains the locations of XLF breaks and a potential cut-off to the XLF at high luminosities.  Population synthesis models indicate that the existence and locations of these breaks likely correspond to important physical and population demographic transitions. For example, the low-$L$ break, $L_{b,1}$, and flattening from $\alpha_1$ to $\alpha_2$ in the young XRB population, may arise at a complex junction where the HMXB populations of Be XRBs and/or wind-fed XRBs dominate at low-$L$ and decline below the level of Roche-lobe overflow HMXBs above $L_{b,1}$ \citep[e.g.,][]{Mis2023}.  Also, the high-$L$ break, $L_{b,2}$, may correspond to a sudden decline in persistent systems above the Eddington limit of a typical neutron star, a feature most pronounced in old LMXB populations \citep[e.g.,][]{Fra2008}. The presence of a cut-off luminosity at $L_c \approx$~few~$\times 10^{40}$~\lum\ is certainly required by our data; however, whether $L_c$ is simply a feature (e.g., another break in the XLF) or a true cut-off remains unclear. The recent comprehensive statistical study of the ULX XLF from \citet{Tra2022}, which is based on $\approx$1500 ULXs detected in the \xmm, \swift, and \chandra\ archive, identified a significant steepening in the HMXB XLF slope above $\approx$10$^{40}$~\lum, in a manner consistent with our data.

%
%
\begin{figure*}
\centerline{
\includegraphics[width=18cm]{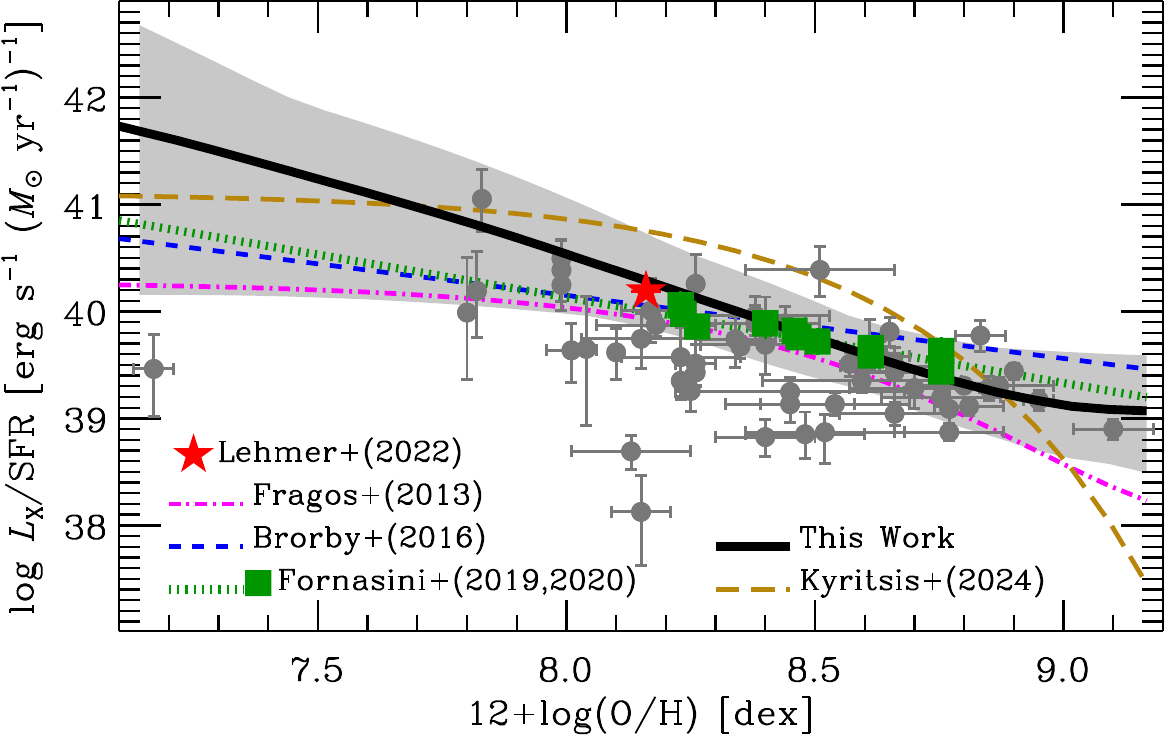}
}
\caption{
XLF model-integrated $L_{\rm X}$/SFR versus metallicity for young stellar populations ($<$100~Myr), as calculated using Equation~\eqref{eqn:lxsz} ({\it solid black curve with gray envelope} representing the 16--84\% confidence region).  The observed quantities for the \nms\ main-sequence and starburst galaxies in our sample are shown as {\it filled dark-gray circles} with 16--84\% confidence intervals error bars.  For comparison, we overlay the observation-based relations derived from \citet{Bro2016} ({\it short-dashed blue\/}) and \citet{Kyr2024} ({\it long-dashed gold\/}), as well as the mean constraint for a sample of 30 high-$z$ analogs from \citet{Leh2022} ({\it red filled star\/}), the X-ray stacking constraints and best-fit relation from \citet{For2019,For2020} ({\it green squares and green dotted curve}), and the binary population synthesis constraint from \citet{Fra2013a} ({long-dashed magenta\/}).
}
\label{fig:lxsz}
\end{figure*}

\subsection{Model-Integrated Scaling Relations}\label{sub:modscal}

Given that our model is composed of fundamental base functions for the evolution of the XRB XLF as a function of age and metallicity, we can integrate our models to derive expected scaling relations, provided assumptions about population SFH and metallicity.  

\subsubsection{The $L_{\rm X}$-SFR-$Z$ Relation}\label{subsublxsz}

To derive the $L_{\rm X}$-SFR-$Z$ relation for young stellar populations (0--100~Myr) implied by our model, we apply the below equation:
\begin{equation}\label{eqn:lxsz}
    L_{\rm X}(Z)/{\rm SFR} = \int_{t=0}^{100~{\rm Myr}} \int_L \frac{1}{\log e} \frac{dN(t,Z)}{d \log L \, dM_\star} \, dL \, dt.
\end{equation}
In Figure~\ref{fig:lxsz}, we show our derived constraint on $L_{\rm X}(Z)/{\rm SFR}$ as a function of metallicity based on Equation~\eqref{eqn:lxsz}, and in Col.~(1)--(4) of Table~\ref{tab:scal} we tabulate the median, 16\%, and 84\% confidence intervals.  We further overlay the $L_{\rm X}/{\rm SFR}$ values derived for each of our galaxies on an individual basis following the methods discussed in $\S$\ref{sec:xray} (see Col.(7) of Table~\ref{tab:xlf} and Col.(10) of Table~\ref{tab:sam}).  We note that the distribution of $L_{\rm X}/{\rm SFR}$ values is found to skew below our best-fit relation.  Such a distribution is expected due to stochastic sampling of the underlying XLF, which results in the galaxy-integrated $L_{\rm X}$ values more likely to be below the XLF-integrated expectation \citep[see, e.g., ][for more detailed discussions]{Gil2004b,Leh2019,Leh2021,Ged2024}.

In Figure~\ref{fig:lxsz}, we also compare our scaling relation result with the population synthesis model predictions from \citet{Fra2013a} ({\it dot-dashed magenta curve\/}) and four independent observational constraints from the literature.   The first of these observational comparisons is from \citet{Leh2022} ({\it red star\/}), which is the average XRB component $L_{\rm X}$[HMXB]/SFR constraint from simultaneous \xray\ spectral fitting of a sample of 30 galaxies, with $D \approx$~200--400~Mpc, selected from SDSS spectra to be in a narrow range of metallicity (\lgoh~$\approx$~8.1--8.2).  The second observational comparison is the relation from \citet{Bro2016} ({\it blue short-dashed curve\/}), which is based on culling samples of 10 blue compact dwarf galaxies from their work, 19 local star-forming galaxies from \citet{Min2012a}, and 10 \xray\ detected extreme metal poor galaxies included in the \citet{Dou2015} low-metallicity sample. The third observational comparison is from the relation derived by \citet{Kyr2024} ({\it gold dot-dashed curve\/}) using \erosita\ all sky survey (eRASS1) stacking analyses of $\approx$19,000 galaxies in 239 distinct regions of SFR-$M_\star$-distance space (for galaxies with $D<200$~Mpc).  Finally, for the fourth comparison, we display the 12 metallicity-binned stacked samples of high-sSFR galaxies from \citet{For2019} and \citet{For2020} that span the redshift range $z \approx$~0.1--2.6 from the \chandra\ Deep Field \citep{For2019} and \chandra\ COSMOS \citep{For2020} surveys ({\it green squares\/}), and the resulting best-fit $L_{\rm X}$-SFR-$Z$ relation to these stacked samples ({\it green dotted curve}). 

We find that our optimized model is in outstanding agreement with the stacked data points from \citet{For2019,For2020} and low-metallicity analog sample constraint from \citet{Leh2022}.  These constraints are based on deep observations of samples that have been carefully selected to have high-sSFRs and vetted for AGN contaminants, and are completely independent of the samples used here.  When comparing between relations, however, we find somewhat poorer agreement, albeit with large uncertainty in the low-metallicity regime \citep[\lgoh~$\simlt 8.0$; see also][for detailed discussions on constraints in this regime]{Ged2024}.  At relatively high metallicity (e.g., \lgoh~$\simgt$~8.8), the \citet{Fra2013b} population synthesis and \citet{Bro2016} blue-compact dwarf curves are deficient and elevated, respectively, over our constraint.  As noted in \citet{Leh2021}, the elevation of the \citet{Bro2016} curve may be enhanced in this regime due to a small number of elevated sources that contain either contamination from LMXBs and/or AGN \citep[e.g., KUG 0842+527 in this sample is a reported radio galaxy with relatively large stellar mass; see][]{Svo2019}.  

In the intermediate-metallicity regime (\lgoh~$\approx$~8.0--8.8), where our model is most tightly constrained, we find agreement between all relations except for that of \citet{Kyr2024}, which reaches up to a factor of $\approx$3 times higher than the other relations.  \citet{Kyr2024} note this elevation and suggest that this is caused by the much broader selection of galaxies in the eRASS1 survey that includes galaxies dominated by populations of relatively young stars and bursty recent SFHs.  This hypothesis is indeed consistent with the age-dependence implied our models, which show that the XRB XLF declines quickly with age from $\approx$3~Myr to 100~Myr (see Fig.~\ref{fig:base} and $\S$\ref{subsublxm} below), the span of time by which calculations of $L_{\rm X}$/SFR average over.  However, given the relatively shallow depth of eRASS1, it is difficult to assess precisely the contributions from undetected \xray\ AGN and hot gas in the \citet{Kyr2024} sample.

%
%
\begin{figure*}
\centerline{
\includegraphics[width=18cm]{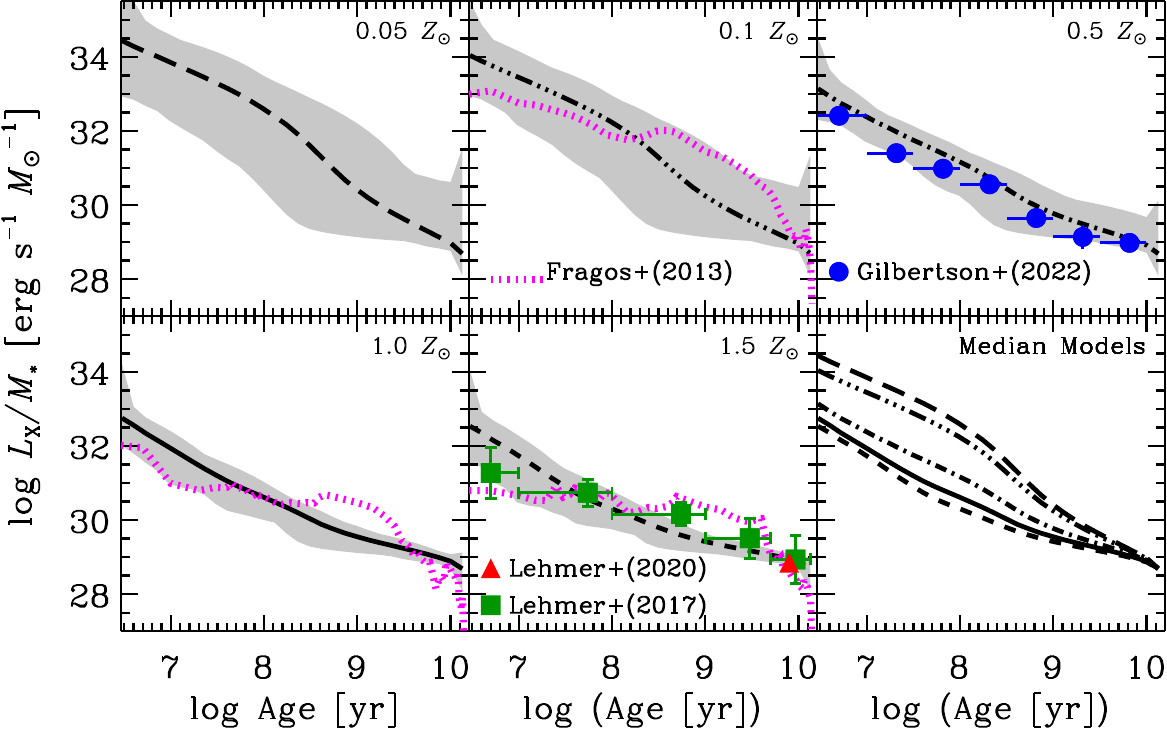}
}
\caption{
XLF model integrated $L_{\rm X}/M_\star$ as a function of age for five metallicity bins ({\it see annotations in upper right of each panel\/}).  The black curves ({\it linestyles varying by metallicity\/}) and gray bands indicate median and 16--84\% ranges, respectively.  The lower-right panel shows models for all five metallicity bins together for comparison.  We overlay comparisons of the binary population synthesis models from \citet{Fra2013a} ({\it dotted magenta curves\/}), the \chandra\ Deep Field statistical constraints for $z \approx$~0--2 normal galaxies from \citet{Gil2022} ({\it blue filled circles} in the 0.5~$Z_\odot$ panel), subgalactic constraints in M51 from \citet{Leh2017} ({\it green filled squares} in the 1.5~$Z_\odot$ panel), and average $L_{\rm X}/M_\star$ values for elliptical galaxies from \citet{Leh2020} ({\it red filled triangle\/} in the 1.5~$Z_\odot$ panel).
}
\label{fig:lxmtz}
\end{figure*}

\subsubsection{The $L_{\rm X}$-$M_\star$-$t$-$Z$ Relation}\label{subsublxm}

Our base function models can also uniquely be integrated over point-source luminosity to provide XRB population $L_{\rm X}/M_\star$ as a function of age and metallicity:
\begin{equation}\label{eqn:lxmtz}
    L_{\rm X}(t,Z)/M_\star = \int_L \frac{1}{\log e} \frac{dN(t,Z)}{d \log L \, dM_\star} \, dL .
\end{equation}
Given the assumptions used to generate our models, the evaluation of Equation~\ref{eqn:lxmtz} provides a prediction of the \xray\ luminosity per stellar mass appropriate for a stellar population with age, $t$, and gas-phase metallicity $Z$.  Thus, both the $L_{\rm X}$ and $M_\star$ term are associated with the population of age $t$, but the $Z$ term is interpreted as the gas-phase metallicity as observed at present.  

In Figure~\ref{fig:lxmtz}, we show $L_{\rm X}(t,Z)/M_\star$ for a continuous age grid, evaluated at five metallicities spanning 0.05--1.5~$Z_\odot$; and we tabulate these results in Col.~(5)--(20) of Table~\ref{tab:scal}.  Uncertainty bands in Figure~\ref{fig:lxmtz} indicate 16--84\% confidence intervals, generated using our MC procedure described in $\S$\ref{sub:fit} and propagated to Equation~\eqref{eqn:lxmtz}.  Our model suggests that the $L_{\rm X}/M_\star$ ratio declines by $\approx$3--5~dex as stellar populations age from 3~Myr to 13.4~Gyr.  At low-metallicity, the magnitude of this evolution is larger and the timescale for X-ray emission to decline following a star-formation event is more delayed compared to high-metallicity populations.  However, we note that the uncertainties on this result are large for $Z \simlt 0.1 Z_\odot$, due to the lack of extreme metal-poor galaxies in the nearby Universe with high-quality XLF constraints.

While there are few constraints on $L_{\rm X}(t,Z)/M_\star$ in the literature, in Figure~\ref{fig:lxmtz}, we include three observational benchmarks to compare with our results.  The first is the \citet{Leh2017} toy-model framework constraints on the age-dependent XLF within M51 (NGC~5194; {\it green squares} in the 1.5~$Z_\odot$ panel), a $\approx$1.5~$Z_\odot$ galaxy that is also included in the sample of the present paper.  \citet{Leh2017} used SFH maps from \citet{Euf2017} and ultradeep (850~ks) \chandra\ imaging to extract subgalactic SFH and XLF information for several regions across M51.  Similar to the techniques adopted in this paper, they developed a parameterized toy model for the age-dependence of the XLF (with resulting $L_{\rm X}/M_\star$) that optimally describes the full suite of subgalactic region XLFs.  The model framework in that paper is somewhat different from that used here, however, the resulting constraints on  $L_{\rm X}/M_\star$ as a function of age are consistent with our model predictions. 

The second observational benchmark is a direct constraint on the average value of $L_{\rm X}/M_\star$ for field LMXBs present within a sample of 24 early-type galaxies \citep[see][; {\it red triangle\/} in 1.5~$Z_\odot$]{Leh2020}.  We place this constraint at a light-weighted age of $8.0 \times 10^9$~yr, based on the SED fit results in \citet{Leh2020}, and a metallicity of 1.5~$Z_\odot$, based on the mass-metallicity relation average value.  The location of this point is in excellent agreement with our model predictions.

The final observational benchmark that we compare to is from the work of \citet{Gil2022} ({\it blue circles} in the 0.5~$Z_\odot$ panel), which utilized a statistical approach to estimate the average galaxy-integrated $L_{\rm X}/M_\star$ as a function of age for 344 $z \approx$~0--2 galaxies located in the \chandra\ Deep Fields.  The statistical approach is similar in nature to that presented in the current paper, using SFH measurements from SED fitting and \chandra\ data to decompose $L_{\rm X}/M_\star$ as a function of age.  The variation of metallicity was not explored explicitly, but the sample average metallicity was determined to be $\approx$0.6~$Z_\odot$.  The comparison of the \citet{Gil2022} results with our model constraints show good agreement across the full range of stellar ages. 

In addition to the observational constraints, we further compare with the theoretical population synthesis models of \citet{Fra2013b}, which were calibrated to match local $L_{\rm X}$[HMXB]/SFR and $L_{\rm X}$[LMXB]/$M_\star$ constraints that were available at the time.  The models
show promising agreement with our observed trends (see {\it dotted magenta curves} in Fig.~\ref{fig:lxmtz}), including both the magnitudes of declines in $L_{\rm X}/M_\star$ as a function of age, as well as the metallicity dependence of the trends.  One notable exception is that the \citet{Fra2013b} models predict a relatively X-ray bright population at ages $\approx$0.3--3~Gyr that is inconsistent with our models.  This disagreement between population synthesis and observational constraints has been previously discussed in \citet{Gil2022}, who attribute the excess to the population synthesis prescriptions applied to binaries with $\approx$1.5--4~$M_\odot$ donor stars. These intermediate-mass stars are expected to go through a short-lived high-accretion state before becoming more traditional LMXBs, a stage of binary evolution that is difficult to model accurately without observational constraints.

\begin{table*}
\renewcommand\thetable{7}
{\small
\begin{center}
\caption{Integrated X-ray Scaling Relations}
\label{tab:scal}
\begin{tabular}{rcc | rccccc}
\hline\hline
 \multicolumn{3}{c}{$\log L_{\rm X}(Z)$/SFR versus $Z$} & \multicolumn{6}{c}{$\log L_{\rm X}(t,Z)/M_\star$ versus $t$ (for selected $Z$)} \\
\vspace{-0.25in} \\
  \multicolumn{3}{c}{\rule{2.5in}{0.01in}} & \multicolumn{6}{c}{\rule{3.5in}{0.01in}} \\
 \multicolumn{1}{c}{$Z$} & \multicolumn{1}{c}{} & \multicolumn{1}{c}{$\log L_{\rm X}/{\rm SFR}$} & \multicolumn{1}{c}{$\log t$} & \multicolumn{5}{c}{ \rule[0.3em]{0.7in}{0.01in} $\log L_{\rm X}/M_\star$ (\lum~$M_\odot^{-1}$) \rule[0.3em]{0.7in}{0.01in} } \\
 \multicolumn{1}{c}{($Z_\odot$)} & \multicolumn{1}{c}{12 + $\log$~(O/H)} & \multicolumn{1}{c}{(\lum~[\sfr]$^{-1}$)} & \multicolumn{1}{c}{(yr)} & \multicolumn{1}{c}{(0.05~$Z_\odot$)} &  \multicolumn{1}{c}{(0.1~$Z_\odot$)} & \multicolumn{1}{c}{(0.5~$Z_\odot$)} & \multicolumn{1}{c}{(1.0~$Z_\odot$)} & \multicolumn{1}{c}{(1.5~$Z_\odot$)} \\
 \multicolumn{1}{c}{(1)} & (2) & (3)--(5) & \multicolumn{1}{c}{(6)} & (7)--(9) & (10)--(12) & (13)--(15) & (16)--(18) & (19)--(21) \\
\hline
0.020 & 6.99 & 41.85$_{-1.71}^{+1.15}$ & 6.48 & 34.44$_{-1.51}^{+2.24}$ & 34.04$_{-1.10}^{+1.78}$ & 33.14$_{-0.84}^{+1.37}$ & 32.75$_{-0.75}^{+1.45}$ & 32.53$_{-0.99}^{+1.51}$ \\
0.024 & 7.07 & 41.77$_{-1.62}^{+1.07}$ & 6.60 & 34.28$_{-1.44}^{+1.31}$ & 33.88$_{-1.04}^{+0.98}$ & 32.93$_{-0.69}^{+0.72}$ & 32.56$_{-0.77}^{+0.49}$ & 32.36$_{-0.92}^{+0.56}$ \\
0.028 & 7.14 & 41.69$_{-1.53}^{+0.99}$ & 6.73 & 34.14$_{-1.46}^{+0.84}$ & 33.74$_{-1.07}^{+0.72}$ & 32.74$_{-0.60}^{+0.57}$ & 32.34$_{-0.79}^{+0.43}$ & 32.15$_{-1.13}^{+0.54}$ \\
0.034 & 7.22 & 41.60$_{-1.45}^{+0.91}$ & 6.85 & 34.01$_{-1.47}^{+0.76}$ & 33.60$_{-1.11}^{+0.72}$ & 32.57$_{-0.64}^{+0.52}$ & 32.16$_{-0.83}^{+0.44}$ & 31.97$_{-1.11}^{+0.54}$ \\
0.040 & 7.29 & 41.51$_{-1.35}^{+0.83}$ & 6.98 & 33.87$_{-1.59}^{+0.74}$ & 33.47$_{-1.30}^{+0.71}$ & 32.40$_{-0.69}^{+0.45}$ & 31.97$_{-0.99}^{+0.46}$ & 31.76$_{-1.29}^{+0.54}$ \\
0.047 & 7.37 & 41.41$_{-1.26}^{+0.76}$ & 7.11 & 33.72$_{-1.64}^{+0.73}$ & 33.33$_{-1.35}^{+0.69}$ & 32.24$_{-0.66}^{+0.37}$ & 31.77$_{-0.88}^{+0.46}$ & 31.56$_{-1.17}^{+0.54}$ \\
0.056 & 7.44 & 41.31$_{-1.18}^{+0.69}$ & 7.23 & 33.59$_{-1.68}^{+0.76}$ & 33.19$_{-1.38}^{+0.67}$ & 32.08$_{-0.64}^{+0.34}$ & 31.58$_{-0.76}^{+0.44}$ & 31.34$_{-1.00}^{+0.54}$ \\
0.067 & 7.52 & 41.22$_{-1.09}^{+0.66}$ & 7.36 & 33.45$_{-1.73}^{+0.80}$ & 33.04$_{-1.42}^{+0.66}$ & 31.93$_{-0.61}^{+0.36}$ & 31.39$_{-0.70}^{+0.39}$ & 31.13$_{-0.92}^{+0.50}$ \\
0.080 & 7.59 & 41.12$_{-1.01}^{+0.65}$ & 7.48 & 33.30$_{-1.83}^{+0.84}$ & 32.90$_{-1.55}^{+0.69}$ & 31.77$_{-0.70}^{+0.38}$ & 31.21$_{-0.78}^{+0.43}$ & 30.93$_{-0.82}^{+0.47}$ \\
0.095 & 7.67 & 41.02$_{-0.92}^{+0.63}$ & 7.61 & 33.15$_{-1.91}^{+0.86}$ & 32.75$_{-1.63}^{+0.72}$ & 31.63$_{-0.86}^{+0.37}$ & 31.05$_{-0.80}^{+0.47}$ & 30.73$_{-0.82}^{+0.52}$ \\
0.113 & 7.74 & 40.92$_{-0.84}^{+0.62}$ & 7.74 & 32.98$_{-1.94}^{+0.88}$ & 32.60$_{-1.68}^{+0.73}$ & 31.48$_{-0.95}^{+0.38}$ & 30.90$_{-0.73}^{+0.47}$ & 30.57$_{-0.73}^{+0.54}$ \\
0.134 & 7.82 & 40.81$_{-0.76}^{+0.59}$ & 7.86 & 32.80$_{-2.05}^{+0.90}$ & 32.43$_{-1.69}^{+0.73}$ & 31.33$_{-0.96}^{+0.40}$ & 30.76$_{-0.67}^{+0.46}$ & 30.43$_{-0.64}^{+0.51}$ \\
0.159 & 7.89 & 40.70$_{-0.68}^{+0.57}$ & 7.99 & 32.61$_{-2.19}^{+0.89}$ & 32.25$_{-1.84}^{+0.72}$ & 31.18$_{-0.92}^{+0.39}$ & 30.63$_{-0.62}^{+0.42}$ & 30.32$_{-0.56}^{+0.48}$ \\
0.189 & 7.97 & 40.59$_{-0.59}^{+0.54}$ & 8.11 & 32.39$_{-2.32}^{+0.94}$ & 32.05$_{-1.98}^{+0.75}$ & 31.02$_{-1.01}^{+0.38}$ & 30.49$_{-0.56}^{+0.37}$ & 30.19$_{-0.52}^{+0.46}$ \\
0.225 & 8.04 & 40.47$_{-0.51}^{+0.51}$ & 8.24 & 32.17$_{-2.42}^{+1.02}$ & 31.83$_{-2.09}^{+0.83}$ & 30.84$_{-1.15}^{+0.44}$ & 30.35$_{-0.71}^{+0.33}$ & 30.07$_{-0.47}^{+0.41}$ \\
0.267 & 8.12 & 40.36$_{-0.46}^{+0.47}$ & 8.37 & 31.91$_{-2.41}^{+1.14}$ & 31.58$_{-2.09}^{+0.93}$ & 30.66$_{-1.20}^{+0.49}$ & 30.20$_{-0.78}^{+0.32}$ & 29.94$_{-0.56}^{+0.40}$ \\
0.317 & 8.19 & 40.24$_{-0.43}^{+0.42}$ & 8.49 & 31.61$_{-2.25}^{+1.29}$ & 31.30$_{-1.95}^{+1.04}$ & 30.46$_{-1.13}^{+0.56}$ & 30.04$_{-0.75}^{+0.34}$ & 29.81$_{-0.55}^{+0.38}$ \\
0.377 & 8.27 & 40.12$_{-0.40}^{+0.39}$ & 8.62 & 31.29$_{-2.02}^{+1.45}$ & 31.01$_{-1.74}^{+1.19}$ & 30.25$_{-1.00}^{+0.63}$ & 29.88$_{-0.66}^{+0.35}$ & 29.69$_{-0.49}^{+0.34}$ \\
0.448 & 8.34 & 40.00$_{-0.40}^{+0.38}$ & 8.74 & 30.98$_{-1.75}^{+1.61}$ & 30.73$_{-1.52}^{+1.31}$ & 30.07$_{-0.87}^{+0.65}$ & 29.76$_{-0.58}^{+0.32}$ & 29.59$_{-0.43}^{+0.30}$ \\
0.533 & 8.42 & 39.88$_{-0.39}^{+0.37}$ & 8.87 & 30.68$_{-1.50}^{+1.73}$ & 30.48$_{-1.30}^{+1.40}$ & 29.92$_{-0.76}^{+0.63}$ & 29.65$_{-0.50}^{+0.30}$ & 29.50$_{-0.37}^{+0.26}$ \\
0.634 & 8.49 & 39.77$_{-0.38}^{+0.34}$ & 8.99 & 30.42$_{-1.28}^{+1.79}$ & 30.26$_{-1.12}^{+1.43}$ & 29.78$_{-0.65}^{+0.62}$ & 29.56$_{-0.44}^{+0.29}$ & 29.42$_{-0.32}^{+0.22}$ \\
0.753 & 8.57 & 39.65$_{-0.35}^{+0.31}$ & 9.12 & 30.19$_{-1.09}^{+1.80}$ & 30.06$_{-0.96}^{+1.42}$ & 29.66$_{-0.56}^{+0.60}$ & 29.47$_{-0.38}^{+0.26}$ & 29.36$_{-0.28}^{+0.19}$ \\
0.895 & 8.64 & 39.54$_{-0.34}^{+0.32}$ & 9.25 & 29.98$_{-0.91}^{+1.75}$ & 29.88$_{-0.80}^{+1.36}$ & 29.55$_{-0.48}^{+0.59}$ & 29.39$_{-0.32}^{+0.24}$ & 29.29$_{-0.25}^{+0.20}$ \\
1.064 & 8.72 & 39.43$_{-0.39}^{+0.35}$ & 9.37 & 29.80$_{-0.75}^{+1.64}$ & 29.71$_{-0.66}^{+1.25}$ & 29.45$_{-0.40}^{+0.63}$ & 29.31$_{-0.29}^{+0.24}$ & 29.23$_{-0.24}^{+0.19}$ \\
1.265 & 8.79 & 39.33$_{-0.41}^{+0.39}$ & 9.50 & 29.63$_{-0.60}^{+1.47}$ & 29.56$_{-0.53}^{+1.25}$ & 29.35$_{-0.37}^{+0.67}$ & 29.24$_{-0.29}^{+0.24}$ & 29.16$_{-0.23}^{+0.18}$ \\
1.503 & 8.87 & 39.24$_{-0.43}^{+0.42}$ & 9.62 & 29.46$_{-0.49}^{+1.38}$ & 29.41$_{-0.47}^{+1.32}$ & 29.25$_{-0.34}^{+0.70}$ & 29.16$_{-0.26}^{+0.22}$ & 29.10$_{-0.22}^{+0.17}$ \\
1.787 & 8.94 & 39.17$_{-0.47}^{+0.47}$ & 9.75 & 29.30$_{-0.42}^{+1.46}$ & 29.26$_{-0.39}^{+1.39}$ & 29.14$_{-0.28}^{+0.72}$ & 29.08$_{-0.22}^{+0.20}$ & 29.04$_{-0.18}^{+0.16}$ \\
2.123 & 9.02 & 39.11$_{-0.49}^{+0.51}$ & 9.88 & 29.14$_{-0.31}^{+1.53}$ & 29.12$_{-0.29}^{+1.45}$ & 29.04$_{-0.21}^{+0.74}$ & 28.99$_{-0.17}^{+0.18}$ & 28.96$_{-0.14}^{+0.15}$ \\
2.524 & 9.09 & 39.08$_{-0.51}^{+0.52}$ & 10.00 & 28.98$_{-0.21}^{+1.64}$ & 28.96$_{-0.21}^{+1.52}$ & 28.92$_{-0.21}^{+0.76}$ & 28.89$_{-0.20}^{+0.19}$ & 28.87$_{-0.19}^{+0.17}$ \\
3.000 & 9.17 & 39.07$_{-0.58}^{+0.52}$ & 10.13 & 28.69$_{-0.59}^{+2.84}$ & 28.69$_{-0.59}^{+2.65}$ & 28.69$_{-0.59}^{+1.43}$ & 28.69$_{-0.59}^{+0.44}$ & 28.69$_{-0.59}^{+0.30}$ \\
\hline
\end{tabular}
\end{center}
\tablecomments{Col.(1) and (2): Metallicity in solar units and oxygen abundance relative to hydrogen (\lgoh), respectively.  Col.(3)--(5): Median, 16\%, and 84\% confidence interval for $\log L_{\rm X}(Z)$/SFR for the population of sources with $<$100~Myr, given the metallicity in Column~(1).  Values of $\log L_{\rm X}(Z)$/SFR were calculated following Eqn.~\eqref{eqn:lxsz}. Col.(6): Stellar population age. Col.(7)--(21): Median, 16\%, and 84\% confidence interval for $\log L_{\rm X}(t,Z)/M_\star$ for the population with age $t$ listed in Column~(4) and metallicity provided in the column header.  Values of $\log L_{\rm X}(t,Z)/M_\star$ were calculated following Eqn.~\eqref{eqn:lxmtz}.}
}
\end{table*}

\subsection{Caveats and Future Improvements of Our XLF Model Framework}\label{sub:cav}

While our model base functions provide a new empirical framework that allows for characterization of XLFs across a broad range of galaxy types, there are still a number of caveats to our model that are present and can be improved upon in future studies.  

{\it Metallicity:} Our study makes use of single-value metallicity measurements for each galaxy that are based on emission-line diagnostics, and they are appropriate for ionized gaseous nebulae that surround young ($\simlt$10~Myr) stellar populations (see $\S$\ref{sec:samp} for details).  As such, these values represent either light-weighted averages over several \HII\ regions or large apertures (or strips from slits), and they do not account for metallicity gradients across the galaxies.    

The use of single-valued metallicities also ignores inherent metallicity histories in galaxies.  In particular, older stellar populations (e.g., $\simgt$1~Gyr) will have lower metallicities than those adopted in this study.  As discussed in $\S$\ref{sec:samp}, in the case of early-type galaxies, adopted metallicities are based on the $M_\star$-$Z$ relation and can be higher than the stellar metallicities by up to an order of magnitude.   Future models would benefit from some self-consistent tracking of metallicity histories along with star-formation histories. 

{\it Supernova remnants:}  Throughout this paper, we have discussed \xray\ point-source populations detected across the galactic footprints as either XRBs associated with the galaxies or foreground/background objects (e.g., AGN).  Our statistical methods allow for the separation of sources that are intrinsic to the galaxies from background sources by modeling these components separately; however, our modeling does not distinguish the nature of the sources that are detected.  While XRBs are expected to dominate the point-source populations intrinsic to the galaxies, the fraction of sources that are supernova remnants is known to climb with decreasing luminosity, becoming non-negligible at $L \simlt 10^{37}$~\lum\ \citep[e.g.,][]{Tul2011,Lon2014}.

While our data are insufficient to differentiate supernova remnants from XRBs, we can infer that supernova remnants would be primarily associated with the young stellar populations ($\simlt$100~Myr) and that they will have their largest impact below $L \simlt 10^{37}$~\lum.  As such, the {\it true} faint-end XLF slopes for young populations of XRBs are expected to be less steep than the values derived for our models.  Future work that involves the direct association and/or \xray\ spectral classification of supernova remnants would be required to clearly disentangle their contributions from XRBs \citep[see, e.g., ][for a such a study in M83]{Hun2021}.

{\it Globular clusters and dynamical formation processes:} Several past studies have shown clearly that LMXBs can form via dynamical interactions (e.g., tidal capture  and multibody exchange with
constituent stars in primordial binaries) in high stellar density environments
like GCs \citep[e.g.,][]{Cla1975,Fab1975,Hil1976}, and
possibly some high-density galactic regions \citep[e.g.,][]{Vos2007,Zha2013}.  For galaxies with high $S_N$ (i.e., GC specific frequency; see $\S$\ref{sec:intro} for discussion), there is evidence that LMXBs form dynamically in GCs can dominate the total LMXB population power output of their host galaxy \citep[see, e.g.,][]{Irw2005,Jue2005,Siv2007,Hum2008,Zha2012,Kim2013}.

Such dynamical LMXB formation pathways are not expected to have formation frequencies that scale directly with the properties of the host galaxy, in the manner that we have explored in this paper.  In the present study, we have removed from our sample galaxies with $S_N \simgt 2$, which are expected to be dominated by LMXB populations that were formed within and/or kicked out of GCs \citep[see][ for motivation]{Leh2020}.  We assume that the LMXB populations in the galaxies in our sample are dominated by field LMXBs that are form in-situ within the galactic stellar population.  Despite our efforts, the level by which the GC LMXB populations contribute to the galaxy sample in this paper is difficult to quantify and may still be significant.  Further detailed studies are required to help interpret this possibility and quantify the impact of GC, or more generally dynamical, LMXB formation.

{\it X-ray binary population variability:} The \chandra\ point-source catalogs that are used in this paper are constructed using (when relevant) merged observational data sets to obtain the deepest possible constraints on the XLFs.  Some of the data sets (e.g., for NGC~5194) were generated using many ObsIDs, with exposures taken weeks-to-years apart.  It has been shown, by several studies of XRBs in the Milky Way, that XRBs can vary in luminosity by orders of magnitude over similar timescales \citep[e.g.,][]{Bell2010}.  As such, the combined observations that make up our source selection and property measurements will be based on average XRB characteristics, and the variation of source luminosities will lead to variations in the XLF across each ObsID.

While a thorough analysis of this issue is of interest, it is both beyond the scope of the current paper, and not expected to have a significant impact on the results presented in this paper.  Investigations of the XLFs of select nearby galaxies with multiple \chandra\ observations have shown that while \xray\ point-sources indeed vary with time, the population XLF shape remains persistent, at a level well below the stochastic variance \citep[see, e.g., ][]{Zez2007,Fri2008,Sel2011,Bin2017}, which is accounted for here in the use of the $C$ statistic.  

{\it XLF Modeling with Continuous Age and Metallicity Variables:} In order to make the empirical problem tractable, our models, as defined in Equations~\eqref{eqn:mod} and \eqref{eqn:par}, are continuous functions of age $t$ and metallicity $Z$.  However, population synthesis modeling has shown evidence for more complex evolution and metallicity dependencies that are not captured in the simple functions used in our work \citep[see, e.g.,][]{Fra2008,Lin2010,Fra2013a,Wik2017}.  Similarly, discontinuities have been observed in the formation timescales of HMXB populations in nearby galaxies, for example, as supergiant donor-star HMXBs decline and Be-HMXBs increase in numbers \citep[see, e.g.,][]{Ant2016,Ant2019,Gar2018,Laz2023}.

While certain features could be included, to some degree, in our model framework, it is uncertain what types of functions would be relevant (including their dependence on both age and metallicity).  As we discuss in $\S$\ref{sub:mod}, our model statistic is acceptable overall, but only marginally so.  It is possible that adding correct features (e.g., discontinuities that are attributed to specific epochs) to our models and/or SFHs could meaningfully improve the fit statistics.

Future versions of this modeling framework would benefit most greatly from the inclusion of workable binary population synthesis models that simultaneously predict the XLF formation and evolution as functions of age and metallicity and provide underlying stellar SED models.  Such models could be calibrated using the full suite of data sets presented in this paper.  At the time of the writing of this paper, there are several binary population synthesis models that are in use, capable of predicting XRB populations \citep[e.g.,][]{Bel2008,Bre2020,Ril2022,Ior2023,Fra2023}; however, no publicly available code can simultaneously predict XRB XLFs, and X-ray--to--IR SEDs from such inherent populations \citep[however, see][for a recent example of steps moving in this direction]{Lec2024}.

%
\section{Summary and Future Constraints}\label{sec:sum}
%

We have presented a new empirical framework for predicting XRB XLFs, given SFHs and metallicities.  Our framework was calibrated using \chandra\ and FUV-to-FIR multiwavelength data sets for a sample of \ngal\ nearby ($D < 40$~Mpc) galaxies that span broad ranges of morphological type, SFR, $M_\star$, SFH, and metallicity.  SED fitting techniques were used to analyze the multiwavelength data and extract SFH information for the galaxies.  XLF data for all galaxies was simultaneously forward-modeled using a SFH and metallicity dependent ``global'' XLF model.  This model self-consistently describes the age evolution of XRB population demographics as a function of metallicity, analogous to the population synthesis SED model base functions.  Below, we list our key findings.

\begin{enumerate}[leftmargin=*]

\item We present new SED fit solutions for the \ngal\ galaxies in our sample using a recently updated version of the \lightning\ SED fitting code (see Fig.~\ref{fig:sed} and Tables~\ref{tab:sed}).  \lightning\ allows for the selection of either \pegase\ single-star stellar models or \bpass\ models that include binary star evolution.  We show that, while both stellar models can provide reasonable characterizations of the FUV-to-FIR SEDs, \pegase\ provides somewhat better statistical modeling of the data in the 1.5--2.5~$\mu$m range.  The resulting \pegase\ and \bpass\ SFHs are similar across the majority of cosmic look-back times, with the exception of the 0.13--2.1~Gyr range, where the \pegase\ SFHs are mildly elevated over \bpass.  The remaining results from this paper are based on \pegase\ SED fit results.

\item We fit the XLFs of all \ngal\ galaxies and derived intrinsic galaxy-integrated \xray\ point-source luminosities, $L_{\rm X}$ (0.5--8~keV).  Using these derived values, along with SFR and $M_\star$ values derived from our SED fits, we computed revised \xray\ scaling relations.  For star-forming active galaxies with sSFR~$> 10^{-10}$~yr$^{-1}$, we find 
$$\log L_{\rm X} = \log {\rm SFR} + (39.39 \pm 0.017)$$
(Fig.~\ref{fig:lxsfr}), which is in agreement with several past studies, but more tightly constrained (see $\S$\ref{sec:xray}).  Such a relation is expected to be driven by HMXB population scalings with SFR.  For the full galaxy sample, we computed relations involving both SFR (for HMXBs) and $M_\star$ (for LMXBs), finding a best-fit relation 
$$L_{\rm X} = \alpha_{\rm LMXB} M_\star + \beta_{\rm HMXB} {\rm SFR}$$
$$\log (\alpha_{\rm LMXB}~[{\rm ergs~s^{-1}~M_\odot^{-1}}])   = 29.957 \pm 0.004 $$
$$\log (\beta_{\rm HMXB}~[{\rm ergs~s^{-1}~(M_\odot~yr)^{-1}}])  = 39.303 \pm 0.004.$$
While this relation provides a good overall characterization of the galaxy-integrated $L_{\rm X}$, given SFR and $M_\star$, substantial scatter remains, in particular for high-sSFR galaxies, that appears to be correlated with metallicity (see Fig.~\ref{fig:lxalbe}).

\item To gain a sense of how the XLF varies with age and metallicity, we inspected co-added XLFs of galaxy subsamples, selected by physical-properties (Fig.~\ref{fig:xlfdep}).  Construction of co-added XLFs from galaxy subsamples selected by sSFR, a proxy for average stellar age, shows that the stellar-mass normalized XLF undergoes both a decline in normalization and an overall change in shape (flattening low-$L$ slope and steepening high-$L$ slope) with decreasing sSFR (increasing stellar age).  When selecting galaxy subsamples with high-sSFR ($\simgt$10$^{-10}$~yr$^{-1}$) in bins of metallicity, we find that the SFR-normalized XLF steepens with increasing metallicity, yielding lower integrated XRB population luminosities (see $\S$\ref{sub:mod} for details).  

\item Motivated by the trends observed for co-added XLFs of galaxy subsamples, we constructed a ``global'' model, detailing the evolution of XLF shape and normalization parameters as functions of age and metallicity (see Eqns.~\eqref{eqn:xlfmod} and \eqref{eqn:par}).  Our model contains 21 total free parameters that we constrained by simultaneously modeling the XLF data for all \ngal\ galaxies, given their SFHs and metallicities (see Table~\ref{tab:xfit}, Fig.~\ref{fig:par}, and $\S$\ref{sub:fit} for details).  Simultaneous application of our model to all XLF \xray\ luminosity bins for all galaxies reveals statistically acceptable characterization for the full galaxy sample ($p_{\rm null} =$~\pnull), albeit with room for improvement in the statistic.  In particular, galaxies with very low-mass ($M_\star \simlt 10^9$~$M_\odot$), which are expected to have bursty SFHs, are the worst fit by our model framework, given the SFH constraints.

\item Integration of our models allow for predictions of galaxy-integrated scaling relations.  Integration over the first 100~Myr and over all XRB luminosities (Eqn.~\eqref{eqn:lxsz}) yields a prediction for $L_{\rm X}$[HMXB]-SFR-$Z$ (see Fig.~\ref{fig:lxsz}).  We find good agreement with past studies based on galaxy-integrated quantities, showing that $L_{\rm X}$[HMXB]/SFR declines with metallicity by $\approx$1.5--2~dex from 0.5--1.5~$Z_\odot$.  Integration over XRB luminosity alone provides unique scalings of $L_{\rm X}$[XRB]$/M_\star$ as functions of age and metallicity (see Fig.~\ref{fig:lxmtz}).  These relations reproduce $L_{\rm X}$[XRB]$/M_\star$ versus age benchmarks for average populations in the \chandra\ Deep Fields, subgalactic regions in M51, and old LMXBs in elliptical galaxies.  Comparison of these relations with past population synthesis results shows similar trends, but tension for ages $\approx$0.3--3~Gyr, where population synthesis models overpredict the power-output (per stellar mass) from these populations.

\end{enumerate}

The framework presented in this paper provides a blueprint for connecting XRB population and galaxy stellar population data with inter-connected model components. Several future avenues can be taken to improve this framework using both expanded data sets and new models. In $\S$\ref{sub:cav}, we discussed several additional model considerations that can be added to this framework in the future. In particular, a binary population synthesis model framework that simultaneously produces XRB predictions and carries out stellar atmosphere calculations that can produce self-consistent SED models is highly desirable: e.g., a fully consistent set of binary population synthesis base functions like those shown in Figure~\ref{fig:xsbase}.

For observational constraints, future studies that include both new XLF data and galaxy-integrated data from galaxy samples where XLFs cannot be extracted directly (e.g., due to large distances or shallow observations) would significantly improve the quality of our constraints.  There are much larger samples of galaxies for which only integrated population characteristics can be obtained that fill in regions of parameter space.  These include (but are not limited to) luminous infrared galaxies \citep[e.g., ][]{Iwa2011,Tor2018}, high-redshift analog galaxies \citep[e.g.,][]{Bas2013a,Bas2016,Pre2013,Bro2014,Bro2017,Leh2022}, Lyman-continuum emitters \citep[e.g.,][]{Ble2019} deep-field galaxy samples \citep[e.g.,][]{Leh2008,Leh2016,Bas2013b,Air2017,For2019,For2020,Gil2022}, and more distant galaxy samples selected over wide-area surveys \citep[e.g.,][]{Vul2021,Sor2022,Kyr2024}.  Such samples span broader ranges of physical parameter space than those constrained here (particularly in the low-metallicity regime).

Expanded XLF constraints can be achieved by broadening the sample and through the use of subgalactic region analyses.  For example, excellent constraints on both SFHs and XLFs are available for Local Group galaxies \citep[e.g.,][]{Ant2016,Ant2019,Gar2018,Laz2023}. The near proximity of these galaxies enable high age resolution on SFHs via color-magnitude diagram fitting, as well as XLF extension to very low luminosities (e.g., to $\simlt$10$^{34}$~\lum). 

For somewhat more distant galaxies ($D \approx$~3--40~Mpc), where XLF analyses is still possible, larger samples of galaxies exist with growing archives of high angular resolution multiwavelength data sets.  In the future, multiwavelength analyses that focus on subgalactic regions, in these samples, can be carried out to more cleanly isolate and constrain SFHs and the \xray\ sources. For example, at the time of this writing, the PHANGS survey is carrying out very large treasury programs with \hst, \jwst, VLT MUSE, ALMA, and other facilities to obtain high-resolution spectrophotometric data sets for a sample of 74 nearby galaxies ($D \approx$3--30~Mpc).  A series of \chandra\ programs, including both a Large ($\approx$1~Ms; PI: Lehmer) and Legacy ($\approx$3~Ms; PI: Mathur) have now been approved by the CXC to provide deep \xray\ coverage for the entire PHANGS sample. With such high-resolution and panchromatic multiwavelength coverage (including both spectral and photometric constraints), many \xray\ sources and background AGN can be classified directly \citep[see, e.g.,][]{Ran2011,Ran2012,Chan2020,Leh2020,Hun2021,Hun2023} and their local properties can be constrained more cleanly.  In particular, such data will allow for higher age resolution constraints on XRB populations in the first $\approx$1~Gyr following a star-formation event, an era where several interesting XRB population transitions are expected to take place.

\vspace{0.1in} {\large {\it Acknowledgements:}}
We thank the anonymous referee for their helpful comments, which have improved the quality of this paper.  We gratefully acknowledge financial support from NASA Astrophysics Data Analysis Program 80NSSC20K0444 (B.D.L., A.A., R.T.E., K.D., E.B.M.) and Chandra X-ray Center grant GO2-23064X (B.D.L., A.A.).  E.B.M. acknowledges support from Penn State ACIS Instrument Team Contract SV4-74018 (issued by the Chandra X-ray Center, which is operated by the Smithsonian Astrophysical Observatory for and on behalf of NASA under contract NAS8-03060). This work has utilized Chandra ACIS Guaranteed Time Observations (GTO) selected by the ACIS Instrument Principal Investigator, Gordon P. Garmire, currently of the Huntingdon Institute for X-ray Astronomy, LLC, which is under contract to the Smithsonian Astrophysical Observatory via Contract SV2-82024.  This work was supported by NASA under award number 80GSFC21M0002 (A.B.Z.).  K.K. is supported by a fellowship program at the Institute of Space Sciences (ICE-CSIC) funded by the program Unidad de Excelencia Mar\'{i}a de Maeztu CEX2020-001058-M.

This publication uses the data from the AstroSat mission of the
Indian Space Research Organisation (ISRO), archived at the Indian Space Science Data Centre (ISSDC)

We acknowledge the use of public data from the Swift data archive.

This work is based in part on observations made with the
Galaxy Evolution Explorer (GALEX). GALEX is a NASA
Small Explorer, whose mission was developed in cooperation
with the Centre National d’Etudes Spatiales (CNES) of France
and the Korean Ministry of Science and Technology. GALEX
is operated for NASA by the California Institute of Technology
under NASA contract NAS5-98034.

This research is based in part on observations made with the NASA/ESA Hubble Space Telescope obtained from the Space Telescope Science Institute, which is operated by the Association of Universities for Research in Astronomy, Inc., under NASA contract NAS 5–26555.  Based in part on observations made with the NASA/ESA Hubble Space Telescope, as obtained from the Hubble Legacy Archive, which is a collaboration between the Space Telescope Science Institute (STScI/NASA), the Space Telescope European Coordinating Facility (ST-ECF/ESAC/ESA) and the Canadian Astronomy Data Centre (CADC/NRC/CSA).

This work is based in part on observations made with the NASA/ESA/CSA James Webb Space Telescope. The data were obtained from the Mikulski Archive for Space Telescopes at the Space Telescope Science Institute, which is operated by the Association of Universities for Research in Astronomy, Inc., under NASA contract NAS 5-03127 for JWST. 

Funding for SDSS-III has been provided by the Alfred P. Sloan Foundation, the Participating Institutions, the National Science Foundation, and the U.S. Department of Energy Office of Science. The SDSS-III web site is http://www.sdss3.org/. SDSS-III is managed by the Astrophysical Research Consortium for the Participating Institutions of the SDSS-III Collaboration including the University of Arizona, the Brazilian Participation Group, Brookhaven National Laboratory, University of Cambridge, Carnegie Mellon University, University of Florida, the French Participation Group, the German Participation Group, Harvard University, the Instituto de Astrofisica de Canarias, the Michigan State/Notre Dame/JINA Participation Group, Johns Hopkins University, Lawrence Berkeley National Laboratory, Max Planck Institute for Astrophysics, Max Planck Institute for Extraterrestrial Physics, New Mexico State University, New York University, Ohio State University, Pennsylvania State University, University of Portsmouth, Princeton University, the Spanish Participation Group, University of Tokyo, University of Utah, Vanderbilt University, University of Virginia, University of Washington, and Yale University. 

This project used public archival data from the Dark Energy Survey (DES) as distributed by the Astro Data Archive at NSF NOIRLab. Funding for the DES Projects has been provided by the US Department of Energy, the US National Science Foundation, the Ministry of Science and Education of Spain, the Science and Technology Facilities Council of the United Kingdom, the Higher Education Funding Council for England, the National Center for Supercomputing Applications at the University of Illinois at Urbana-Champaign, the Kavli Institute for Cosmological Physics at the University of Chicago, Center for Cosmology and Astro-Particle Physics at the Ohio State University, the Mitchell Institute for Fundamental Physics and Astronomy at Texas A\&M University, Financiadora de Estudos e Projetos, Fundação Carlos Chagas Filho de Amparo à Pesquisa do Estado do Rio de Janeiro, Conselho Nacional de Desenvolvimento Científico e Tecnológico and the Ministério da Ciência, Tecnologia e Inovação, the Deutsche Forschungsgemeinschaft and the Collaborating Institutions in the Dark Energy Survey.

The Collaborating Institutions are Argonne National Laboratory, the University of California at Santa Cruz, the University of Cambridge, Centro de Investigaciones Enérgeticas, 22 Medioambientales y Tecnológicas- Madrid, the University of Chicago, University College London, the DES-Brazil Consortium, the University of Edinburgh, the Eidgenössische Technische Hochschule (ETH) Zürich, Fermi National Accelerator Laboratory, the University of Illinois at Urbana-Champaign, the Institut de Ciències de l’Espai (IEEC/CSIC), the Institut de Física d’Altes Energies, Lawrence Berkeley National Laboratory, the Ludwig-Maximilians Universität München and the associated Excellence Cluster Universe, the University of Michigan, the  NSF NOIRLab, the University of Nottingham, the Ohio State University, the OzDES Membership Consortium, the University of Pennsylvania, the University of Portsmouth, SLAC National Accelerator Laboratory, Stanford University, the University of Sussex, and Texas A\&M University.

Based on observations at  NSF Cerro Tololo Inter-American Observatory, a program of NOIRLab (NOIRLab Prop. 2012B-0001; PI J. Frieman), which is managed by the Association of Universities for Research in Astronomy (AURA) under a cooperative agreement with the U.S. National Science Foundation.

We utilized data from the Pan-STARRS1 Surveys (PS1).  The PS1 and the PS1 public science archive have been made possible through contributions by the Institute for Astronomy, the University of Hawaii, the Pan-STARRS Project Office, the Max-Planck Society and its participating institutes, the Max Planck Institute for Astronomy, Heidelberg and the Max Planck Institute for Extraterrestrial Physics, Garching, The Johns Hopkins University, Durham University, the University of Edinburgh, the Queen's University Belfast, the Harvard-Smithsonian Center for Astrophysics, the Las Cumbres Observatory Global Telescope Network Incorporated, the National Central University of Taiwan, the Space Telescope Science Institute, the National Aeronautics and Space Administration under Grant No. NNX08AR22G issued through the Planetary Science Division of the NASA Science Mission Directorate, the National Science Foundation Grant No. AST-1238877, the University of Maryland, Eotvos Lorand University (ELTE), the Los Alamos National Laboratory, and the Gordon and Betty Moore Foundation.

This publication makes use of data products from the Two
Micron All Sky Survey, which is a joint project of the
University of Massachusetts and the Infrared Processing and
Analysis Center/California Institute of Technology, funded by
the National Aeronautics and Space Administration and the
National Science Foundation.

This research has made use of the NASA/IPAC Extragalactic Database (NED),
which is operated by the Jet Propulsion Laboratory, California Institute of Technology, under contract with NASA.  This research has made use of the NASA/IPAC Infrared Science Archive, which is funded by the National Aeronautics and Space Administration and operated by the California Institute of Technology. 

This work is based in part on observations made with the
Spitzer Space Telescope, which is operated by the Jet
Propulsion Laboratory, California Institute of Technology
under a contract with NASA.

This publication makes use of data products from the Wide-field Infrared Survey Explorer (WISE), which is a joint project
of the University of California, Los Angeles, and the Jet
Propulsion Laboratory/California Institute of Technology,
funded by the National Aeronautics and Space Administration.

Herschel is an ESA space observatory with science instruments provided by European-led Principal Investigator consortia and with important participation from NASA. 

\facilities{IRSA, MAST (HLSP, HLA), NED, Astro Data Archive, {\it Chandra}, {\it AstroSAT}, {\it GALEX}, {\it Swift}, {\it HST} (ACS, WFPC2, WFC3), Sloan, PanSTARRS, Blanco (DECam), WIYN:0.9m, CTIO:2MASS, {\it JWST}, {\it WISE}, {\it Herschel}}

\software{{\ttfamily Cloudy} \citep[v22.01][]{Fer2017}, {\ttfamily CIAO} \citep[v4.15][]{Fru2006}, {\ttfamily Acis Extract} \citep[v.2023aug14][]{Bro2010,Bro2012}, {\ttfamily MARX} \citep[v5.5.1][]{Dav2012}, {\ttfamily DS9} \citep[][]{Joy2003}, {\ttfamily Lightning} \citep[v2024.1][]{Euf2017,Doo2023,Mon2024}, {\ttfamily BPASS} \citep[v2.1][]{Eld2017},  {\ttfamily P\'EGASE} \citep{Fio1997}, {\ttfamily Sherpa} \citep[v4.13.0][]{Bur2021}, {\ttfamily XSPEC} \citep{Arn1996}}

\bibliography{mybib.bib}{}

\begin{thebibliography}{}
\expandafter\ifx\csname natexlab\endcsname\relax\def\natexlab#1{#1}\fi
\providecommand{\url}[1]{\href{#1}{#1}}
\providecommand{\dodoi}[1]{doi:~\href{http://doi.org/#1}{\nolinkurl{#1}}}
\providecommand{\doeprint}[1]{\href{http://ascl.net/#1}{\nolinkurl{http://ascl.net/#1}}}
\providecommand{\doarXiv}[1]{\href{https://arxiv.org/abs/#1}{\nolinkurl{https://arxiv.org/abs/#1}}}

\bibitem[{{Aird} {et~al.}(2017){Aird}, {Coil}, \& {Georgakakis}}]{Air2017}
{Aird}, J., {Coil}, A.~L., \& {Georgakakis}, A. 2017, \mnras, 465, 3390, \dodoi{10.1093/mnras/stw2932}

\bibitem[{{Aird} {et~al.}(2019){Aird}, {Coil}, \& {Georgakakis}}]{Air2019}
---. 2019, \mnras, 484, 4360, \dodoi{10.1093/mnras/stz125}

\bibitem[{{Allende Prieto} {et~al.}(2001){Allende Prieto}, {Lambert}, \& {Asplund}}]{All2001}
{Allende Prieto}, C., {Lambert}, D.~L., \& {Asplund}, M. 2001, \apjl, 556, L63, \dodoi{10.1086/322874}

\bibitem[{{Antoniou} \& {Zezas}(2016)}]{Ant2016}
{Antoniou}, V., \& {Zezas}, A. 2016, \mnras, 459, 528, \dodoi{10.1093/mnras/stw167}

\bibitem[{{Antoniou} {et~al.}(2019){Antoniou}, {Zezas}, {Drake}, {Badenes}, {Haberl}, {Wright}, {Hong}, {Di Stefano}, {Gaetz}, {Long}, {Plucinsky}, {Sasaki}, {Williams}, {Winkler}, \& {SMC XVP Collaboration}}]{Ant2019}
{Antoniou}, V., {Zezas}, A., {Drake}, J.~J., {et~al.} 2019, \apj, 887, 20, \dodoi{10.3847/1538-4357/ab4a7a}

\bibitem[{{Arnaud}(1996)}]{Arn1996}
{Arnaud}, K.~A. 1996, in Astronomical Society of the Pacific Conference Series, Vol. 101, Astronomical Data Analysis Software and Systems V, ed. G.~H. {Jacoby} \& J.~{Barnes}, 17

\bibitem[{{Asplund} {et~al.}(2009){Asplund}, {Grevesse}, {Sauval}, \& {Scott}}]{Asp2009}
{Asplund}, M., {Grevesse}, N., {Sauval}, A.~J., \& {Scott}, P. 2009, \araa, 47, 481, \dodoi{10.1146/annurev.astro.46.060407.145222}

\bibitem[{{Basu-Zych} {et~al.}(2016){Basu-Zych}, {Lehmer}, {Fragos}, {Hornschemeier}, {Yukita}, {Zezas}, \& {Ptak}}]{Bas2016}
{Basu-Zych}, A.~R., {Lehmer}, B., {Fragos}, T., {et~al.} 2016, \apj, 818, 140, \dodoi{10.3847/0004-637X/818/2/140}

\bibitem[{{Basu-Zych} {et~al.}(2013{\natexlab{a}}){Basu-Zych}, {Lehmer}, {Hornschemeier}, {Gon{\c{c}}alves}, {Fragos}, {Heckman}, {Overzier}, {Ptak}, \& {Schiminovich}}]{Bas2013a}
{Basu-Zych}, A.~R., {Lehmer}, B.~D., {Hornschemeier}, A.~E., {et~al.} 2013{\natexlab{a}}, \apj, 774, 152, \dodoi{10.1088/0004-637X/774/2/152}

\bibitem[{{Basu-Zych} {et~al.}(2013{\natexlab{b}}){Basu-Zych}, {Lehmer}, {Hornschemeier}, {Bouwens}, {Fragos}, {Oesch}, {Belczynski}, {Brandt}, {Kalogera}, {Luo}, {Miller}, {Mullaney}, {Tzanavaris}, {Xue}, \& {Zezas}}]{Bas2013b}
---. 2013{\natexlab{b}}, \apj, 762, 45, \dodoi{10.1088/0004-637X/762/1/45}

\bibitem[{{Bavera} {et~al.}(2022{\natexlab{a}}){Bavera}, {Franciolini}, {Cusin}, {Riotto}, {Zevin}, \& {Fragos}}]{Bav2022b}
{Bavera}, S.~S., {Franciolini}, G., {Cusin}, G., {et~al.} 2022{\natexlab{a}}, \aap, 660, A26, \dodoi{10.1051/0004-6361/202142208}

\bibitem[{{Bavera} {et~al.}(2022{\natexlab{b}}){Bavera}, {Fragos}, {Zapartas}, {Ramirez-Ruiz}, {Marchant}, {Kelley}, {Zevin}, {Andrews}, {Coughlin}, {Dotter}, {Kovlakas}, {Misra}, {Serra-Perez}, {Qin}, {Rocha}, {Rom{\'a}n-Garza}, {Tran}, \& {Xing}}]{Bav2022a}
{Bavera}, S.~S., {Fragos}, T., {Zapartas}, E., {et~al.} 2022{\natexlab{b}}, \aap, 657, L8, \dodoi{10.1051/0004-6361/202141979}

\bibitem[{{Bavera} {et~al.}(2023){Bavera}, {Fragos}, {Zapartas}, {Andrews}, {Kalogera}, {Berry}, {Kruckow}, {Dotter}, {Kovlakas}, {Misra}, {Rocha}, {Srivastava}, {Sun}, \& {Xing}}]{Bav2023}
---. 2023, Nature Astronomy, 7, 1090, \dodoi{10.1038/s41550-023-02018-5}

\bibitem[{{Belczynski} {et~al.}(2010){Belczynski}, {Dominik}, {Bulik}, {O'Shaughnessy}, {Fryer}, \& {Holz}}]{Bel2010}
{Belczynski}, K., {Dominik}, M., {Bulik}, T., {et~al.} 2010, \apjl, 715, L138, \dodoi{10.1088/2041-8205/715/2/L138}

\bibitem[{{Belczynski} {et~al.}(2008){Belczynski}, {Kalogera}, {Rasio}, {Taam}, {Zezas}, {Bulik}, {Maccarone}, \& {Ivanova}}]{Bel2008}
{Belczynski}, K., {Kalogera}, V., {Rasio}, F.~A., {et~al.} 2008, \apjs, 174, 223, \dodoi{10.1086/521026}

\bibitem[{{Belloni}(2010)}]{Bell2010}
{Belloni}, T.~M. 2010, in Lecture Notes in Physics, Berlin Springer Verlag, ed. T.~{Belloni}, Vol. 794, 53, \dodoi{10.1007/978-3-540-76937-8_3}

\bibitem[{{Berg} {et~al.}(2012){Berg}, {Skillman}, {Marble}, {van Zee}, {Engelbracht}, {Lee}, {Kennicutt}, {Calzetti}, {Dale}, \& {Johnson}}]{Berg2012}
{Berg}, D.~A., {Skillman}, E.~D., {Marble}, A.~R., {et~al.} 2012, \apj, 754, 98, \dodoi{10.1088/0004-637X/754/2/98}

\bibitem[{{Binder} {et~al.}(2017){Binder}, {Gross}, {Williams}, {Eracleous}, {Gaetz}, {Plucinsky}, \& {Skillman}}]{Bin2017}
{Binder}, B., {Gross}, J., {Williams}, B.~F., {et~al.} 2017, \apj, 834, 128, \dodoi{10.3847/1538-4357/834/2/128}

\bibitem[{{Binder} {et~al.}(2023){Binder}, {Anderson}, {Garofali}, {Lazzarini}, \& {Williams}}]{Bin2023}
{Binder}, B.~A., {Anderson}, A.~K., {Garofali}, K., {Lazzarini}, M., \& {Williams}, B.~F. 2023, \mnras, 522, 5669, \dodoi{10.1093/mnras/stad1368}

\bibitem[{{Binder} {et~al.}(2024){Binder}, {Williams}, {Payne}, {Eracleous}, {Belles}, \& {Williams}}]{Bin2024}
{Binder}, B.~A., {Williams}, R., {Payne}, J., {et~al.} 2024, \apj, 969, 97, \dodoi{10.3847/1538-4357/ad46d9}

\bibitem[{{Bluem} {et~al.}(2019){Bluem}, {Kaaret}, {Prestwich}, \& {Brorby}}]{Ble2019}
{Bluem}, J., {Kaaret}, P., {Prestwich}, A., \& {Brorby}, M. 2019, \mnras, 487, 4093, \dodoi{10.1093/mnras/stz1574}

\bibitem[{{Boroson} {et~al.}(2011){Boroson}, {Kim}, \& {Fabbiano}}]{Bor2011}
{Boroson}, B., {Kim}, D.-W., \& {Fabbiano}, G. 2011, \apj, 729, 12, \dodoi{10.1088/0004-637X/729/1/12}

\bibitem[{{Breivik} {et~al.}(2020){Breivik}, {Coughlin}, {Zevin}, {Rodriguez}, {Kremer}, {Ye}, {Andrews}, {Kurkowski}, {Digman}, {Larson}, \& {Rasio}}]{Bre2020}
{Breivik}, K., {Coughlin}, S., {Zevin}, M., {et~al.} 2020, \apj, 898, 71, \dodoi{10.3847/1538-4357/ab9d85}

\bibitem[{{Bresolin} {et~al.}(2009){Bresolin}, {Ryan-Weber}, {Kennicutt}, \& {Goddard}}]{Bre2009}
{Bresolin}, F., {Ryan-Weber}, E., {Kennicutt}, R.~C., \& {Goddard}, Q. 2009, \apj, 695, 580, \dodoi{10.1088/0004-637X/695/1/580}

\bibitem[{{Broos} {et~al.}(2012){Broos}, {Townsley}, {Getman}, \& {Bauer}}]{Bro2012}
{Broos}, P., {Townsley}, L., {Getman}, K., \& {Bauer}, F. 2012, {AE: ACIS Extract}, Astrophysics Source Code Library, record ascl:1203.001.
\newblock \doeprint{1203.001}

\bibitem[{{Broos} {et~al.}(2010){Broos}, {Townsley}, {Feigelson}, {Getman}, {Bauer}, \& {Garmire}}]{Bro2010}
{Broos}, P.~S., {Townsley}, L.~K., {Feigelson}, E.~D., {et~al.} 2010, \apj, 714, 1582, \dodoi{10.1088/0004-637X/714/2/1582}

\bibitem[{{Brorby} \& {Kaaret}(2017)}]{Bro2017}
{Brorby}, M., \& {Kaaret}, P. 2017, \mnras, 470, 606, \dodoi{10.1093/mnras/stx1286}

\bibitem[{{Brorby} {et~al.}(2014){Brorby}, {Kaaret}, \& {Prestwich}}]{Bro2014}
{Brorby}, M., {Kaaret}, P., \& {Prestwich}, A. 2014, \mnras, 441, 2346, \dodoi{10.1093/mnras/stu736}

\bibitem[{{Brorby} {et~al.}(2016){Brorby}, {Kaaret}, {Prestwich}, \& {Mirabel}}]{Bro2016}
{Brorby}, M., {Kaaret}, P., {Prestwich}, A., \& {Mirabel}, I.~F. 2016, \mnras, 457, 4081, \dodoi{10.1093/mnras/stw284}

\bibitem[{Burke {et~al.}(2021)Burke, Laurino, wmclaugh, dtnguyen2, Marie-Terrell, Günther, Siemiginowska, Budynkiewicz, Aldcroft, Deil, Sipőcz, Buchner, Laginja, Leinweber, nplee, \& Todd}]{Bur2021}
Burke, D., Laurino, O., wmclaugh, {et~al.} 2021, sherpa/sherpa: Sherpa 4.13.0, 4.13.0,  Zenodo, \dodoi{10.5281/zenodo.4428938}

\bibitem[{{Calzetti} {et~al.}(2000){Calzetti}, {Armus}, {Bohlin}, {Kinney}, {Koornneef}, \& {Storchi-Bergmann}}]{Cal2000}
{Calzetti}, D., {Armus}, L., {Bohlin}, R.~C., {et~al.} 2000, \apj, 533, 682, \dodoi{10.1086/308692}

\bibitem[{{Cash}(1979)}]{Cas1979}
{Cash}, W. 1979, \apj, 228, 939, \dodoi{10.1086/156922}

\bibitem[{{Chabrier}(2003)}]{Cha2003}
{Chabrier}, G. 2003, \pasp, 115, 763, \dodoi{10.1086/376392}

\bibitem[{{Chandar} {et~al.}(2020){Chandar}, {Johns}, {Mok}, {Prestwich}, {Gallo}, \& {Hunt}}]{Chan2020}
{Chandar}, R., {Johns}, P., {Mok}, A., {et~al.} 2020, \apj, 890, 150, \dodoi{10.3847/1538-4357/ab6b27}

\bibitem[{{Clark} {et~al.}(1978){Clark}, {Doxsey}, {Li}, {Jernigan}, \& {van Paradijs}}]{Cla1978}
{Clark}, G., {Doxsey}, R., {Li}, F., {Jernigan}, J.~G., \& {van Paradijs}, J. 1978, \apjl, 221, L37, \dodoi{10.1086/182660}

\bibitem[{{Clark} {et~al.}(1975){Clark}, {Markert}, \& {Li}}]{Cla1975}
{Clark}, G.~W., {Markert}, T.~H., \& {Li}, F.~K. 1975, \apjl, 199, L93, \dodoi{10.1086/181856}

\bibitem[{{Colbert} {et~al.}(2004){Colbert}, {Heckman}, {Ptak}, {Strickland}, \& {Weaver}}]{Col2004}
{Colbert}, E. J.~M., {Heckman}, T.~M., {Ptak}, A.~F., {Strickland}, D.~K., \& {Weaver}, K.~A. 2004, \apj, 602, 231, \dodoi{10.1086/380899}

\bibitem[{{Croxall} {et~al.}(2009){Croxall}, {van Zee}, {Lee}, {Skillman}, {Lee}, {C{\^o}t{\'e}}, {Kennicutt}, \& {Miller}}]{Cro2009}
{Croxall}, K.~V., {van Zee}, L., {Lee}, H., {et~al.} 2009, \apj, 705, 723, \dodoi{10.1088/0004-637X/705/1/723}

\bibitem[{{Curti} {et~al.}(2017){Curti}, {Cresci}, {Mannucci}, {Marconi}, {Maiolino}, \& {Esposito}}]{Cur2017}
{Curti}, M., {Cresci}, G., {Mannucci}, F., {et~al.} 2017, \mnras, 465, 1384, \dodoi{10.1093/mnras/stw2766}

\bibitem[{{Davies} {et~al.}(2017){Davies}, {Kudritzki}, {Lardo}, {Bergemann}, {Beasor}, {Plez}, {Evans}, {Bastian}, \& {Patrick}}]{Dav2017}
{Davies}, B., {Kudritzki}, R.-P., {Lardo}, C., {et~al.} 2017, \apj, 847, 112, \dodoi{10.3847/1538-4357/aa89ed}

\bibitem[{{Davis} {et~al.}(2012){Davis}, {Bautz}, {Dewey}, {Heilmann}, {Houck}, {Huenemoerder}, {Marshall}, {Nowak}, {Schattenburg}, {Schulz}, \& {Smith}}]{Dav2012}
{Davis}, J.~E., {Bautz}, M.~W., {Dewey}, D., {et~al.} 2012, in Society of Photo-Optical Instrumentation Engineers (SPIE) Conference Series, Vol. 8443, Space Telescopes and Instrumentation 2012: Ultraviolet to Gamma Ray, ed. T.~{Takahashi}, S.~S. {Murray}, \& J.-W.~A. {den Herder}, 84431A, \dodoi{10.1117/12.926937}

\bibitem[{{Doore} {et~al.}(2023){Doore}, {Monson}, {Eufrasio}, {Lehmer}, {Garofali}, \& {Basu-Zych}}]{Doo2023}
{Doore}, K., {Monson}, E.~B., {Eufrasio}, R.~T., {et~al.} 2023, \apjs, 266, 39, \dodoi{10.3847/1538-4365/accc29}

\bibitem[{{Douna} {et~al.}(2015){Douna}, {Pellizza}, {Mirabel}, \& {Pedrosa}}]{Dou2015}
{Douna}, V.~M., {Pellizza}, L.~J., {Mirabel}, I.~F., \& {Pedrosa}, S.~E. 2015, \aap, 579, A44, \dodoi{10.1051/0004-6361/201525617}

\bibitem[{{Draine} \& {Li}(2007)}]{Dra2007}
{Draine}, B.~T., \& {Li}, A. 2007, \apj, 657, 810, \dodoi{10.1086/511055}

\bibitem[{{Dray}(2006)}]{Dra2006}
{Dray}, L.~M. 2006, \mnras, 370, 2079, \dodoi{10.1111/j.1365-2966.2006.10635.x}

\bibitem[{{Eldridge} {et~al.}(2017){Eldridge}, {Stanway}, {Xiao}, {McClelland}, {Taylor}, {Ng}, {Greis}, \& {Bray}}]{Eld2017}
{Eldridge}, J.~J., {Stanway}, E.~R., {Xiao}, L., {et~al.} 2017, \pasa, 34, e058, \dodoi{10.1017/pasa.2017.51}

\bibitem[{{Engelbracht} {et~al.}(2008){Engelbracht}, {Rieke}, {Gordon}, {Smith}, {Werner}, {Moustakas}, {Willmer}, \& {Vanzi}}]{Eng2008}
{Engelbracht}, C.~W., {Rieke}, G.~H., {Gordon}, K.~D., {et~al.} 2008, \apj, 678, 804, \dodoi{10.1086/529513}

\bibitem[{{Esteban} {et~al.}(2014){Esteban}, {Garc{\'\i}a-Rojas}, {Carigi}, {Peimbert}, {Bresolin}, {L{\'o}pez-S{\'a}nchez}, \& {Mesa-Delgado}}]{Est2014}
{Esteban}, C., {Garc{\'\i}a-Rojas}, J., {Carigi}, L., {et~al.} 2014, \mnras, 443, 624, \dodoi{10.1093/mnras/stu1177}

\bibitem[{{Eufrasio} {et~al.}(2017){Eufrasio}, {Lehmer}, {Zezas}, {Dwek}, {Arendt}, {Basu-Zych}, {Wiklind}, {Yukita}, {Fragos}, {Hornschemeier}, {Markwardt}, {Ptak}, \& {Tzanavaris}}]{Euf2017}
{Eufrasio}, R.~T., {Lehmer}, B.~D., {Zezas}, A., {et~al.} 2017, \apj, 851, 10, \dodoi{10.3847/1538-4357/aa9569}

\bibitem[{{Fabbiano}(1989)}]{Fab1989}
{Fabbiano}, G. 1989, \araa, 27, 87, \dodoi{10.1146/annurev.aa.27.090189.000511}

\bibitem[{{Fabbiano}(2006)}]{Fab2006}
---. 2006, \araa, 44, 323, \dodoi{10.1146/annurev.astro.44.051905.092519}

\bibitem[{{Fabbiano}(2019)}]{Fab2019}
---. 2019, {X-Rays from Galaxies}, ed. B.~{Wilkes} \& W.~{Tucker}, 7--1, \dodoi{10.1088/2514-3433/ab43dcch7}

\bibitem[{{Fabbiano} {et~al.}(1982){Fabbiano}, {Feigelson}, \& {Zamorani}}]{Fab1982}
{Fabbiano}, G., {Feigelson}, E., \& {Zamorani}, G. 1982, \apj, 256, 397, \dodoi{10.1086/159917}

\bibitem[{{Fabian} {et~al.}(1975){Fabian}, {Pringle}, \& {Rees}}]{Fab1975}
{Fabian}, A.~C., {Pringle}, J.~E., \& {Rees}, M.~J. 1975, \mnras, 172, 15, \dodoi{10.1093/mnras/172.1.15P}

\bibitem[{{Ferland}(1993)}]{Fer1993}
{Ferland}, G.~J. 1993, {Hazy, A Brief Introduction to Cloudy 84}

\bibitem[{{Ferland} {et~al.}(2013){Ferland}, {Porter}, {van Hoof}, {Williams}, {Abel}, {Lykins}, {Shaw}, {Henney}, \& {Stancil}}]{Fer2013}
{Ferland}, G.~J., {Porter}, R.~L., {van Hoof}, P.~A.~M., {et~al.} 2013, \rmxaa, 49, 137, \dodoi{10.48550/arXiv.1302.4485}

\bibitem[{{Ferland} {et~al.}(2017){Ferland}, {Chatzikos}, {Guzm{\'a}n}, {Lykins}, {van Hoof}, {Williams}, {Abel}, {Badnell}, {Keenan}, {Porter}, \& {Stancil}}]{Fer2017}
{Ferland}, G.~J., {Chatzikos}, M., {Guzm{\'a}n}, F., {et~al.} 2017, \rmxaa, 53, 385, \dodoi{10.48550/arXiv.1705.10877}

\bibitem[{{Fioc} \& {Rocca-Volmerange}(1997)}]{Fio1997}
{Fioc}, M., \& {Rocca-Volmerange}, B. 1997, \aap, 326, 950.
\newblock \doarXiv{astro-ph/9707017}

\bibitem[{{Fitzpatrick}(1999)}]{Fit1999}
{Fitzpatrick}, E.~L. 1999, \pasp, 111, 63, \dodoi{10.1086/316293}

\bibitem[{{Foreman-Mackey} {et~al.}(2013){Foreman-Mackey}, {Hogg}, {Lang}, \& {Goodman}}]{For2013}
{Foreman-Mackey}, D., {Hogg}, D.~W., {Lang}, D., \& {Goodman}, J. 2013, \pasp, 125, 306, \dodoi{10.1086/670067}

\bibitem[{{Fornasini} {et~al.}(2020){Fornasini}, {Civano}, \& {Suh}}]{For2020}
{Fornasini}, F.~M., {Civano}, F., \& {Suh}, H. 2020, \mnras, 495, 771, \dodoi{10.1093/mnras/staa1211}

\bibitem[{{Fornasini} {et~al.}(2019){Fornasini}, {Kriek}, {Sanders}, {Shivaei}, {Civano}, {Reddy}, {Shapley}, {Coil}, {Mobasher}, {Siana}, {Aird}, {Azadi}, {Freeman}, {Leung}, {Price}, {Fetherolf}, {Zick}, \& {Barro}}]{For2019}
{Fornasini}, F.~M., {Kriek}, M., {Sanders}, R.~L., {et~al.} 2019, \apj, 885, 65, \dodoi{10.3847/1538-4357/ab4653}

\bibitem[{{Fragos} {et~al.}(2013{\natexlab{a}}){Fragos}, {Lehmer}, {Naoz}, {Zezas}, \& {Basu-Zych}}]{Fra2013b}
{Fragos}, T., {Lehmer}, B.~D., {Naoz}, S., {Zezas}, A., \& {Basu-Zych}, A. 2013{\natexlab{a}}, \apjl, 776, L31, \dodoi{10.1088/2041-8205/776/2/L31}

\bibitem[{{Fragos} {et~al.}(2008){Fragos}, {Kalogera}, {Belczynski}, {Fabbiano}, {Kim}, {Brassington}, {Angelini}, {Davies}, {Gallagher}, {King}, {Pellegrini}, {Trinchieri}, {Zepf}, {Kundu}, \& {Zezas}}]{Fra2008}
{Fragos}, T., {Kalogera}, V., {Belczynski}, K., {et~al.} 2008, \apj, 683, 346, \dodoi{10.1086/588456}

\bibitem[{{Fragos} {et~al.}(2013{\natexlab{b}}){Fragos}, {Lehmer}, {Tremmel}, {Tzanavaris}, {Basu-Zych}, {Belczynski}, {Hornschemeier}, {Jenkins}, {Kalogera}, {Ptak}, \& {Zezas}}]{Fra2013a}
{Fragos}, T., {Lehmer}, B., {Tremmel}, M., {et~al.} 2013{\natexlab{b}}, \apj, 764, 41, \dodoi{10.1088/0004-637X/764/1/41}

\bibitem[{{Fragos} {et~al.}(2023){Fragos}, {Andrews}, {Bavera}, {Berry}, {Coughlin}, {Dotter}, {Giri}, {Kalogera}, {Katsaggelos}, {Kovlakas}, {Lalvani}, {Misra}, {Srivastava}, {Qin}, {Rocha}, {Rom{\'a}n-Garza}, {Serra}, {Stahle}, {Sun}, {Teng}, {Trajcevski}, {Tran}, {Xing}, {Zapartas}, \& {Zevin}}]{Fra2023}
{Fragos}, T., {Andrews}, J.~J., {Bavera}, S.~S., {et~al.} 2023, \apjs, 264, 45, \dodoi{10.3847/1538-4365/ac90c1}

\bibitem[{{Freedman} {et~al.}(2001){Freedman}, {Madore}, {Gibson}, {Ferrarese}, {Kelson}, {Sakai}, {Mould}, {Kennicutt}, {Ford}, {Graham}, {Huchra}, {Hughes}, {Illingworth}, {Macri}, \& {Stetson}}]{Fre2001}
{Freedman}, W.~L., {Madore}, B.~F., {Gibson}, B.~K., {et~al.} 2001, \apj, 553, 47, \dodoi{10.1086/320638}

\bibitem[{{Fridriksson} {et~al.}(2008){Fridriksson}, {Homan}, {Lewin}, {Kong}, \& {Pooley}}]{Fri2008}
{Fridriksson}, J.~K., {Homan}, J., {Lewin}, W. H.~G., {Kong}, A. K.~H., \& {Pooley}, D. 2008, \apjs, 177, 465, \dodoi{10.1086/588817}

\bibitem[{{Fruscione} {et~al.}(2006){Fruscione}, {McDowell}, {Allen}, {Brickhouse}, {Burke}, {Davis}, {Durham}, {Elvis}, {Galle}, {Harris}, {Huenemoerder}, {Houck}, {Ishibashi}, {Karovska}, {Nicastro}, {Noble}, {Nowak}, {Primini}, {Siemiginowska}, {Smith}, \& {Wise}}]{Fru2006}
{Fruscione}, A., {McDowell}, J.~C., {Allen}, G.~E., {et~al.} 2006, in Society of Photo-Optical Instrumentation Engineers (SPIE) Conference Series, Vol. 6270, Society of Photo-Optical Instrumentation Engineers (SPIE) Conference Series, ed. D.~R. {Silva} \& R.~E. {Doxsey}, 62701V, \dodoi{10.1117/12.671760}

\bibitem[{{Gallazzi} {et~al.}(2005){Gallazzi}, {Charlot}, {Brinchmann}, {White}, \& {Tremonti}}]{Gal2005}
{Gallazzi}, A., {Charlot}, S., {Brinchmann}, J., {White}, S. D.~M., \& {Tremonti}, C.~A. 2005, \mnras, 362, 41, \dodoi{10.1111/j.1365-2966.2005.09321.x}

\bibitem[{{Ganss} {et~al.}(2022){Ganss}, {Pledger}, {Sansom}, {James}, {Puls}, \& {Habergham-Mawson}}]{Gan2022}
{Ganss}, R., {Pledger}, J.~L., {Sansom}, A.~E., {et~al.} 2022, \mnras, 512, 1541, \dodoi{10.1093/mnras/stac625}

\bibitem[{{Garofali} {et~al.}(2018){Garofali}, {Williams}, {Hillis}, {Gilbert}, {Dolphin}, {Eracleous}, \& {Binder}}]{Gar2018}
{Garofali}, K., {Williams}, B.~F., {Hillis}, T., {et~al.} 2018, \mnras, 479, 3526, \dodoi{10.1093/mnras/sty1612}

\bibitem[{{Garofali} {et~al.}(2024){Garofali}, {Basu-Zych}, {Johnson}, {Tzanavaris}, {Jaskot}, {Richardson}, {Lehmer}, {Yukita}, {Hodges-Kluck}, {Hornschemeier}, {Ptak}, \& {Vulic}}]{Gar2024}
{Garofali}, K., {Basu-Zych}, A.~R., {Johnson}, B.~D., {et~al.} 2024, \apj, 960, 13, \dodoi{10.3847/1538-4357/ad0a6a}

\bibitem[{{Geda} {et~al.}(2024){Geda}, {Goulding}, {Lehmer}, {Greene}, \& {Kulkarni}}]{Ged2024}
{Geda}, R., {Goulding}, A.~D., {Lehmer}, B.~D., {Greene}, J.~E., \& {Kulkarni}, A. 2024, arXiv e-prints, arXiv:2401.14477, \dodoi{10.48550/arXiv.2401.14477}

\bibitem[{{Ghosh} \& {White}(2001)}]{Gho2001}
{Ghosh}, P., \& {White}, N.~E. 2001, \apjl, 559, L97, \dodoi{10.1086/323641}

\bibitem[{{Gilbertson} {et~al.}(2022){Gilbertson}, {Lehmer}, {Doore}, {Eufrasio}, {Basu-Zych}, {Brandt}, {Fragos}, {Garofali}, {Kovlakas}, {Luo}, {Tozzi}, {Vito}, {Williams}, \& {Xue}}]{Gil2022}
{Gilbertson}, W., {Lehmer}, B.~D., {Doore}, K., {et~al.} 2022, \apj, 926, 28, \dodoi{10.3847/1538-4357/ac4049}

\bibitem[{{Gilfanov}(2004)}]{Gil2004b}
{Gilfanov}, M. 2004, \mnras, 349, 146, \dodoi{10.1111/j.1365-2966.2004.07473.x}

\bibitem[{{Gilfanov} {et~al.}(2022){Gilfanov}, {Fabbiano}, {Lehmer}, \& {Zezas}}]{Gil2023}
{Gilfanov}, M., {Fabbiano}, G., {Lehmer}, B., \& {Zezas}, A. 2022, in Handbook of X-ray and Gamma-ray Astrophysics, 105, \dodoi{10.1007/978-981-16-4544-0_108-1}

\bibitem[{{Gilfanov} {et~al.}(2004){Gilfanov}, {Grimm}, \& {Sunyaev}}]{Gil2004a}
{Gilfanov}, M., {Grimm}, H.~J., \& {Sunyaev}, R. 2004, \mnras, 351, 1365, \dodoi{10.1111/j.1365-2966.2004.07874.x}

\bibitem[{{G{\'o}mez-Gonz{\'a}lez} {et~al.}(2021){G{\'o}mez-Gonz{\'a}lez}, {Mayya}, {Toal{\'a}}, {Arthur}, {Zaragoza-Cardiel}, \& {Guerrero}}]{Gom2021}
{G{\'o}mez-Gonz{\'a}lez}, V.~M.~A., {Mayya}, Y.~D., {Toal{\'a}}, J.~A., {et~al.} 2021, \mnras, 500, 2076, \dodoi{10.1093/mnras/staa3304}

\bibitem[{{Goodman} \& {Weare}(2010)}]{Goo2010}
{Goodman}, J., \& {Weare}, J. 2010, Communications in Applied Mathematics and Computational Science, 5, 65, \dodoi{10.2140/camcos.2010.5.65}

\bibitem[{{Grimm} {et~al.}(2003){Grimm}, {Gilfanov}, \& {Sunyaev}}]{Gri2003}
{Grimm}, H.~J., {Gilfanov}, M., \& {Sunyaev}, R. 2003, \mnras, 339, 793, \dodoi{10.1046/j.1365-8711.2003.06224.x}

\bibitem[{{Groves} {et~al.}(2023){Groves}, {Kreckel}, {Santoro}, {Belfiore}, {Zavodnik}, {Congiu}, {Egorov}, {Emsellem}, {Grasha}, {Leroy}, {Scheuermann}, {Schinnerer}, {Watkins}, {Barnes}, {Bigiel}, {Dale}, {Glover}, {Pessa}, {Sanchez-Blazquez}, \& {Williams}}]{Gro2023}
{Groves}, B., {Kreckel}, K., {Santoro}, F., {et~al.} 2023, \mnras, 520, 4902, \dodoi{10.1093/mnras/stad114}

\bibitem[{{Harris} {et~al.}(2013){Harris}, {Harris}, \& {Alessi}}]{Har2013}
{Harris}, W.~E., {Harris}, G. L.~H., \& {Alessi}, M. 2013, \apj, 772, 82, \dodoi{10.1088/0004-637X/772/2/82}

\bibitem[{{Hassani} {et~al.}(2024){Hassani}, {Rosolowsky}, {Koch}, {Postma}, {Nofech}, {Corbould}, {Thilker}, {Leroy}, {Schinnerer}, {Belfiore}, {Bigiel}, {Boquien}, {Chevance}, {Dale}, {Egorov}, {Emsellem}, {Glover}, {Grasha}, {Groves}, {Henny}, {Kim}, {Klessen}, {Kreckel}, {Kruijssen}, {Lee}, {Lopez}, {Neumann}, {Pan}, {Sandstrom}, {Sarbadhicary}, {Sun}, \& {Williams}}]{Has2024}
{Hassani}, H., {Rosolowsky}, E., {Koch}, E.~W., {et~al.} 2024, \apjs, 271, 2, \dodoi{10.3847/1538-4365/ad152c}

\bibitem[{Hastings(1970)}]{Has1970}
Hastings, W.~K. 1970, Biometrika, 57, 97, \dodoi{10.1093/biomet/57.1.97}

\bibitem[{{Hills}(1976)}]{Hil1976}
{Hills}, J.~G. 1976, \mnras, 175, 1P, \dodoi{10.1093/mnras/175.1.1P}

\bibitem[{{Hu} {et~al.}(2018){Hu}, {Wang}, {Lin}, {Kong}, {Cheng}, {Fan}, {Fang}, {Lin}, {Mao}, {Wang}, {Zhou}, {Zhou}, {Zhu}, \& {Zou}}]{Hu2018}
{Hu}, N., {Wang}, E., {Lin}, Z., {et~al.} 2018, \apj, 854, 68, \dodoi{10.3847/1538-4357/aaa6ca}

\bibitem[{{Humphrey} \& {Buote}(2008)}]{Hum2008}
{Humphrey}, P.~J., \& {Buote}, D.~A. 2008, \apj, 689, 983, \dodoi{10.1086/592590}

\bibitem[{{Hunt} {et~al.}(2023){Hunt}, {Chandar}, {Gallo}, {Floyd}, {Maccarone}, \& {Thilker}}]{Hun2023}
{Hunt}, Q., {Chandar}, R., {Gallo}, E., {et~al.} 2023, \apj, 953, 126, \dodoi{10.3847/1538-4357/ace162}

\bibitem[{{Hunt} {et~al.}(2021){Hunt}, {Gallo}, {Chandar}, {Johns Mulia}, {Mok}, {Prestwich}, \& {Liu}}]{Hun2021}
{Hunt}, Q., {Gallo}, E., {Chandar}, R., {et~al.} 2021, \apj, 912, 31, \dodoi{10.3847/1538-4357/abe531}

\bibitem[{{Iorio} {et~al.}(2023){Iorio}, {Mapelli}, {Costa}, {Spera}, {Escobar}, {Sgalletta}, {Trani}, {Korb}, {Santoliquido}, {Dall'Amico}, {Gaspari}, \& {Bressan}}]{Ior2023}
{Iorio}, G., {Mapelli}, M., {Costa}, G., {et~al.} 2023, \mnras, 524, 426, \dodoi{10.1093/mnras/stad1630}

\bibitem[{{Irwin}(2005)}]{Irw2005}
{Irwin}, J.~A. 2005, \apj, 631, 511, \dodoi{10.1086/432611}

\bibitem[{{Iwasawa} {et~al.}(2011){Iwasawa}, {Sanders}, {Teng}, {U}, {Armus}, {Evans}, {Howell}, {Komossa}, {Mazzarella}, {Petric}, {Surace}, {Vavilkin}, {Veilleux}, \& {Trentham}}]{Iwa2011}
{Iwasawa}, K., {Sanders}, D.~B., {Teng}, S.~H., {et~al.} 2011, \aap, 529, A106, \dodoi{10.1051/0004-6361/201015264}

\bibitem[{{Izotov} \& {Thuan}(2007)}]{Izo2007}
{Izotov}, Y.~I., \& {Thuan}, T.~X. 2007, \apj, 665, 1115, \dodoi{10.1086/519922}

\bibitem[{{Jarrett} {et~al.}(2003){Jarrett}, {Chester}, {Cutri}, {Schneider}, \& {Huchra}}]{Jar2003}
{Jarrett}, T.~H., {Chester}, T., {Cutri}, R., {Schneider}, S.~E., \& {Huchra}, J.~P. 2003, \aj, 125, 525, \dodoi{10.1086/345794}

\bibitem[{{Joye} \& {Mandel}(2003)}]{Joy2003}
{Joye}, W.~A., \& {Mandel}, E. 2003, in Astronomical Society of the Pacific Conference Series, Vol. 295, Astronomical Data Analysis Software and Systems XII, ed. H.~E. {Payne}, R.~I. {Jedrzejewski}, \& R.~N. {Hook}, 489

\bibitem[{{Juett}(2005)}]{Jue2005}
{Juett}, A.~M. 2005, \apjl, 621, L25, \dodoi{10.1086/428905}

\bibitem[{{Kaastra}(2017)}]{Kaa2017}
{Kaastra}, J.~S. 2017, \aap, 605, A51, \dodoi{10.1051/0004-6361/201629319}

\bibitem[{{Kalogera} \& {Webbink}(1998)}]{Kal1998}
{Kalogera}, V., \& {Webbink}, R.~F. 1998, \apj, 493, 351, \dodoi{10.1086/305085}

\bibitem[{{Kennicutt} {et~al.}(2003){Kennicutt}, {Armus}, {Bendo}, {Calzetti}, {Dale}, {Draine}, {Engelbracht}, {Gordon}, {Grauer}, {Helou}, {Hollenbach}, {Jarrett}, {Kewley}, {Leitherer}, {Li}, {Malhotra}, {Regan}, {Rieke}, {Rieke}, {Roussel}, {Smith}, {Thornley}, \& {Walter}}]{Ken2003}
{Kennicutt}, Robert~C., J., {Armus}, L., {Bendo}, G., {et~al.} 2003, \pasp, 115, 928, \dodoi{10.1086/376941}

\bibitem[{{Kewley} \& {Ellison}(2008)}]{Kew2008}
{Kewley}, L.~J., \& {Ellison}, S.~L. 2008, \apj, 681, 1183, \dodoi{10.1086/587500}

\bibitem[{{Kilgard} {et~al.}(2005){Kilgard}, {Cowan}, {Garcia}, {Kaaret}, {Krauss}, {McDowell}, {Prestwich}, {Primini}, {Stockdale}, {Trinchieri}, {Ward}, \& {Zezas}}]{Kil2005}
{Kilgard}, R.~E., {Cowan}, J.~J., {Garcia}, M.~R., {et~al.} 2005, \apjs, 159, 214, \dodoi{10.1086/430443}

\bibitem[{{Kim} \& {Fabbiano}(2004)}]{Kim2004}
{Kim}, D.-W., \& {Fabbiano}, G. 2004, \apj, 611, 846, \dodoi{10.1086/422210}

\bibitem[{{Kim} {et~al.}(2013){Kim}, {Fabbiano}, {Ivanova}, {Fragos}, {Jord{\'a}n}, {Sivakoff}, \& {Voss}}]{Kim2013}
{Kim}, D.~W., {Fabbiano}, G., {Ivanova}, N., {et~al.} 2013, \apj, 764, 98, \dodoi{10.1088/0004-637X/764/1/98}

\bibitem[{{Kim} {et~al.}(1992){Kim}, {Fabbiano}, \& {Trinchieri}}]{Kim1992}
{Kim}, D.~W., {Fabbiano}, G., \& {Trinchieri}, G. 1992, \apj, 393, 134, \dodoi{10.1086/171492}

\bibitem[{{Kim} {et~al.}(2007){Kim}, {Wilkes}, {Kim}, {Green}, {Barkhouse}, {Lee}, {Silverman}, \& {Tananbaum}}]{Kim2007}
{Kim}, M., {Wilkes}, B.~J., {Kim}, D.-W., {et~al.} 2007, \apj, 659, 29, \dodoi{10.1086/511630}

\bibitem[{{Kotko} \& {Belczynski}(2024)}]{Kot2024}
{Kotko}, I., \& {Belczynski}, K. 2024, \aap, 683, A85, \dodoi{10.1051/0004-6361/202346880}

\bibitem[{{Kouroumpatzakis} {et~al.}(2020){Kouroumpatzakis}, {Zezas}, {Sell}, {Kovlakas}, {Bonfini}, {Willner}, {Ashby}, {Maragkoudakis}, \& {Jarrett}}]{Kou2020}
{Kouroumpatzakis}, K., {Zezas}, A., {Sell}, P., {et~al.} 2020, \mnras, 494, 5967, \dodoi{10.1093/mnras/staa1063}

\bibitem[{{Kovlakas} {et~al.}(2022){Kovlakas}, {Fragos}, {Schaerer}, \& {Mesinger}}]{Kov2022}
{Kovlakas}, K., {Fragos}, T., {Schaerer}, D., \& {Mesinger}, A. 2022, \aap, 665, A28, \dodoi{10.1051/0004-6361/202244252}

\bibitem[{{Kovlakas} {et~al.}(2020){Kovlakas}, {Zezas}, {Andrews}, {Basu-Zych}, {Fragos}, {Hornschemeier}, {Lehmer}, \& {Ptak}}]{Kov2020}
{Kovlakas}, K., {Zezas}, A., {Andrews}, J.~J., {et~al.} 2020, \mnras, 498, 4790, \dodoi{10.1093/mnras/staa2481}

\bibitem[{{Kovlakas} {et~al.}(2021){Kovlakas}, {Zezas}, {Andrews}, {Basu-Zych}, {Fragos}, {Hornschemeier}, {Kouroumpatzakis}, {Lehmer}, \& {Ptak}}]{Kov2021}
---. 2021, \mnras, 506, 1896, \dodoi{10.1093/mnras/stab1799}

\bibitem[{{Kroupa}(2001)}]{Kro2001}
{Kroupa}, P. 2001, \mnras, 322, 231, \dodoi{10.1046/j.1365-8711.2001.04022.x}

\bibitem[{{Kyritsis} {et~al.}(2024){Kyritsis}, {Zezas}, {Haberl}, {Weber}, {Basu-Zych}, {Vulic}, {Maitra}, {H{\"a}mmerich}, {Wilms}, {Sasaki}, {Hornschemeier}, {Ptak}, {Merloni}, \& {Comparat}}]{Kyr2024}
{Kyritsis}, E., {Zezas}, A., {Haberl}, F., {et~al.} 2024, arXiv e-prints, arXiv:2402.12367, \dodoi{10.48550/arXiv.2402.12367}

\bibitem[{{Lazzarini} {et~al.}(2021){Lazzarini}, {Williams}, {Durbin}, {Dalcanton}, {Antoniou}, {Binder}, {Eracleous}, {Plucinsky}, {Sasaki}, \& {Vulic}}]{Laz2021}
{Lazzarini}, M., {Williams}, B.~F., {Durbin}, M., {et~al.} 2021, \apj, 906, 120, \dodoi{10.3847/1538-4357/abccca}

\bibitem[{{Lazzarini} {et~al.}(2023){Lazzarini}, {Hinton}, {Shariat}, {Williams}, {Garofali}, {Dalcanton}, {Durbin}, {Antoniou}, {Binder}, {Eracleous}, {Vulic}, {Yang}, {Wik}, {Gasca}, \& {Kuauhtzin}}]{Laz2023}
{Lazzarini}, M., {Hinton}, K., {Shariat}, C., {et~al.} 2023, \apj, 952, 114, \dodoi{10.3847/1538-4357/acdbc8}

\bibitem[{{Lecroq} {et~al.}(2024){Lecroq}, {Charlot}, {Bressan}, {Bruzual}, {Costa}, {Iorio}, {Spera}, {Mapelli}, {Chen}, {Chevallard}, \& {Dall'Amico}}]{Lec2024}
{Lecroq}, M., {Charlot}, S., {Bressan}, A., {et~al.} 2024, \mnras, 527, 9480, \dodoi{10.1093/mnras/stad3838}

\bibitem[{{Lee} {et~al.}(2009){Lee}, {Gil de Paz}, {Tremonti}, {Kennicutt}, {Salim}, {Bothwell}, {Calzetti}, {Dalcanton}, {Dale}, {Engelbracht}, {Funes}, {Johnson}, {Sakai}, {Skillman}, {van Zee}, {Walter}, \& {Weisz}}]{Lee2009}
{Lee}, J.~C., {Gil de Paz}, A., {Tremonti}, C., {et~al.} 2009, \apj, 706, 599, \dodoi{10.1088/0004-637X/706/1/599}

\bibitem[{{Lee} {et~al.}(2023){Lee}, {Sandstrom}, {Leroy}, {Thilker}, {Schinnerer}, {Rosolowsky}, {Larson}, {Egorov}, {Williams}, {Schmidt}, {Emsellem}, {Anand}, {Barnes}, {Belfiore}, {Be{\v{s}}li{\'c}}, {Bigiel}, {Blanc}, {Bolatto}, {Boquien}, {den Brok}, {Cao}, {Chandar}, {Chastenet}, {Chevance}, {Chiang}, {Congiu}, {Dale}, {Deger}, {Eibensteiner}, {Faesi}, {Glover}, {Grasha}, {Groves}, {Hassani}, {Henny}, {Henshaw}, {Hoyer}, {Hughes}, {Jeffreson}, {Jim{\'e}nez-Donaire}, {Kim}, {Kim}, {Klessen}, {Koch}, {Kreckel}, {Kruijssen}, {Li}, {Liu}, {Lopez}, {Maschmann}, {Chen}, {Meidt}, {Murphy}, {Neumann}, {Neumayer}, {Pan}, {Pessa}, {Pety}, {Querejeta}, {Pinna}, {Rodr{\'\i}guez}, {Saito}, {S{\'a}nchez-Bl{\'a}zquez}, {Santoro}, {Sardone}, {Smith}, {Sormani}, {Scheuermann}, {Stuber}, {Sutter}, {Sun}, {Teng}, {Tre{\ss}}, {Usero}, {Watkins}, {Whitmore}, \& {Razza}}]{Lee2023}
{Lee}, J.~C., {Sandstrom}, K.~M., {Leroy}, A.~K., {et~al.} 2023, \apjl, 944, L17, \dodoi{10.3847/2041-8213/acaaae}

\bibitem[{{Lehmer} {et~al.}(2010){Lehmer}, {Alexander}, {Bauer}, {Brandt}, {Goulding}, {Jenkins}, {Ptak}, \& {Roberts}}]{Leh2010}
{Lehmer}, B.~D., {Alexander}, D.~M., {Bauer}, F.~E., {et~al.} 2010, \apj, 724, 559, \dodoi{10.1088/0004-637X/724/1/559}

\bibitem[{{Lehmer} {et~al.}(2022){Lehmer}, {Eufrasio}, {Basu-Zych}, {Garofali}, {Gilbertson}, {Mesinger}, \& {Yukita}}]{Leh2022}
{Lehmer}, B.~D., {Eufrasio}, R.~T., {Basu-Zych}, A., {et~al.} 2022, \apj, 930, 135, \dodoi{10.3847/1538-4357/ac63a7}

\bibitem[{{Lehmer} {et~al.}(2008){Lehmer}, {Brandt}, {Alexander}, {Bell}, {Hornschemeier}, {McIntosh}, {Bauer}, {Gilli}, {Mainieri}, {Schneider}, {Silverman}, {Steffen}, {Tozzi}, \& {Wolf}}]{Leh2008}
{Lehmer}, B.~D., {Brandt}, W.~N., {Alexander}, D.~M., {et~al.} 2008, \apj, 681, 1163, \dodoi{10.1086/588459}

\bibitem[{{Lehmer} {et~al.}(2012){Lehmer}, {Xue}, {Brandt}, {Alexander}, {Bauer}, {Brusa}, {Comastri}, {Gilli}, {Hornschemeier}, {Luo}, {Paolillo}, {Ptak}, {Shemmer}, {Schneider}, {Tozzi}, \& {Vignali}}]{Leh2012}
{Lehmer}, B.~D., {Xue}, Y.~Q., {Brandt}, W.~N., {et~al.} 2012, \apj, 752, 46, \dodoi{10.1088/0004-637X/752/1/46}

\bibitem[{{Lehmer} {et~al.}(2016){Lehmer}, {Basu-Zych}, {Mineo}, {Brandt}, {Eufrasio}, {Fragos}, {Hornschemeier}, {Luo}, {Xue}, {Bauer}, {Gilfanov}, {Ranalli}, {Schneider}, {Shemmer}, {Tozzi}, {Trump}, {Vignali}, {Wang}, {Yukita}, \& {Zezas}}]{Leh2016}
{Lehmer}, B.~D., {Basu-Zych}, A.~R., {Mineo}, S., {et~al.} 2016, \apj, 825, 7, \dodoi{10.3847/0004-637X/825/1/7}

\bibitem[{{Lehmer} {et~al.}(2017){Lehmer}, {Eufrasio}, {Markwardt}, {Zezas}, {Basu-Zych}, {Fragos}, {Hornschemeier}, {Ptak}, {Tzanavaris}, \& {Yukita}}]{Leh2017}
{Lehmer}, B.~D., {Eufrasio}, R.~T., {Markwardt}, L., {et~al.} 2017, \apj, 851, 11, \dodoi{10.3847/1538-4357/aa9578}

\bibitem[{{Lehmer} {et~al.}(2019){Lehmer}, {Eufrasio}, {Tzanavaris}, {Basu-Zych}, {Fragos}, {Prestwich}, {Yukita}, {Zezas}, {Hornschemeier}, \& {Ptak}}]{Leh2019}
{Lehmer}, B.~D., {Eufrasio}, R.~T., {Tzanavaris}, P., {et~al.} 2019, \apjs, 243, 3, \dodoi{10.3847/1538-4365/ab22a8}

\bibitem[{{Lehmer} {et~al.}(2020){Lehmer}, {Ferrell}, {Doore}, {Eufrasio}, {Monson}, {Alexander}, {Basu-Zych}, {Brandt}, {Sivakoff}, {Tzanavaris}, {Yukita}, {Fragos}, \& {Ptak}}]{Leh2020}
{Lehmer}, B.~D., {Ferrell}, A.~P., {Doore}, K., {et~al.} 2020, \apjs, 248, 31, \dodoi{10.3847/1538-4365/ab9175}

\bibitem[{{Lehmer} {et~al.}(2021){Lehmer}, {Eufrasio}, {Basu-Zych}, {Doore}, {Fragos}, {Garofali}, {Kovlakas}, {Williams}, {Zezas}, \& {Santana-Silva}}]{Leh2021}
{Lehmer}, B.~D., {Eufrasio}, R.~T., {Basu-Zych}, A., {et~al.} 2021, \apj, 907, 17, \dodoi{10.3847/1538-4357/abcec1}

\bibitem[{{Linden} {et~al.}(2010){Linden}, {Kalogera}, {Sepinsky}, {Prestwich}, {Zezas}, \& {Gallagher}}]{Lin2010}
{Linden}, T., {Kalogera}, V., {Sepinsky}, J.~F., {et~al.} 2010, \apj, 725, 1984, \dodoi{10.1088/0004-637X/725/2/1984}

\bibitem[{{Liu} {et~al.}(2024){Liu}, {Sartorio}, {Izzard}, \& {Fialkov}}]{Liu2024}
{Liu}, B., {Sartorio}, N.~S., {Izzard}, R.~G., \& {Fialkov}, A. 2024, \mnras, 527, 5023, \dodoi{10.1093/mnras/stad3475}

\bibitem[{{Long} {et~al.}(2014){Long}, {Kuntz}, {Blair}, {Godfrey}, {Plucinsky}, {Soria}, {Stockdale}, \& {Winkler}}]{Lon2014}
{Long}, K.~S., {Kuntz}, K.~D., {Blair}, W.~P., {et~al.} 2014, \apjs, 212, 21, \dodoi{10.1088/0067-0049/212/2/21}

\bibitem[{{Looser} {et~al.}(2024){Looser}, {D'Eugenio}, {Piotrowska}, {Belfiore}, {Maiolino}, {Cappellari}, {Baker}, \& {Tacchella}}]{Loo2024}
{Looser}, T.~J., {D'Eugenio}, F., {Piotrowska}, J.~M., {et~al.} 2024, arXiv e-prints, arXiv:2401.08769, \dodoi{10.48550/arXiv.2401.08769}

\bibitem[{{Madau} \& {Fragos}(2017)}]{Mad2017}
{Madau}, P., \& {Fragos}, T. 2017, \apj, 840, 39, \dodoi{10.3847/1538-4357/aa6af9}

\bibitem[{{Madden} {et~al.}(2013){Madden}, {R{\'e}my-Ruyer}, {Galametz}, {Cormier}, {Lebouteiller}, {Galliano}, {Hony}, {Bendo}, {Smith}, {Pohlen}, {Roussel}, {Sauvage}, {Wu}, {Sturm}, {Poglitsch}, {Contursi}, {Doublier}, {Baes}, {Barlow}, {Boselli}, {Boquien}, {Carlson}, {Ciesla}, {Cooray}, {Cortese}, {de Looze}, {Irwin}, {Isaak}, {Kamenetzky}, {Karczewski}, {Lu}, {MacHattie}, {O'Halloran}, {Parkin}, {Rangwala}, {Schirm}, {Schulz}, {Spinoglio}, {Vaccari}, {Wilson}, \& {Wozniak}}]{Mad2013}
{Madden}, S.~C., {R{\'e}my-Ruyer}, A., {Galametz}, M., {et~al.} 2013, \pasp, 125, 600, \dodoi{10.1086/671138}

\bibitem[{{Maiolino} \& {Mannucci}(2019)}]{Mai2019}
{Maiolino}, R., \& {Mannucci}, F. 2019, \aapr, 27, 3, \dodoi{10.1007/s00159-018-0112-2}

\bibitem[{{Mapelli} {et~al.}(2010){Mapelli}, {Ripamonti}, {Zampieri}, {Colpi}, \& {Bressan}}]{Map2010}
{Mapelli}, M., {Ripamonti}, E., {Zampieri}, L., {Colpi}, M., \& {Bressan}, A. 2010, \mnras, 408, 234, \dodoi{10.1111/j.1365-2966.2010.17048.x}

\bibitem[{{Markwardt}(2009)}]{Mar2009}
{Markwardt}, C.~B. 2009, in Astronomical Society of the Pacific Conference Series, Vol. 411, Astronomical Data Analysis Software and Systems XVIII, ed. D.~A. {Bohlender}, D.~{Durand}, \& P.~{Dowler}, 251, \dodoi{10.48550/arXiv.0902.2850}

\bibitem[{{McQuinn} {et~al.}(2016){McQuinn}, {Skillman}, {Dolphin}, {Berg}, \& {Kennicutt}}]{McQ2016}
{McQuinn}, K. B.~W., {Skillman}, E.~D., {Dolphin}, A.~E., {Berg}, D., \& {Kennicutt}, R. 2016, \aj, 152, 144, \dodoi{10.3847/0004-6256/152/5/144}

\bibitem[{{McQuinn} {et~al.}(2018){McQuinn}, {Skillman}, {Heilman}, {Mitchell}, \& {Kelley}}]{McQ2018}
{McQuinn}, K. B.~W., {Skillman}, E.~D., {Heilman}, T.~N., {Mitchell}, N.~P., \& {Kelley}, T. 2018, \mnras, 477, 3164, \dodoi{10.1093/mnras/sty839}

\bibitem[{{McQuinn} {et~al.}(2019){McQuinn}, {van Zee}, \& {Skillman}}]{McQ2019}
{McQuinn}, K. B.~W., {van Zee}, L., \& {Skillman}, E.~D. 2019, \apj, 886, 74, \dodoi{10.3847/1538-4357/ab4c37}

\bibitem[{{Mesinger} {et~al.}(2013){Mesinger}, {Ferrara}, \& {Spiegel}}]{Mes2013}
{Mesinger}, A., {Ferrara}, A., \& {Spiegel}, D.~S. 2013, \mnras, 431, 621, \dodoi{10.1093/mnras/stt198}

\bibitem[{{Mineo} {et~al.}(2012){Mineo}, {Gilfanov}, \& {Sunyaev}}]{Min2012a}
{Mineo}, S., {Gilfanov}, M., \& {Sunyaev}, R. 2012, \mnras, 419, 2095, \dodoi{10.1111/j.1365-2966.2011.19862.x}

\bibitem[{{Misra} {et~al.}(2023){Misra}, {Kovlakas}, {Fragos}, {Lazzarini}, {Bavera}, {Lehmer}, {Zezas}, {Zapartas}, {Xing}, {Andrews}, {Dotter}, {Rocha}, {Srivastava}, \& {Sun}}]{Mis2023}
{Misra}, D., {Kovlakas}, K., {Fragos}, T., {et~al.} 2023, \aap, 672, A99, \dodoi{10.1051/0004-6361/202244929}

\bibitem[{{Misra} {et~al.}(2024){Misra}, {Kovlakas}, {Fragos}, {Andrews}, {Bavera}, {Zapartas}, {Xing}, {Dotter}, {Rocha}, {Srivastava}, \& {Sun}}]{Mis2024}
---. 2024, \aap, 682, A69, \dodoi{10.1051/0004-6361/202347880}

\bibitem[{{Monreal-Ibero} {et~al.}(2012){Monreal-Ibero}, {Walsh}, \& {V{\'\i}lchez}}]{Mon2012}
{Monreal-Ibero}, A., {Walsh}, J.~R., \& {V{\'\i}lchez}, J.~M. 2012, \aap, 544, A60, \dodoi{10.1051/0004-6361/201219543}

\bibitem[{{Monson} {et~al.}(2023){Monson}, {Doore}, {Eufrasio}, {Lehmer}, {Alexander}, {Harrison}, {Kubo}, {Saez}, \& {Umehata}}]{Mon2023}
{Monson}, E.~B., {Doore}, K., {Eufrasio}, R.~T., {et~al.} 2023, \apj, 951, 15, \dodoi{10.3847/1538-4357/acd449}

\bibitem[{{Monson} {et~al.}(2024)}]{Mon2024}
{Monson}, E.~B., {et~al.} 2024, \apj, in-prep, 1.
\newblock \doarXiv{2400.0000}

\bibitem[{{Moustakas} \& {Kennicutt}(2006)}]{Mou2006}
{Moustakas}, J., \& {Kennicutt}, Robert~C., J. 2006, \apj, 651, 155, \dodoi{10.1086/507570}

\bibitem[{{Moustakas} {et~al.}(2010){Moustakas}, {Kennicutt}, {Tremonti}, {Dale}, {Smith}, \& {Calzetti}}]{Mou2010}
{Moustakas}, J., {Kennicutt}, Robert~C., J., {Tremonti}, C.~A., {et~al.} 2010, \apjs, 190, 233, \dodoi{10.1088/0067-0049/190/2/233}

\bibitem[{{Mu{\~n}oz} {et~al.}(2022){Mu{\~n}oz}, {Qin}, {Mesinger}, {Murray}, {Greig}, \& {Mason}}]{Mun2022}
{Mu{\~n}oz}, J.~B., {Qin}, Y., {Mesinger}, A., {et~al.} 2022, \mnras, 511, 3657, \dodoi{10.1093/mnras/stac185}

\bibitem[{{Nataf}(2015)}]{Nat2015}
{Nataf}, D.~M. 2015, \mnras, 449, 1171, \dodoi{10.1093/mnras/stv156}

\bibitem[{{Noll} {et~al.}(2009){Noll}, {Burgarella}, {Giovannoli}, {Buat}, {Marcillac}, \& {Mu{\~n}oz-Mateos}}]{Nol2009}
{Noll}, S., {Burgarella}, D., {Giovannoli}, E., {et~al.} 2009, \aap, 507, 1793, \dodoi{10.1051/0004-6361/200912497}

\bibitem[{{Olivier} {et~al.}(2021){Olivier}, {Berg}, {Chisholm}, {Erb}, {Pogge}, \& {Skillman}}]{Oli2021}
{Olivier}, G.~M., {Berg}, D.~A., {Chisholm}, J., {et~al.} 2021, arXiv e-prints, arXiv:2109.06725.
\newblock \doarXiv{2109.06725}

\bibitem[{{Pacucci} {et~al.}(2014){Pacucci}, {Mesinger}, {Mineo}, \& {Ferrara}}]{Pac2014}
{Pacucci}, F., {Mesinger}, A., {Mineo}, S., \& {Ferrara}, A. 2014, \mnras, 443, 678, \dodoi{10.1093/mnras/stu1240}

\bibitem[{{Panter} {et~al.}(2008){Panter}, {Jimenez}, {Heavens}, \& {Charlot}}]{Pan2008}
{Panter}, B., {Jimenez}, R., {Heavens}, A.~F., \& {Charlot}, S. 2008, \mnras, 391, 1117, \dodoi{10.1111/j.1365-2966.2008.13981.x}

\bibitem[{{Persic} \& {Rephaeli}(2007)}]{Per2007}
{Persic}, M., \& {Rephaeli}, Y. 2007, \aap, 463, 481, \dodoi{10.1051/0004-6361:20054146}

\bibitem[{{Pettini} \& {Pagel}(2004)}]{Pet2004}
{Pettini}, M., \& {Pagel}, B. E.~J. 2004, \mnras, 348, L59, \dodoi{10.1111/j.1365-2966.2004.07591.x}

\bibitem[{{Pilyugin} \& {Grebel}(2016)}]{Pil2016}
{Pilyugin}, L.~S., \& {Grebel}, E.~K. 2016, \mnras, 457, 3678, \dodoi{10.1093/mnras/stw238}

\bibitem[{{Pilyugin} {et~al.}(2014){Pilyugin}, {Grebel}, \& {Kniazev}}]{Pil2014}
{Pilyugin}, L.~S., {Grebel}, E.~K., \& {Kniazev}, A.~Y. 2014, \aj, 147, 131, \dodoi{10.1088/0004-6256/147/6/131}

\bibitem[{{Pilyugin} \& {Thuan}(2007)}]{Pil2007}
{Pilyugin}, L.~S., \& {Thuan}, T.~X. 2007, \apj, 669, 299, \dodoi{10.1086/521597}

\bibitem[{{Prestwich} {et~al.}(2013){Prestwich}, {Tsantaki}, {Zezas}, {Jackson}, {Roberts}, {Foltz}, {Linden}, \& {Kalogera}}]{Pre2013}
{Prestwich}, A.~H., {Tsantaki}, M., {Zezas}, A., {et~al.} 2013, \apj, 769, 92, \dodoi{10.1088/0004-637X/769/2/92}

\bibitem[{{Ranalli} {et~al.}(2003){Ranalli}, {Comastri}, \& {Setti}}]{Ran2003}
{Ranalli}, P., {Comastri}, A., \& {Setti}, G. 2003, \aap, 399, 39, \dodoi{10.1051/0004-6361:20021600}

\bibitem[{{Rangelov} {et~al.}(2012){Rangelov}, {Chandar}, {Prestwich}, \& {Whitmore}}]{Ran2012}
{Rangelov}, B., {Chandar}, R., {Prestwich}, A., \& {Whitmore}, B.~C. 2012, \apj, 758, 99, \dodoi{10.1088/0004-637X/758/2/99}

\bibitem[{{Rangelov} {et~al.}(2011){Rangelov}, {Prestwich}, \& {Chandar}}]{Ran2011}
{Rangelov}, B., {Prestwich}, A.~H., \& {Chandar}, R. 2011, \apj, 741, 86, \dodoi{10.1088/0004-637X/741/2/86}

\bibitem[{{Riley} {et~al.}(2022){Riley}, {Agrawal}, {Barrett}, {Boyett}, {Broekgaarden}, {Chattopadhyay}, {Gaebel}, {Gittins}, {Hirai}, {Howitt}, {Justham}, {Khandelwal}, {Kummer}, {Lau}, {Mandel}, {de Mink}, {Neijssel}, {Riley}, {van Son}, {Stevenson}, {Vigna-G{\'o}mez}, {Vinciguerra}, {Wagg}, {Willcox}, \& {Team Compas}}]{Ril2022}
{Riley}, J., {Agrawal}, P., {Barrett}, J.~W., {et~al.} 2022, \apjs, 258, 34, \dodoi{10.3847/1538-4365/ac416c}

\bibitem[{{Sabbi} {et~al.}(2018){Sabbi}, {Calzetti}, {Ubeda}, {Adamo}, {Cignoni}, {Thilker}, {Aloisi}, {Elmegreen}, {Elmegreen}, {Gouliermis}, {Grebel}, {Messa}, {Smith}, {Tosi}, {Dolphin}, {Andrews}, {Ashworth}, {Bright}, {Brown}, {Chandar}, {Christian}, {Clayton}, {Cook}, {Dale}, {de Mink}, {Dobbs}, {Evans}, {Fumagalli}, {Gallagher}, {Grasha}, {Herrero}, {Hunter}, {Johnson}, {Kahre}, {Kennicutt}, {Kim}, {Krumholz}, {Lee}, {Lennon}, {Martin}, {Nair}, {Nota}, {{\"O}stlin}, {Pellerin}, {Prieto}, {Regan}, {Ryon}, {Sacchi}, {Schaerer}, {Schiminovich}, {Shabani}, {Van Dyk}, {Walterbos}, {Whitmore}, \& {Wofford}}]{Sab2018}
{Sabbi}, E., {Calzetti}, D., {Ubeda}, L., {et~al.} 2018, \apjs, 235, 23, \dodoi{10.3847/1538-4365/aaa8e5}

\bibitem[{{Sacchi} {et~al.}(2016){Sacchi}, {Annibali}, {Cignoni}, {Aloisi}, {Sohn}, {Tosi}, {van der Marel}, {Grocholski}, \& {James}}]{Sac2016}
{Sacchi}, E., {Annibali}, F., {Cignoni}, M., {et~al.} 2016, \apj, 830, 3, \dodoi{10.3847/0004-637X/830/1/3}

\bibitem[{{Schaerer} {et~al.}(2019){Schaerer}, {Fragos}, \& {Izotov}}]{Sch2019}
{Schaerer}, D., {Fragos}, T., \& {Izotov}, Y.~I. 2019, \aap, 622, L10, \dodoi{10.1051/0004-6361/201935005}

\bibitem[{{Schlafly} \& {Finkbeiner}(2011)}]{Sch2011}
{Schlafly}, E.~F., \& {Finkbeiner}, D.~P. 2011, \apj, 737, 103, \dodoi{10.1088/0004-637X/737/2/103}

\bibitem[{{Schlegel} {et~al.}(1998){Schlegel}, {Finkbeiner}, \& {Davis}}]{Sch1998}
{Schlegel}, D.~J., {Finkbeiner}, D.~P., \& {Davis}, M. 1998, \apj, 500, 525, \dodoi{10.1086/305772}

\bibitem[{{Sell} {et~al.}(2011){Sell}, {Pooley}, {Zezas}, {Heinz}, {Homan}, \& {Lewin}}]{Sel2011}
{Sell}, P.~H., {Pooley}, D., {Zezas}, A., {et~al.} 2011, \apj, 735, 26, \dodoi{10.1088/0004-637X/735/1/26}

\bibitem[{{Senchyna} {et~al.}(2019){Senchyna}, {Stark}, {Chevallard}, {Charlot}, {Jones}, \& {Vidal-Garc{\'\i}a}}]{Sen2019}
{Senchyna}, P., {Stark}, D.~P., {Chevallard}, J., {et~al.} 2019, \mnras, 488, 3492, \dodoi{10.1093/mnras/stz1907}

\bibitem[{{Shi} {et~al.}(2005){Shi}, {Kong}, {Li}, \& {Cheng}}]{Shi2005}
{Shi}, F., {Kong}, X., {Li}, C., \& {Cheng}, F.~Z. 2005, \aap, 437, 849, \dodoi{10.1051/0004-6361:20041945}

\bibitem[{{Shtykovskiy} \& {Gilfanov}(2007)}]{Sht2007}
{Shtykovskiy}, P.~E., \& {Gilfanov}, M.~R. 2007, Astronomy Letters, 33, 437, \dodoi{10.1134/S106377370707002X}

\bibitem[{{Simmonds} {et~al.}(2021){Simmonds}, {Schaerer}, \& {Verhamme}}]{Sim2021}
{Simmonds}, C., {Schaerer}, D., \& {Verhamme}, A. 2021, \aap, 656, A127, \dodoi{10.1051/0004-6361/202141856}

\bibitem[{{Sivakoff} {et~al.}(2007){Sivakoff}, {Jord{\'a}n}, {Sarazin}, {Blakeslee}, {C{\^o}t{\'e}}, {Ferrarese}, {Juett}, {Mei}, \& {Peng}}]{Siv2007}
{Sivakoff}, G.~R., {Jord{\'a}n}, A., {Sarazin}, C.~L., {et~al.} 2007, \apj, 660, 1246, \dodoi{10.1086/513094}

\bibitem[{{Skillman} {et~al.}(2003){Skillman}, {C{\^o}t{\'e}}, \& {Miller}}]{Ski2003}
{Skillman}, E.~D., {C{\^o}t{\'e}}, S., \& {Miller}, B.~W. 2003, \aj, 125, 610, \dodoi{10.1086/345965}

\bibitem[{{Soria} \& {Wu}(2003)}]{Sor2003}
{Soria}, R., \& {Wu}, K. 2003, \aap, 410, 53, \dodoi{10.1051/0004-6361:20031074}

\bibitem[{{Soria} {et~al.}(2022){Soria}, {Kolehmainen}, {Graham}, {Swartz}, {Yukita}, {Motch}, {Jarrett}, {Miller-Jones}, {Plotkin}, {Maccarone}, {Ferrarese}, {Guest}, \& {Lan{\c{c}}on}}]{Sor2022}
{Soria}, R., {Kolehmainen}, M., {Graham}, A.~W., {et~al.} 2022, \mnras, 512, 3284, \dodoi{10.1093/mnras/stac148}

\bibitem[{{Spergel} {et~al.}(2003){Spergel}, {Verde}, {Peiris}, {Komatsu}, {Nolta}, {Bennett}, {Halpern}, {Hinshaw}, {Jarosik}, {Kogut}, {Limon}, {Meyer}, {Page}, {Tucker}, {Weiland}, {Wollack}, \& {Wright}}]{Spe2003}
{Spergel}, D.~N., {Verde}, L., {Peiris}, H.~V., {et~al.} 2003, \apjs, 148, 175, \dodoi{10.1086/377226}

\bibitem[{{Stark} {et~al.}(2017){Stark}, {Ellis}, {Charlot}, {Chevallard}, {Tang}, {Belli}, {Zitrin}, {Mainali}, {Gutkin}, {Vidal-Garc{\'\i}a}, {Bouwens}, \& {Oesch}}]{Sta2017}
{Stark}, D.~P., {Ellis}, R.~S., {Charlot}, S., {et~al.} 2017, \mnras, 464, 469, \dodoi{10.1093/mnras/stw2233}

\bibitem[{{Svoboda} {et~al.}(2019){Svoboda}, {Douna}, {Orlitov{\'a}}, \& {Ehle}}]{Svo2019}
{Svoboda}, J., {Douna}, V., {Orlitov{\'a}}, I., \& {Ehle}, M. 2019, \apj, 880, 144, \dodoi{10.3847/1538-4357/ab2b39}

\bibitem[{{Taddia} {et~al.}(2015){Taddia}, {Sollerman}, {Fremling}, {Pastorello}, {Leloudas}, {Fransson}, {Nyholm}, {Stritzinger}, {Ergon}, {Roy}, \& {Migotto}}]{Tad2015}
{Taddia}, F., {Sollerman}, J., {Fremling}, C., {et~al.} 2015, \aap, 580, A131, \dodoi{10.1051/0004-6361/201525989}

\bibitem[{{Tennant} {et~al.}(2001){Tennant}, {Wu}, {Ghosh}, {Kolodziejczak}, \& {Swartz}}]{Ten2001}
{Tennant}, A.~F., {Wu}, K., {Ghosh}, K.~K., {Kolodziejczak}, J.~J., \& {Swartz}, D.~A. 2001, \apjl, 549, L43, \dodoi{10.1086/319145}

\bibitem[{{Torres-Alb{\`a}} {et~al.}(2018){Torres-Alb{\`a}}, {Iwasawa}, {D{\'\i}az-Santos}, {Charmandaris}, {Ricci}, {Chu}, {Sanders}, {Armus}, {Barcos-Mu{\~n}oz}, {Evans}, {Howell}, {Inami}, {Linden}, {Medling}, {Privon}, {U}, \& {Yoon}}]{Tor2018}
{Torres-Alb{\`a}}, N., {Iwasawa}, K., {D{\'\i}az-Santos}, T., {et~al.} 2018, \aap, 620, A140, \dodoi{10.1051/0004-6361/201834105}

\bibitem[{{Tranin} {et~al.}(2022){Tranin}, {Godet}, {Webb}, \& {Primorac}}]{Tra2022}
{Tranin}, H., {Godet}, O., {Webb}, N., \& {Primorac}, D. 2022, \aap, 657, A138, \dodoi{10.1051/0004-6361/202141259}

\bibitem[{{T{\"u}llmann} {et~al.}(2011){T{\"u}llmann}, {Gaetz}, {Plucinsky}, {Kuntz}, {Williams}, {Pietsch}, {Haberl}, {Long}, {Blair}, {Sasaki}, {Winkler}, {Challis}, {Pannuti}, {Edgar}, {Helfand}, {Hughes}, {Kirshner}, {Mazeh}, \& {Shporer}}]{Tul2011}
{T{\"u}llmann}, R., {Gaetz}, T.~J., {Plucinsky}, P.~P., {et~al.} 2011, \apjs, 193, 31, \dodoi{10.1088/0067-0049/193/2/31}

\bibitem[{{Tully} {et~al.}(2013){Tully}, {Courtois}, {Dolphin}, {Fisher}, {H{\'e}raudeau}, {Jacobs}, {Karachentsev}, {Makarov}, {Makarova}, {Mitronova}, {Rizzi}, {Shaya}, {Sorce}, \& {Wu}}]{Tul2013}
{Tully}, R.~B., {Courtois}, H.~M., {Dolphin}, A.~E., {et~al.} 2013, \aj, 146, 86, \dodoi{10.1088/0004-6256/146/4/86}

\bibitem[{{Vladutescu-Zopp} {et~al.}(2023){Vladutescu-Zopp}, {Biffi}, \& {Dolag}}]{Vla2023}
{Vladutescu-Zopp}, S., {Biffi}, V., \& {Dolag}, K. 2023, \aap, 669, A34, \dodoi{10.1051/0004-6361/202244726}

\bibitem[{{Voss} \& {Gilfanov}(2007)}]{Vos2007}
{Voss}, R., \& {Gilfanov}, M. 2007, \mnras, 380, 1685, \dodoi{10.1111/j.1365-2966.2007.12223.x}

\bibitem[{{Vulic} {et~al.}(2021){Vulic}, {Hornschemeier}, {Haberl}, {Basu-Zych}, {Kyritsis}, {Zezas}, {Salvato}, {Ptak}, {Bogdan}, {Kovlakas}, {Wilms}, {Sasaki}, {Liu}, {Merloni}, {Dwelly}, {Brunner}, {Lamer}, {Maitra}, {Nandra}, \& {Santangelo}}]{Vul2021}
{Vulic}, N., {Hornschemeier}, A.~E., {Haberl}, F., {et~al.} 2021, arXiv e-prints, arXiv:2106.14526.
\newblock \doarXiv{2106.14526}

\bibitem[{{Walsh} \& {Roy}(1997)}]{Wal1997}
{Walsh}, J.~R., \& {Roy}, J.~R. 1997, \mnras, 288, 726, \dodoi{10.1093/mnras/288.3.726}

\bibitem[{{Walton} {et~al.}(2022){Walton}, {Mackenzie}, {Gully}, {Patel}, {Roberts}, {Earnshaw}, \& {Mateos}}]{Wal2022}
{Walton}, D.~J., {Mackenzie}, A.~D.~A., {Gully}, H., {et~al.} 2022, \mnras, 509, 1587, \dodoi{10.1093/mnras/stab3001}

\bibitem[{{Wang} {et~al.}(2024){Wang}, {Diaz}, {Kamieneski}, {Harrington}, {Yun}, {Foo}, {Frye}, {Jimenez-Andrade}, {Liu}, {Lowenthal}, {Pampliega}, {Pascale}, {Vishwas}, \& {Gurwell}}]{Wan2024}
{Wang}, Q.~D., {Diaz}, C.~G., {Kamieneski}, P.~S., {et~al.} 2024, \mnras, 527, 10584, \dodoi{10.1093/mnras/stad3827}

\bibitem[{{West} {et~al.}(2023){West}, {Garofali}, {Lehmer}, {Prestwich}, {Eufrasio}, {Luangtip}, {Roberts}, \& {Zezas}}]{Wes2023}
{West}, L., {Garofali}, K., {Lehmer}, B.~D., {et~al.} 2023, \apj, 952, 22, \dodoi{10.3847/1538-4357/acd9aa}

\bibitem[{{Wiktorowicz} {et~al.}(2017){Wiktorowicz}, {Sobolewska}, {Lasota}, \& {Belczynski}}]{Wik2017}
{Wiktorowicz}, G., {Sobolewska}, M., {Lasota}, J.-P., \& {Belczynski}, K. 2017, \apj, 846, 17, \dodoi{10.3847/1538-4357/aa821d}

\bibitem[{{Wiktorowicz} {et~al.}(2019){Wiktorowicz}, {Wyrzykowski}, {Chruslinska}, {Klencki}, {Rybicki}, \& {Belczynski}}]{Wik2019}
{Wiktorowicz}, G., {Wyrzykowski}, {\L}., {Chruslinska}, M., {et~al.} 2019, \apj, 885, 1, \dodoi{10.3847/1538-4357/ab45e6}

\bibitem[{{Wolter} {et~al.}(2018){Wolter}, {Fruscione}, \& {Mapelli}}]{Wol2018}
{Wolter}, A., {Fruscione}, A., \& {Mapelli}, M. 2018, \apj, 863, 43, \dodoi{10.3847/1538-4357/aacb34}

\bibitem[{{Wolter} {et~al.}(1999){Wolter}, {Trinchieri}, \& {Iovino}}]{Wol1999}
{Wolter}, A., {Trinchieri}, G., \& {Iovino}, A. 1999, \aap, 342, 41, \dodoi{10.48550/arXiv.astro-ph/9811157}

\bibitem[{{Wright} {et~al.}(2010){Wright}, {Eisenhardt}, {Mainzer}, {Ressler}, {Cutri}, {Jarrett}, {Kirkpatrick}, {Padgett}, {McMillan}, {Skrutskie}, {Stanford}, {Cohen}, {Walker}, {Mather}, {Leisawitz}, {Gautier}, {McLean}, {Benford}, {Lonsdale}, {Blain}, {Mendez}, {Irace}, {Duval}, {Liu}, {Royer}, {Heinrichsen}, {Howard}, {Shannon}, {Kendall}, {Walsh}, {Larsen}, {Cardon}, {Schick}, {Schwalm}, {Abid}, {Fabinsky}, {Naes}, \& {Tsai}}]{Wri2010}
{Wright}, E.~L., {Eisenhardt}, P. R.~M., {Mainzer}, A.~K., {et~al.} 2010, \aj, 140, 1868, \dodoi{10.1088/0004-6256/140/6/1868}

\bibitem[{{Xing} {et~al.}(2023){Xing}, {Bavera}, {Fragos}, {Kruckow}, {Rom{\'a}n-Garza}, {Andrews}, {Dotter}, {Kovlakas}, {Misra}, {Srivastava}, {Rocha}, {Sun}, \& {Zapartas}}]{Xin2023}
{Xing}, Z., {Bavera}, S.~S., {Fragos}, T., {et~al.} 2023, arXiv e-prints, arXiv:2309.09600, \dodoi{10.48550/arXiv.2309.09600}

\bibitem[{{Zapartas} {et~al.}(2021){Zapartas}, {Renzo}, {Fragos}, {Dotter}, {Andrews}, {Bavera}, {Coughlin}, {Misra}, {Kovlakas}, {Rom{\'a}n-Garza}, {Serra}, {Qin}, {Rocha}, {Tran}, \& {Xing}}]{Zap2021}
{Zapartas}, E., {Renzo}, M., {Fragos}, T., {et~al.} 2021, \aap, 656, L19, \dodoi{10.1051/0004-6361/202141506}

\bibitem[{{Zezas} {et~al.}(2007){Zezas}, {Fabbiano}, {Baldi}, {Schweizer}, {King}, {Rots}, \& {Ponman}}]{Zez2007}
{Zezas}, A., {Fabbiano}, G., {Baldi}, A., {et~al.} 2007, \apj, 661, 135, \dodoi{10.1086/513091}

\bibitem[{{Zezas} {et~al.}(2002){Zezas}, {Fabbiano}, {Rots}, \& {Murray}}]{Zez2002}
{Zezas}, A., {Fabbiano}, G., {Rots}, A.~H., \& {Murray}, S.~S. 2002, \apj, 577, 710, \dodoi{10.1086/342160}

\bibitem[{{Zhang} {et~al.}(2012){Zhang}, {Gilfanov}, \& {Bogd{\'a}n}}]{Zha2012}
{Zhang}, Z., {Gilfanov}, M., \& {Bogd{\'a}n}, {\'A}. 2012, \aap, 546, A36, \dodoi{10.1051/0004-6361/201219015}

\bibitem[{{Zhang} {et~al.}(2013){Zhang}, {Gilfanov}, \& {Bogd{\'a}n}}]{Zha2013}
---. 2013, \aap, 556, A9, \dodoi{10.1051/0004-6361/201220685}

\bibitem[{{Zuo} {et~al.}(2014){Zuo}, {Li}, \& {Gu}}]{Zuo2014}
{Zuo}, Z.-Y., {Li}, X.-D., \& {Gu}, Q.-S. 2014, \mnras, 437, 1187, \dodoi{10.1093/mnras/stt1918}

\end{thebibliography}
\bibliographystyle{aasjournal}

\appendix
\counterwithin{figure}{section}

\section{SED Fitting Result Comparison between \pegase\ and \bpass}\label{sec:appA}

As outlined in $\S$\ref{sub:sed}, we fit the FUV-to-FIR photometry of all galaxies in our sample using both \pegase\ and \bpass\ models.  Here we provide basic systematic comparisons of the fitting results between these model assumptions for the \ngal\ galaxies in this paper.

\begin{figure*}
\centerline{
\includegraphics[width=4.4cm]{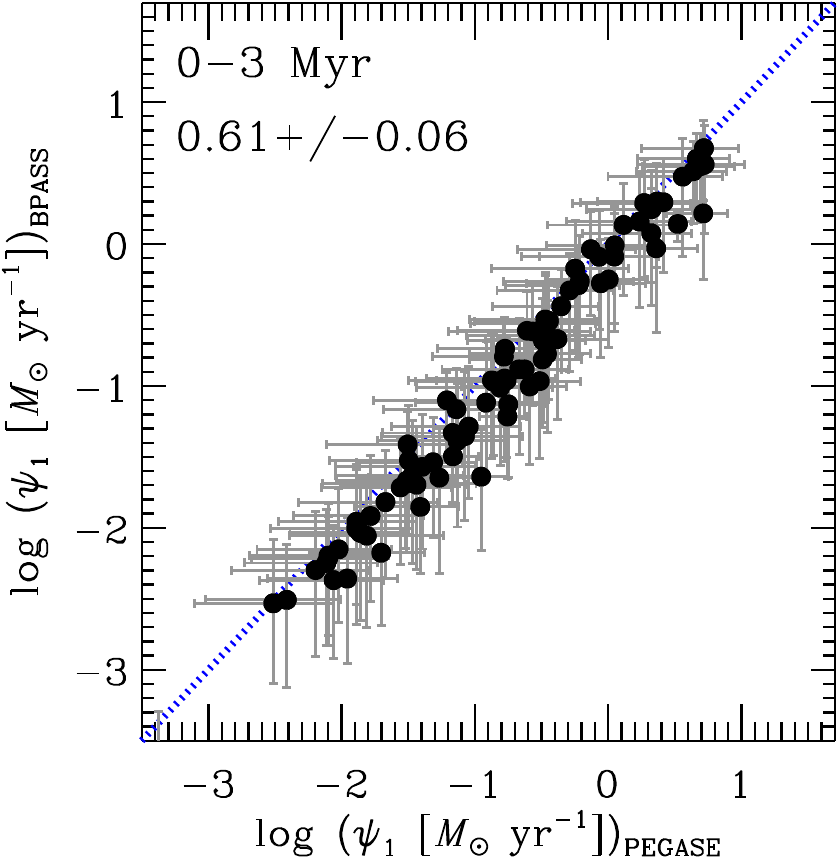}
\includegraphics[width=4.4cm]{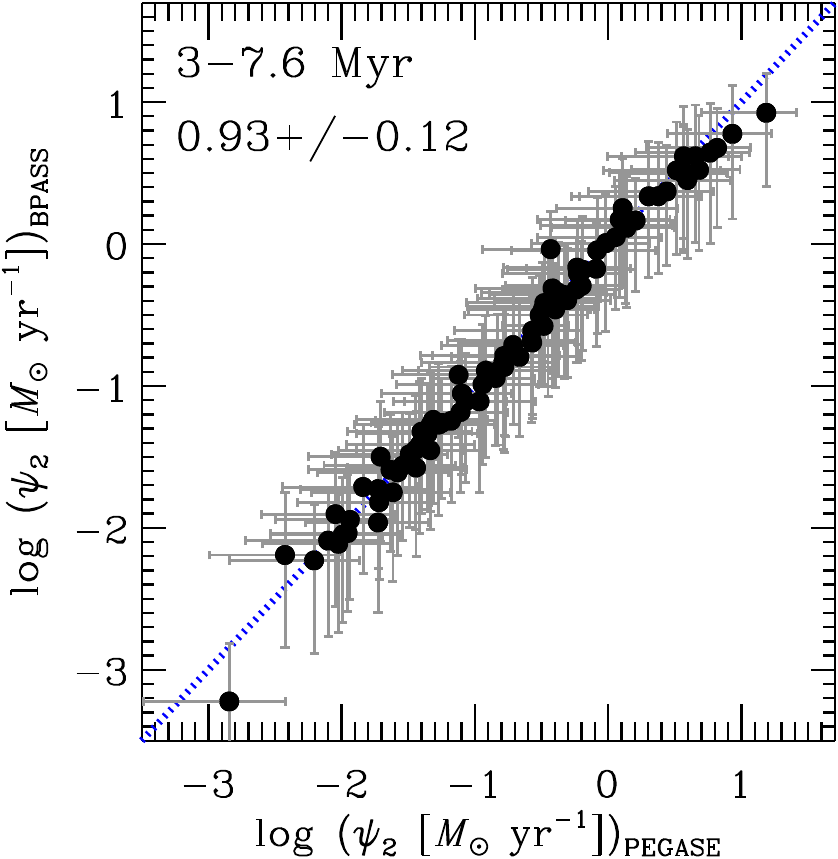}
\includegraphics[width=4.4cm]{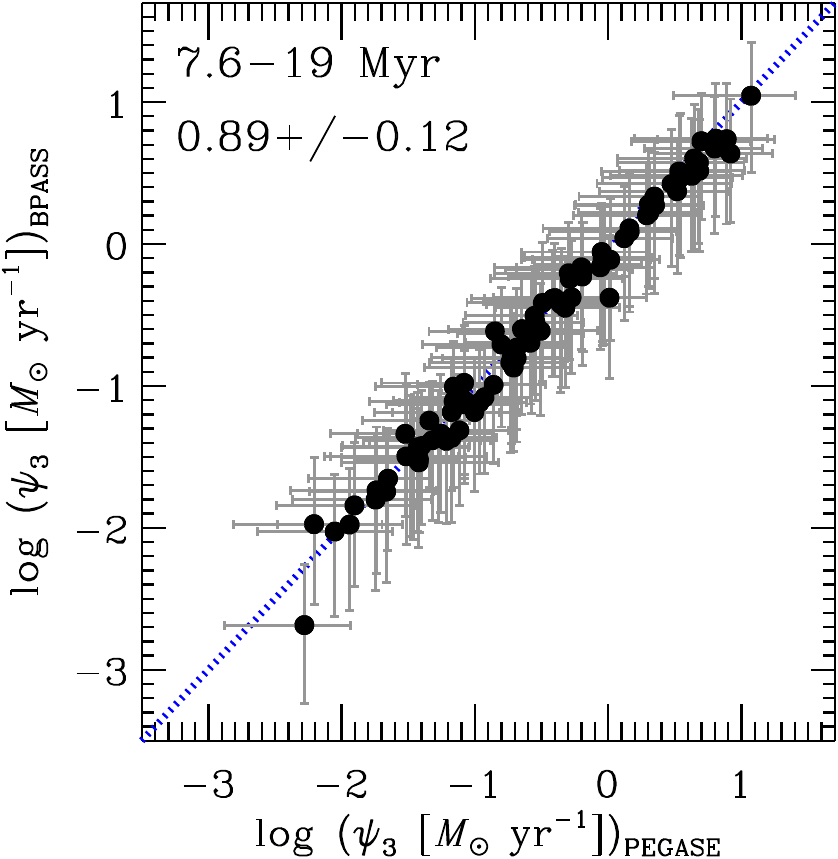}
\includegraphics[width=4.4cm]{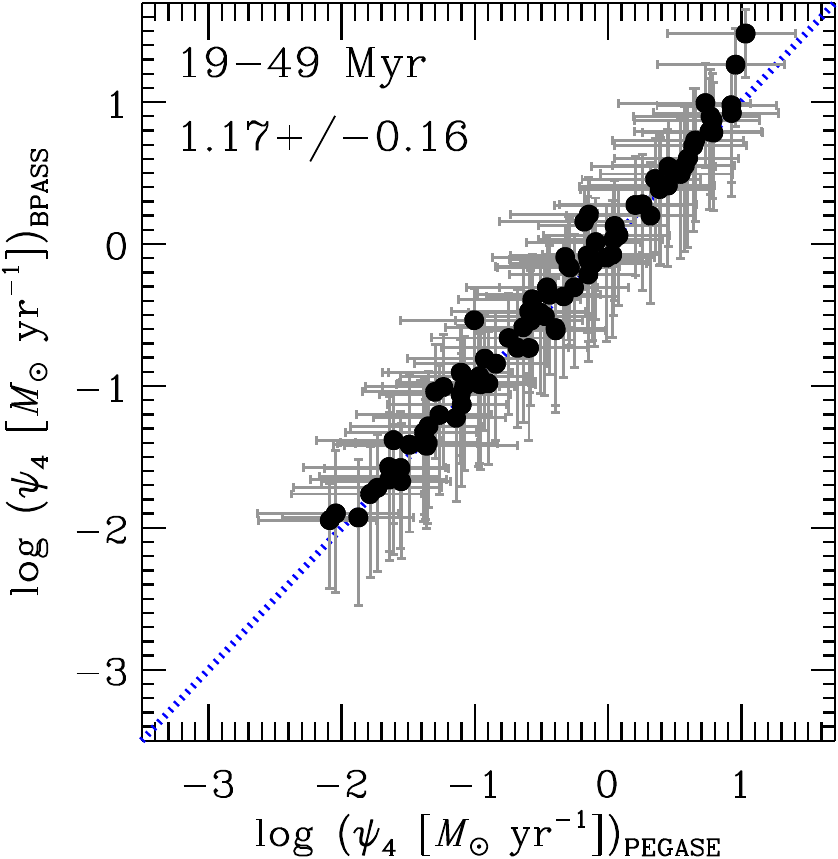}
}
\centerline{
\includegraphics[width=4.4cm]{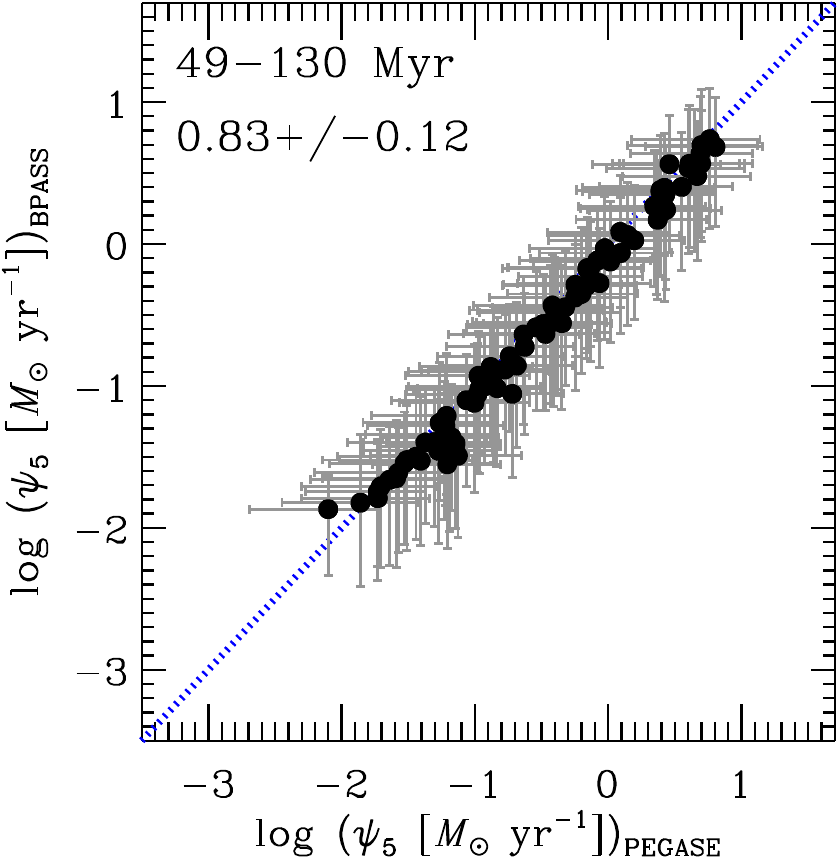}
\includegraphics[width=4.4cm]{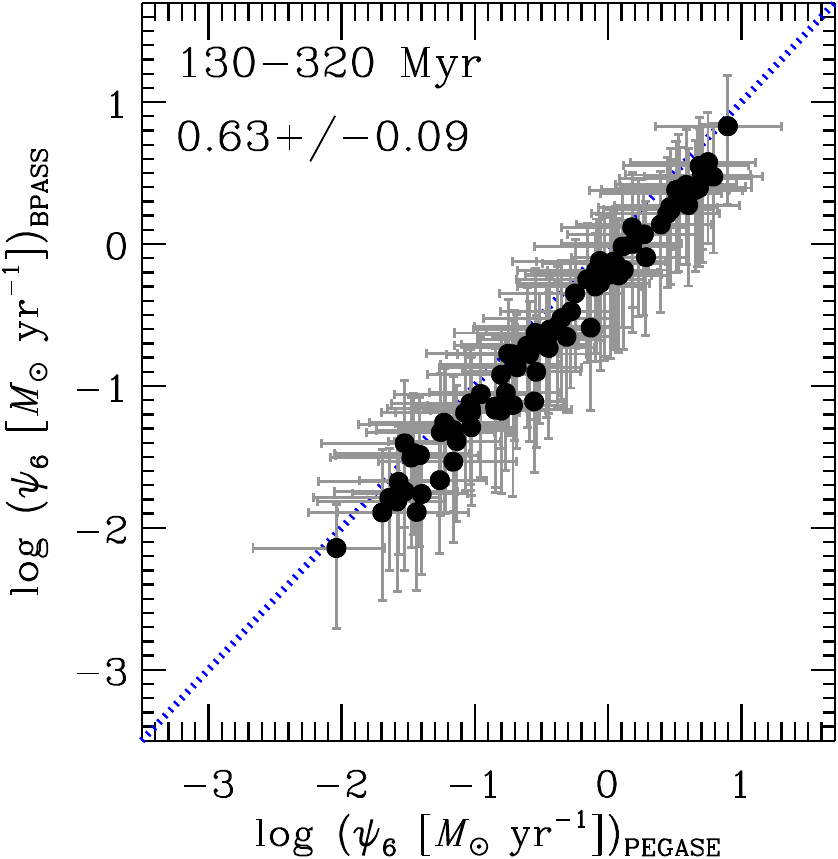}
\includegraphics[width=4.4cm]{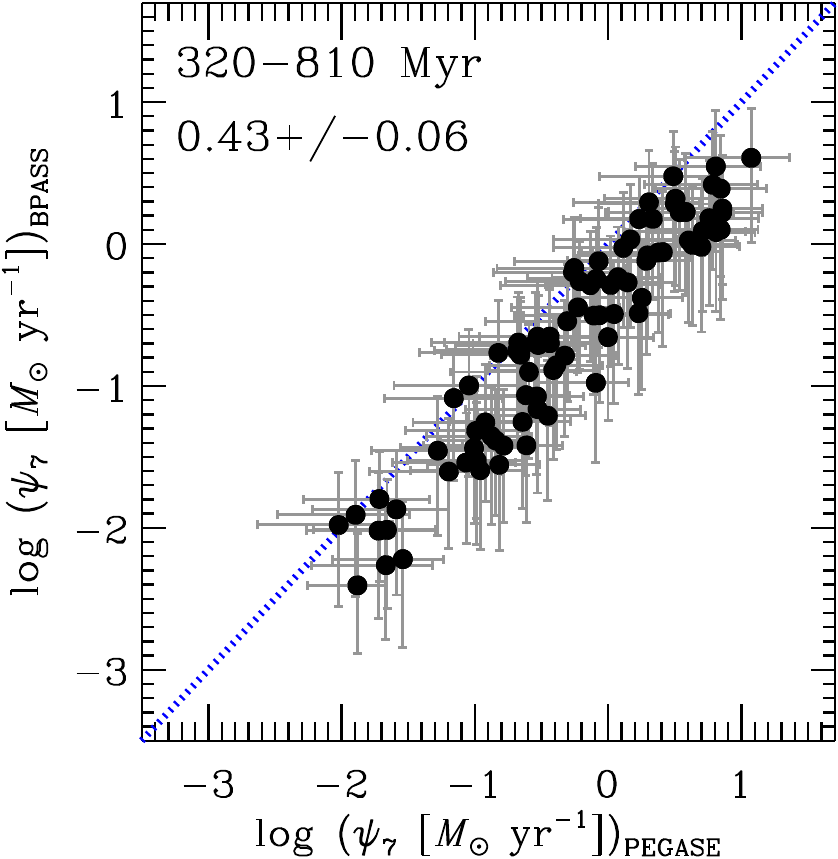}
\includegraphics[width=4.4cm]{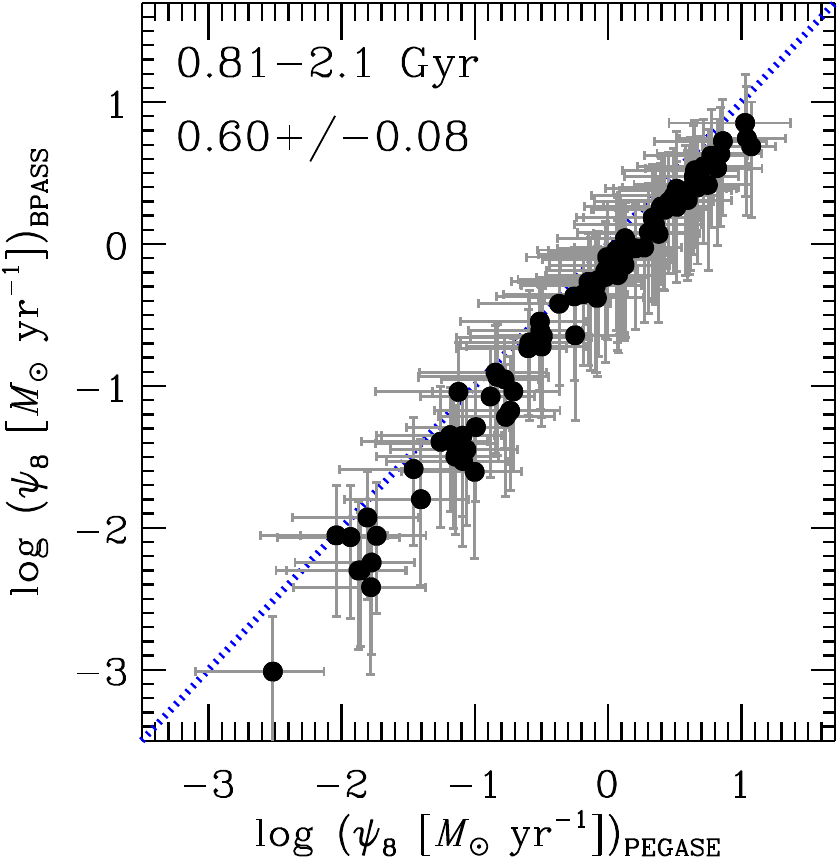}
}
\centerline{
\includegraphics[width=4.4cm]{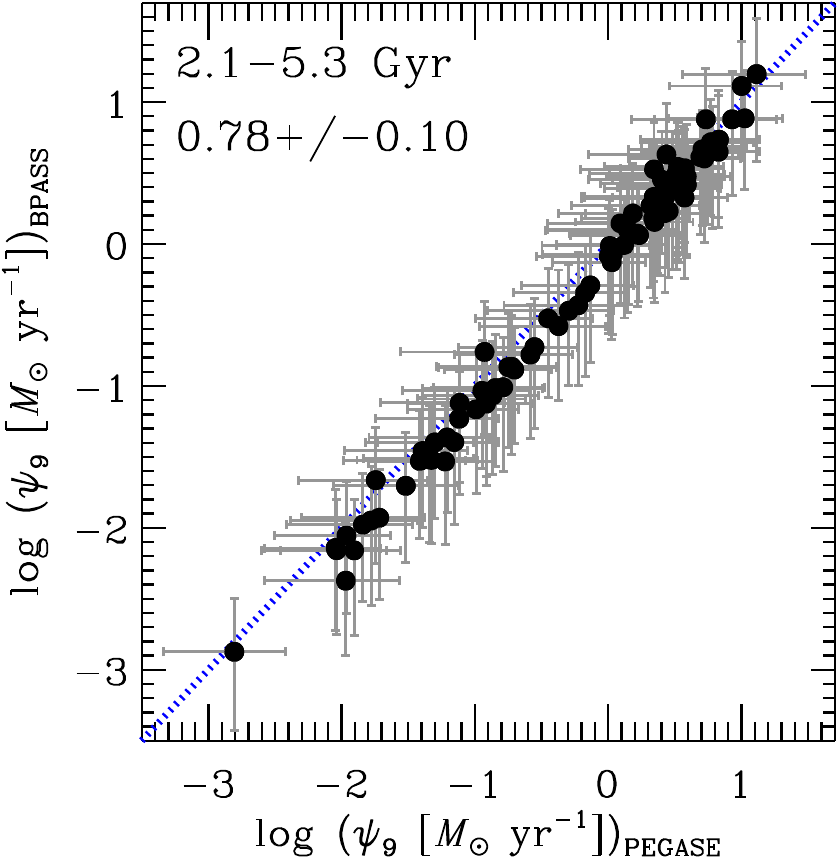}
\includegraphics[width=4.4cm]{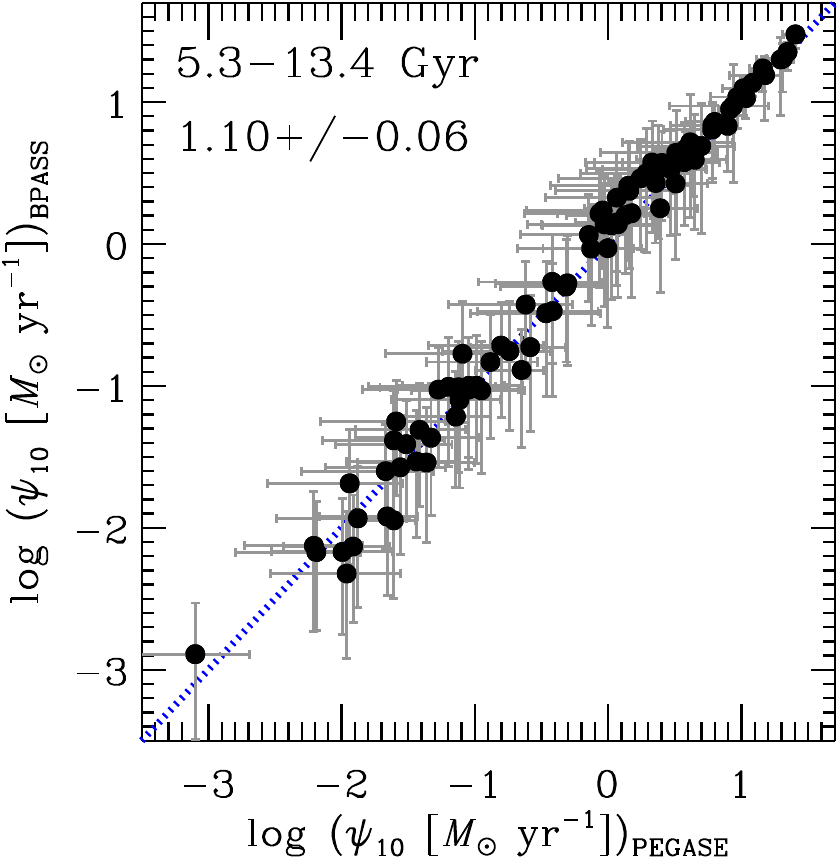}
\includegraphics[width=4.4cm]{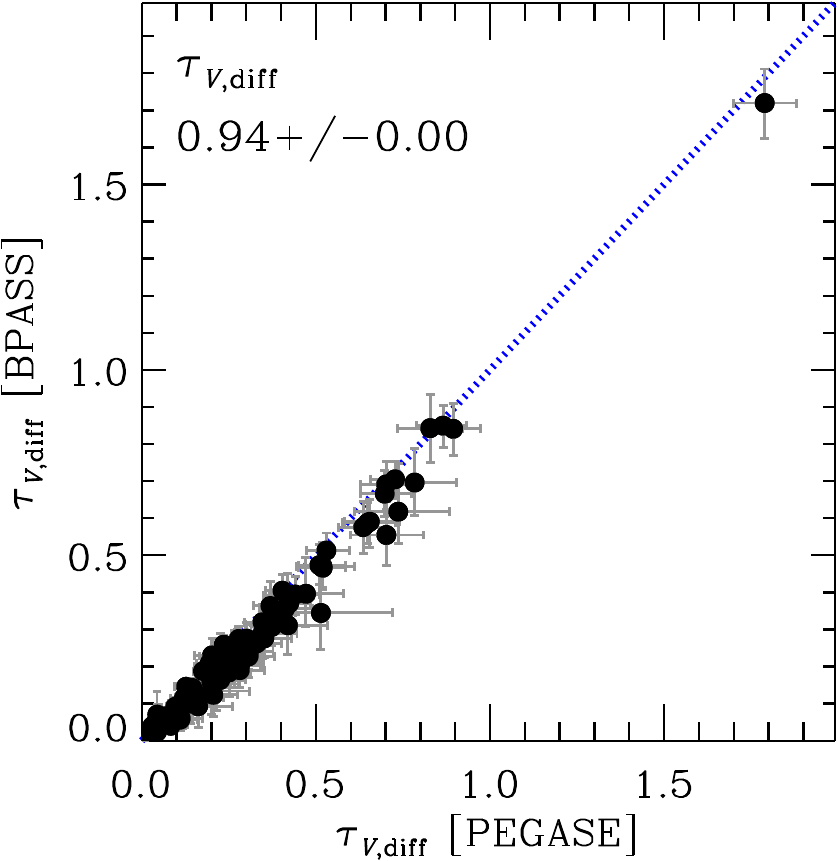}
\includegraphics[width=4.4cm]{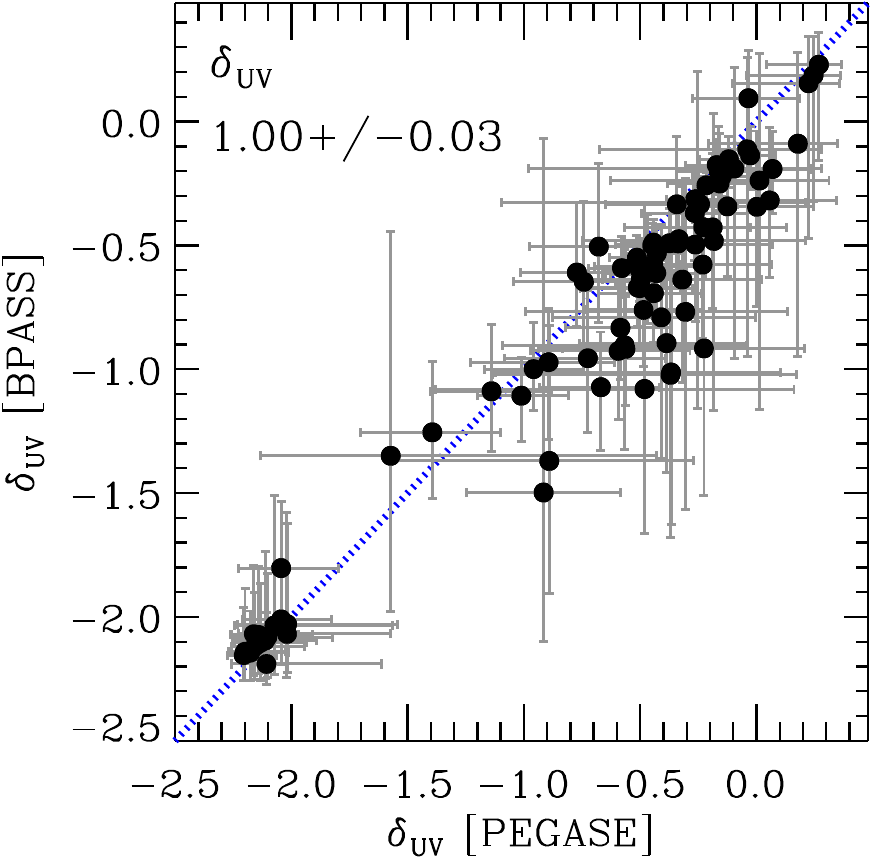}
}
\centerline{
\includegraphics[width=4.4cm]{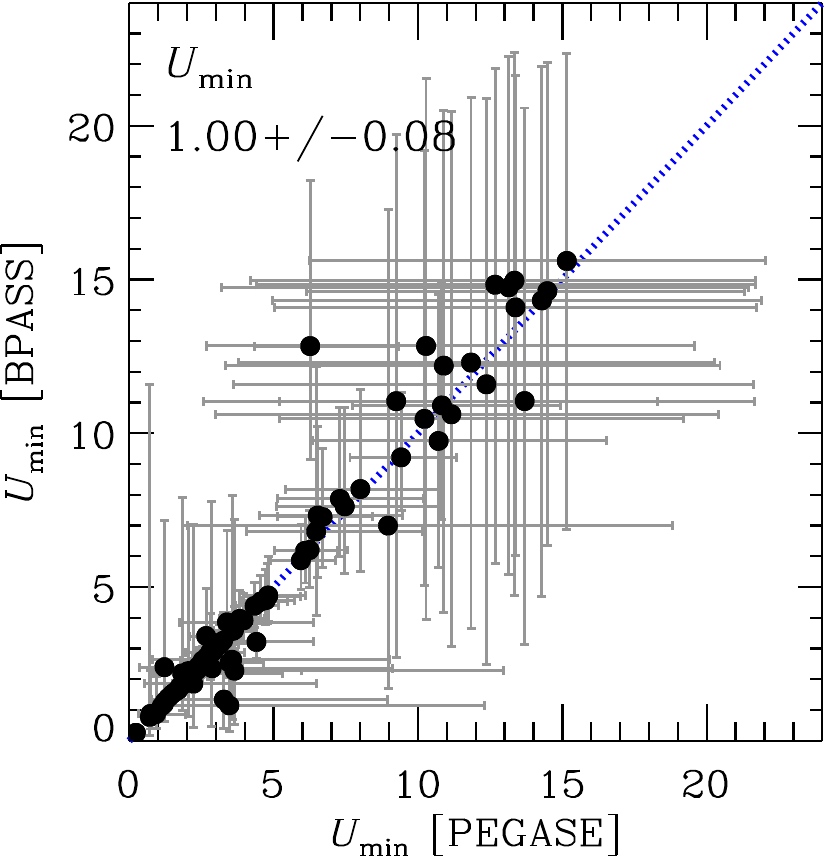}
\includegraphics[width=4.4cm]{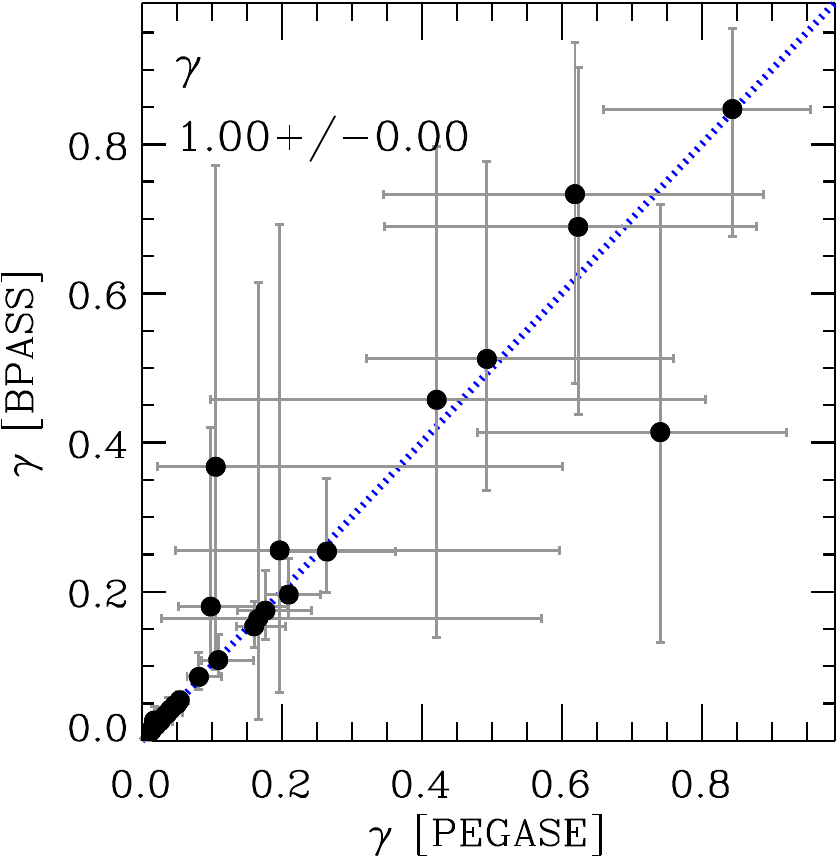}
\includegraphics[width=4.4cm]{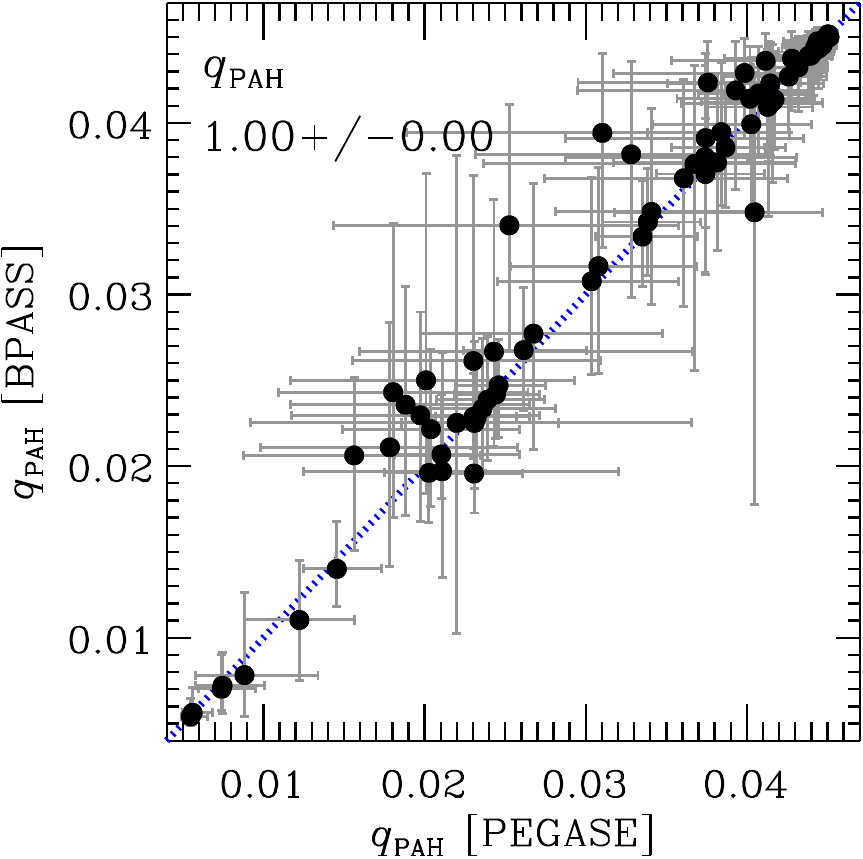}
\includegraphics[width=4.4cm]{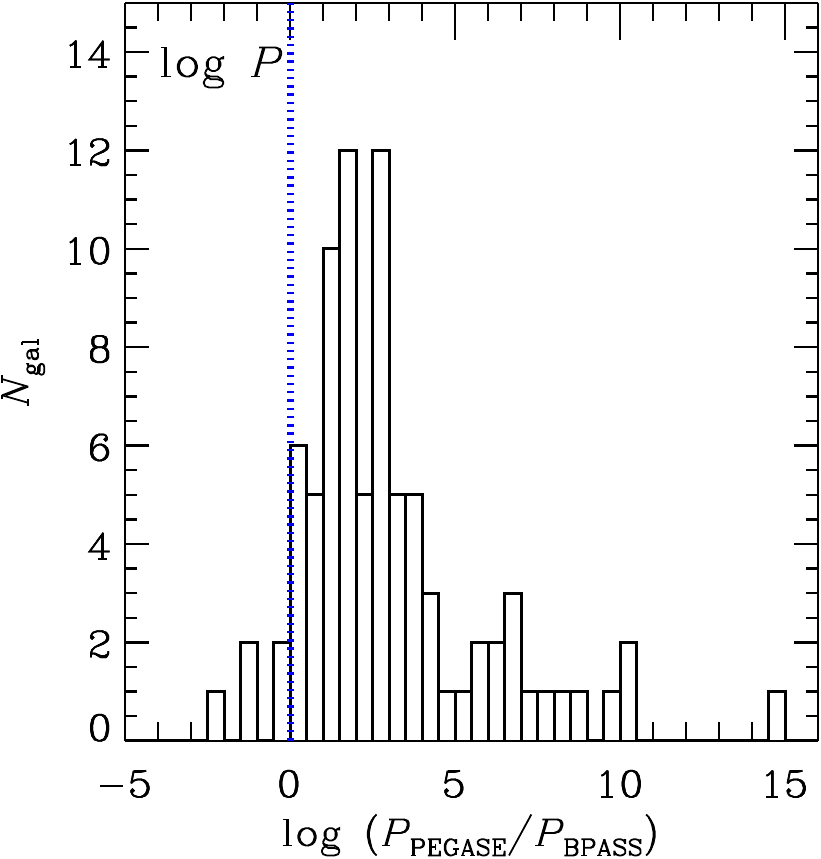}
}
\caption{
({\it first 15 panels\/}) Comparison of 15 derived parameters from SED fitting assuming \bpass\ (abscissa axes) and \pegase\ (ordinate axes).  Parameter names are annotated in the upper left-hand corner of each plot; parameter definitions can be found in Table~\ref{tab:sed} and $\S$\ref{sub:fit}. In each panel, we further annotate the mean parameter ratios (\bpass/\pegase) and their 1$\sigma$ uncertainties.  We find near unity agreement for several parameters with the exception of $\psi_1$, and $\psi_6$--$\psi_8$, which deviate by several sigma.
({\it final panel\/}) Distribution of the logarithm of the posterior probability ratio (\pegase/\bpass) for the full sample.  The distribution skews to values greater than zero, indicating that \pegase-based fits provide better characterizations of the SEDs in our sample.
}
\label{fig:bpasspeg}
\end{figure*}

Figure~\ref{fig:bpasspeg} shows graphical comparisons of the values for all 15 parameters used in our SED fitting procedure (see Table~\ref{tab:sed} for definitions of parameters and priors) for \pegase\ and \bpass.  Also, the lower-right panel of Figure~\ref{fig:bpasspeg} provides a histogram of the fit quality comparison, quantified as the ratio of the maximum posterior values from \pegase\ and \bpass, $\log (P_{\rm PEGASE}/P_{\rm BPASS})$.  Larger values of this ratio indicate better agreement between the best-fit models and the data.  

Overall, we find that the fit quality with the stellar population synthesis models of \pegase\ is statistically preferred over that of \bpass.  In each of the 15 parameter comparison plots in Figure~\ref{fig:bpasspeg}, we can see that the best-fit parameters are very similar between \pegase\ and \bpass, with many parameter comparisons being statistically indistinguishable from the overlaid one-to-one lines ({\it blue dotted lines in each panel\/}).  To clarify the comparisons, we have annotated in each panel the mean relation between \pegase\ and \bpass\ fit results, along with the errors on the means.  We find reasonable, near unity agreement for many parameters, with the exception of $\psi_1$ and $\psi_6$--$\psi_8$, which show statistically significant and systematic disagreements at the sample level.  

From the SEDs presented in Figure~\ref{fig:sed} (see extended materials for plots of all galaxies), comparison of best-fit models shows that the \bpass\ model emission lines are systematically stronger than those produced by the \pegase\ models.  This is primarily due to the elevated intrinsic extreme UV ($\lambda < 912$~\AA) continuum from binary populations that enhance the ionizing photon flux in \bpass.  From the residuals in Figure~\ref{fig:sed}, the most significant systematic differences in fit quality between \pegase\ and \bpass\ are the relatively low residuals (i.e., data values below model predictions) 1--4$\mu$m.  In this wavelength range specifically, \bpass\ predicts large contributions from AGB stars for the stellar populations formed at $\approx$0.13--2.1~Gyr (citations), which impact the shapes of the stellar SED models associated with $\psi_6$--$\psi_8$.  Thus, the disagreement between the \pegase\ and \bpass\ fits can be primarily attributed to the prescriptions for AGB stars.  We expect that these disagreements would persist for single-star \pegase/\bpass\ comparisons, and the binary-star aspect to the \bpass\ models yields no significant differences for the constraints used in this paper.  In the future, the combination of broad-band SED constraints and spectral constraints from emission lines would be helpful to further distinguish SFH comparisons between \pegase\ and \bpass, as the extreme UV ($<$912~\AA) and emission-line strengths from the \bpass\ models appear to be systematically stronger due to the binary aspect of these models (see, e.g., the best-fit models in Figure~\ref{fig:sed}).

%
%

We note
that the many of the data sets for our galaxies were processed in these previous
works and when we re-use those products here.  Galaxies unique to this paper were processed using the same procedures, but with more recent
calibrations, based on {\ttfamily CIAO} v.4.15 with {\ttfamily CALDB} v.4.10.7 and {\ttfamily Acis Extract} (hereafter, {\ttfamily AE}) v.2023aug14, which used {\ttfamily MARX} v.5.5.1.  In Table~\ref{tab:xray}, we provide a full list of \chandra\ ObsIDs and exposure times used in this work, and we detail the analyses of these data in the sections below.

\section{{\itshape Chandra} Data Analysis and Catalog Construction}\label{sec:appB}

\begin{table*}
\renewcommand\thetable{B1}
{\small
\begin{center}
\caption{\chandra\ Advanced CCD Imaging Spectrometer (ACIS) Observation Log}
\begin{tabular}{lcccccccc}
\hline\hline
& \multicolumn{2}{c}{\sc Aim Point} & {\sc Obs. Start} & {\sc Exposure}$^a$ & {\sc Flaring}$^b$ &  $\Delta \alpha$ & $\Delta \delta$  & {\sc Obs.} \\
\multicolumn{1}{c}{\sc Obs. ID} & $\alpha_{\rm J2000}$ & $\delta_{\rm J2000}$ & (UT) & (ks) & {\sc Intervals} & (arcsec) & (arcsec)  & {\sc Mode}$^c$ \\
\hline\hline
\multicolumn{9}{c}{{\bf NGC0024 }} \\
\hline
9547$^d$ & 00 09 57.79 & $-$24 57 47.44 & 2008-10-13T05:43:53 & 43 & \ldots & \ldots & \ldots & F \\
\hline
\multicolumn{9}{c}{{\bf NGC0337 }} \\
\hline
12979$^d$ & 00 59 49.29 & $-$07 34 28.15 & 2011-07-19T23:07:02 & 10 & \ldots & \ldots & \ldots & F \\
\hline
\multicolumn{9}{c}{{\bf NGC0584 }} \\
\hline
12175$^d$ & 01 31 20.38 & $-$06 51 38.45 & 2010-09-07T01:40:53 & 10 & \ldots & \ldots & \ldots & V \\
\hline
\multicolumn{9}{c}{{\bf NGC0625 }} \\
\hline
4746$^d$ & 01 35 03.91 & $-$41 26 12.60 & 2004-03-20T13:59:08 & 60 & \ldots & \ldots & \ldots & V \\
\hline
\multicolumn{9}{c}{{\bf NGC0628 }} \\
\hline
14801  & 01 36 47.41 & +15 45 32.58 & 2013-08-21T15:40:51 & 10 & \ldots & $+$0.05 & $+$0.01 & V \\
16000  & 01 36 47.37 & +15 45 31.61 & 2013-09-21T06:40:27 & 40 & \ldots & $+$0.56 & $-$0.24 & V \\
16001  & 01 36 47.39 & +15 45 29.57 & 2013-10-07T23:56:17 & 15 & \ldots & $+$0.24 & $-$0.07 & V \\
16002  & 01 36 48.85 & +15 45 26.66 & 2013-11-14T20:10:48 & 38 & \ldots & $+$0.08 & $+$0.16 & V \\
16003  & 01 36 48.89 & +15 45 28.36 & 2013-12-15T15:55:42 & 40 & \ldots & $+$0.04 & $-$0.11 & V \\
16484  & 01 36 47.38 & +15 45 29.36 & 2013-10-10T14:31:23 & 15 & \ldots & $+$0.45 & $+$0.14 & V \\
16485  & 01 36 47.39 & +15 45 29.44 & 2013-10-11T11:13:35 & 9 & \ldots & $+$0.32 & $+$0.06 & V \\
2057  & 01 36 40.35 & +15 48 17.73 & 2001-06-19T19:03:09 & 46 & 1, 0.5 & $-$0.05 & $-$0.05 & F \\
2058$^d$ & 01 36 36.11 & +15 46 51.99 & 2001-10-19T04:08:30 & 46 & \ldots & \ldots & \ldots & F \\
4753  & 01 36 51.21 & +15 45 12.44 & 2003-11-20T04:14:02 & 5 & \ldots & $-$0.10 & $-$0.03 & F \\
4754  & 01 36 51.51 & +15 45 12.89 & 2003-12-29T13:07:58 & 5 & \ldots & $+$0.09 & $+$0.07 & F \\
Merged$^e$  &01 36 44.82 & +15 46 11.67 & & 269 & 1, 0.5 &  \ldots & \ldots & \ldots \\
\hline
\end{tabular}
\label{tab:xray}
\end{center}
Note.---The full version of this table contains entries for all \ngal\ galaxies and 324 ObsIDs, and is available in the electronic edition.  An abbreviated version of the table is displayed here to illustrate its form and content.  \\
$^a$ All observations were continuous. These times have been corrected for removed data that were affected by high background.\\
$^b$ Number of flaring intervals and their combined duration.  These intervals were rejected from further analyses. \\
$^c$ The observing mode (F=Faint mode; V=Very Faint mode).\\
$^d$ Indicates Obs.~ID by which all other observations are reprojected to for alignment purposes.  This Obs.~ID was chosen for reprojection as it had the longest initial exposure time, before flaring intervals were removed.\\
$^e$ Aim point represents exposure-time weighted value.
}
\end{table*}

We made use of all ACIS observations with aim points that were offset by less
than 5~arcmin from the adopted coordinates of the galactic centers.  In Table~\ref{tab:xray}, we provide a detailed observational log of the \chandra\ observations that were used in our analyses.
Each observation was reprocessed using {\ttfamily chandra\_repro} script, bad
pixels and columns were identified and removed, and time intervals with high
background levels ($>$3$\sigma$ above the average level) were discarded.  When
more than one ObsID was available for a given galaxy, merged products (events lists, images, and exposure maps) were created and utilized.  Such ObsIDs were
co-aligned to the deepest ObsID using {\ttfamily wcs\_match} and {\ttfamily
wcs\_update} scripts, and then merged using the {\ttfamily merge\_obs} script.

Using the final events lists and astrometric solutions (merged when relevant,
single otherwise), we constructed images, exposure maps, and exposure-weighted
PSF maps with 90\% enclosed counts fractions (ECF).  We searched the 0.5--7~keV
images for point-sources using {\ttfamily wavdetect} at a false-positive
probability threshold of $1 \times 10^{-6}$ over seven wavelet scales from 1--8
pixels in a $\sqrt{2}$ sequence (i.e., 1, $\sqrt{2}$, 2, 2$\sqrt{2}$, 4,
4$\sqrt{2}$, and 8 pixels).  The {\ttfamily wavdetect} catalogs were then used
as input for more detailed point-source analyses performed using {\ttfamily AE}.
For each
point source and ObsID, {\ttfamily AE} performs detailed modeling of the local
PSF, using {\ttfamily MARX} ray-tracing simulations, identifies cases where
PSFs overlap, and extracts photometry and events while mitigating the effects
of source confusion.  {\ttfamily AE} further appropriately combines events
lists for unique sources when more than one ObsID is available, performs basic
spectral fitting using {\ttfamily xspec}, and culls the properties of the
sources into a source catalog.  In these procedures, {\ttfamily AE} properly tracks and combines the Redistribution Matrix Files (RMFs) and Auxiliary Response Files (ARFs) for all ObsIDs of a given data set.

We chose to perform {\ttfamily xspec} spectral fitting for point-sources with
$>$20 net counts. We adopted an absorbed power-law model that included both a fixed Galactic absorption component and a free variable intrinsic absorption
component ({\ttfamily TBABS $\times$ TBABS $\times$ POW} in {\ttfamily xspec}).
The free parameters include the intrinsic column density, $N_{\rm H, int}$, and photon index, $\Gamma$.  The Galactic absorption column, $N_{\rm H, Gal}$, for
each source was fixed to the value appropriate for the location of each galaxy,
as derived by Dickey \& Lockman~(1990).\footnote{Galactic column density values
were extracted using the {\ttfamily colden} tool at \url{http://cxc.harvard.edu/toolkit/colden.jsp}.}   In cases where the fits were
highly degenerate and not well constrained, we chose to fix $N_{\rm H, int} =
0$~cm$^{-2}$ and $\Gamma = 1.7$ while varying the power-law normalization.  Fluxes and luminosities were calculated for all sources based on their best-fit models.

For sources with $\simlt$20 net counts, we adopted {\ttfamily
wavdetect} net count rates (corrected for the ECF) converted to fluxes assuming
the average count-rate--to--flux ratio of the brighter \xray\ sources that had X-ray spectral fits available.  This choice was motivated by the fact that {\ttfamily AE} performs photometry using 90\% ECF apertures, that are often larger than the {\ttfamily wavdetect} extraction areas, which are based on wavelets of varying scales. This can sometimes lead {\ttfamily AE} to provide relatively low signal-to-noise ratio estimates and large uncertainties, when a small number of source counts are only detected from the core of the PSF. In general, we find good agreement between  {\ttfamily AE} and {\ttfamily wavdetect} source counts for sources with $>$20 net counts, except for sources in crowded areas, where  {\ttfamily AE} provides much more careful decomposition of source counts.

In Table~\ref{tab:cat}, we present the \xray\ point-source catalogs for the objects detected in our samples.  We provide X-ray point-source information for \nxps\ \xray\ detected point-sources, including \nx\ sources (Flag = 1) that are within the galactic footprints as defined in Table~\ref{tab:sam}.

\begin{table*}
\renewcommand\thetable{B2}
{\footnotesize
\begin{center}
\caption{X-ray point-source catalog and properties}
\label{tab:cat}
\begin{tabular}{llccccccccc}
\hline\hline
 &  &  &  & $\theta$ & $N_{\rm FB}$ & $N_{\rm H}$  & & $\log F_{\rm FB}$ & $\log L_{\rm FB}$  & Location \\
 \multicolumn{1}{c}{\sc Galaxy} & \multicolumn{1}{c}{\sc ID} & $\alpha_{\rm J2000}$  & $\delta_{\rm J2000}$ & (arcmin) & (counts) & ($10^{22}$~cm$^{-2}$) & $\Gamma$ & (\flux) & (\lum)  & Flag \\
 \multicolumn{1}{c}{(1)} & \multicolumn{1}{c}{(2)} & (3) & (4) & (5) & (6)--(7) & (8)--(9) & (10)--(11)  & (12) & (13) & (14) \\
\hline\hline
NGC0024 & 1 & 00 09 44.06 & $-$24 58 16.38 & 2.9 & 27.1$\pm$5.4 & 0.021 & 1.7 &  $-$14.2 & 37.6& 3 \\
  & 2 & 00 09 44.73 & $-$24 59 03.40 & 3.0 & 15.8$\pm$4.1 & 0.021 & 1.7 &  $-$14.5 & 37.3& 3 \\
  & 3 & 00 09 45.89 & $-$24 56 00.46 & 3.0 & 66.0$\pm$9.9 & 0.194$\pm$0.326 & 1.85$\pm$0.77 &  $-$13.9 & 37.9& 3 \\
  & 4 & 00 09 48.20 & $-$24 58 58.92 & 2.2 & 33.0$\pm$7.3 & 0.100$\pm$0.322 & 1.88$\pm$0.94 &  $-$14.2 & 37.6& 3 \\
  & 5 & 00 09 49.87 & $-$24 57 42.08 & 1.5 & 7.7$\pm$4.3 & 0.021 & 1.7 &  $-$14.8 & 37.0& 3 \\
\\
  & 6 & 00 09 50.25 & $-$25 00 02.24 & 2.7 & 5.3$\pm$2.4 & 0.021 & 1.7 &  $-$14.9 & 36.9& 3 \\
  & 7 & 00 09 51.27 & $-$24 59 28.45 & 2.1 & 9.8$\pm$4.6 & 0.021 & 1.7 &  $-$14.6 & 37.2& 3 \\
  & 8 & 00 09 53.62 & $-$24 58 32.01 & 1.0 & 36.0$\pm$7.5 & 0.226$\pm$0.484 & 1.68$\pm$0.97 &  $-$14.1 & 37.7& 1 \\
  & 9 & 00 09 54.63 & $-$24 56 57.60 & 0.9 & 23.0$\pm$6.3 & 1.657$\pm$4.077 & 0.98$\pm$1.76 &  $-$14.0 & 37.8& 3 \\
  & 10 & 00 09 54.85 & $-$24 57 58.96 & 0.4 & 6.6$\pm$2.6 & 0.021 & 1.7 &  $-$14.9 & 36.9& 1 \\
\\
  & 11 & 00 09 55.22 & $-$24 57 49.50 & 0.3 & 4.6$\pm$2.2 & 0.021 & 1.7 &  $-$15.0 & 36.8& 1 \\
  & 12 & 00 09 55.82 & $-$24 59 29.46 & 1.7 & 31.9$\pm$7.2 & 0.230$\pm$0.413 & $<$2.41 &  $-$14.3 & 37.5& 3 \\
  & 13 & 00 09 56.19 & $-$24 58 02.13 & 0.3 & 13.4$\pm$3.7 & 0.021 & 1.7 &  $-$14.6 & 37.2& 1 \\
  & 14 & 00 09 56.27 & $-$24 57 33.72 & 0.2 & 24.0$\pm$6.4 & 1.396$\pm$0.887 & $<$2.75 &  $-$14.3 & 37.5& 1 \\
  & 15 & 00 09 56.27 & $-$24 57 57.28 & 0.2 & 9.6$\pm$3.2 & 0.021 & 1.7 &  $-$14.7 & 37.1& 1 \\
\\
  & 16 & 00 09 57.31 & $-$24 57 42.01 & 0.2 & 25.4$\pm$5.1 & 0.021 & 1.7 &  $-$14.3 & 37.5& 1 \\
  & 17 & 00 09 58.90 & $-$24 56 57.15 & 1.0 & 8.5$\pm$3.0 & 0.021 & 1.7 &  $-$14.8 & 37.0& 1 \\
  & 18 & 00 10 00.94 & $-$24 57 27.96 & 1.0 & 19.1$\pm$5.9 & 0.021 & 1.7 &  $-$14.5 & 37.4& 3 \\
  & 19 & 00 10 03.29 & $-$24 57 30.24 & 1.6 & 19.5$\pm$5.9 & 0.021 & 1.7 &  $-$14.4 & 37.4& 3 \\
  & 20 & 00 10 03.50 & $-$24 55 28.14 & 2.8 & 17.0$\pm$5.7 & 0.021 & 1.7 &  $-$14.5 & 37.3& 3 \\
\\
\hline
\end{tabular}
\end{center}
\tablecomments{Col.(1): Name of host galaxy. Col.(2): point-source identification number within the galaxy. Col.(3) and (4): Right ascension and declination of the point source. Col.(5): Offset of the point source with respect to the average aim point of  the \chandra\ observations. Col.(6) and (7): 0.5--7~keV net counts (i.e., background subtracted) and 1$\sigma$ errors. Col.(8)--(9) and (10)--(11): Best-fit column density $N_{\rm H}$ and photon index $\Gamma$, respectively, along with their respective 90\% confidence uncertainties, based on spectral fits to an absorbed power-law model ({\ttfamily TBABS$_{\rm Gal} \times$ TBABS $\times$ POW} in {\ttfamily xspec}).  For sources with small numbers of counts ($<$20 net counts), we adopted only Galactic absorption appropriate for each galaxy and a photon index of $\Gamma = 1.7$.  Col.(12) and (13): the respective 0.5--8~keV flux and luminosity of the source. Col.(14): Flag indicating the location of the source within the galaxy.  Flag=1 indicates the source is within the galactic footprint adopted in Table~\ref{tab:sam}, and outside a central region of avoidance, if applicable.  All XLF calculations are based on Flag=1 sources.  Flag=2 indicates that the source is located in the central region of avoidance due to either the presence of an AGN or very high levels of source confusion.  Flag=3 indicates that the source is outside the galactic footprint of the galaxy. The full version of this table is available in the electronic edition and contains \nxps\ point-sources (rows), including all \nx\ sources that were used in our XLF analyses (i.e., Flag=1).  An abbreviated version of the table is displayed here to illustrate its form and content.}
}
\end{table*}

\end{document}